\pdfoutput=1

\documentclass[11pt,twoside,a4paper,cmspaper,final,collab]{cms-tdr}
\def\svnVersion{406189}\def\cmsCernNoTag{CERN-EP/2016-256}\def\cmsCernDate{\today}\def\cmsMessage{Published in the Journal of High Energy Physics as \href{http://dx.doi.org/10.1007/JHEP04(2017)022}{\doi{10.1007/JHEP04(2017)022.}}}

\begin{document}\cmsNoteHeader{SMP-14-013}

\hyphenation{had-ron-i-za-tion}
\hyphenation{cal-or-i-me-ter}
\hyphenation{de-vices}
 \RCS$Revision: 387237 $
\RCS$HeadURL: svn+ssh://svn.cern.ch/reps/tdr2/papers/SMP-14-013/trunk/SMP-14-013.tex $
\RCS$Id: SMP-14-013.tex 387237 2017-02-14 18:56:16Z bbilin $

\newlength\cmsFigWidth
\ifthenelse{\boolean{cms@external}}{\setlength\cmsFigWidth{0.85\columnwidth}}{\setlength\cmsFigWidth{0.4\textwidth}}
\ifthenelse{\boolean{cms@external}}{\providecommand{\cmsLeft}{top\xspace}}{\providecommand{\cmsLeft}{left\xspace}}
\ifthenelse{\boolean{cms@external}}{\providecommand{\cmsRight}{bottom\xspace}}{\providecommand{\cmsRight}{right\xspace}}
\providecommand{\cmsTable}[1]{\resizebox{\textwidth}{!}{#1}}

\providecommand{\MGaMC}{\textsc{mg5}\_a\textsc{mc}\xspace}
 \newcommand{\ydiff}{\ensuremath{y_{\text{diff}}}\xspace}
 \newcommand{\ysum}{\ensuremath{y_{\text{sum}}}\xspace}
\newcommand{\zjets}{\ensuremath{\cPZ+\text{jets}}\xspace}
\newcommand{\zlljets}{\ensuremath{\cPZ~(\to \ell\ell)+\text{jets}}\xspace}
 \newcommand{\alphasunc}[2][]{#1}
\newcommand{\plotstdcapt}{The lower panels show the ratios of the theoretical predictions to the measurements. Error bars around the experimental points show the statistical uncertainty, while the cross-hatched bands indicate the statistical and systematic uncertainties added in quadrature. The boxes around the \MGaMC + \PYTHIA~8 to measurement ratio represent the uncertainty on the prediction, including statistical, theoretical (from scale variations)\alphasunc{, $\alpha_S$}, and PDF uncertainties.  The  dark green area represents
the statistical and theoretical uncertainties only, while the light green area represents the statistical uncertainty alone.}

 \cmsNoteHeader{SMP-14-013}
 \title{Measurements of differential production cross sections for a Z boson in association with jets in pp collisions at $\sqrt{s} = 8\TeV$}

  \date{\today}

\abstract{Cross sections for the production of a Z boson in association with jets in proton-proton collisions at a centre-of-mass energy of $\sqrt{s}=8$\TeV are measured using a data sample collected by the CMS experiment at the LHC corresponding to 19.6\fbinv. Differential cross sections are presented as functions of up to three observables that describe the jet kinematics and the jet activity. Correlations between the azimuthal directions and the rapidities of the jets and
the Z boson are studied in detail.  The predictions of a number of multileg generators with leading or next-to-leading order accuracy are compared with the measurements. The comparison shows the importance of including multi-parton contributions in the matrix elements and the improvement in the predictions when next-to-leading order terms are included.}

  \hypersetup{%
pdfauthor={CMS Collaboration},%
pdftitle={Measurements of the differential production cross sections for a Z boson in association with jets in pp collisions at sqrt(s) = 8 TeV},%
pdfsubject={CMS},%
pdfkeywords={CMS, physics, vector boson, differential cross section, angular correlation, Z boson, LHC, 8 TeV}}

\maketitle

 \section{Introduction}

The high centre-of-mass energy of proton-proton (pp) collisions at the CERN LHC produces events with large jet transverse momenta (\pt) and high jet multiplicities in association with a $\cPZ/\gamma^{*}$ boson. For convenience $\cPZ/\gamma^{*}$ is denoted as $\cPZ$.  The selection of events in which the $\cPZ$ boson decays into two oppositely charged electrons or muons provides a signal sample that is not significantly contaminated by background
processes. This decay channel can be reconstructed with high efficiency due to the presence of charged leptons in the final state and is well suited for the validation of calculations within the framework of perturbative quantum chromodynamics (QCD). Furthermore, the production of massive vector bosons with jets is an important background to a number of standard model (SM) processes (single top and \ttbar production, vector boson fusion, $\PW\PW$ scattering, Higgs boson
production) as well as searches for physics beyond the SM. A good understanding of this background is vital to these searches and measurements.
Perturbative QCD calculations of the differential cross sections involve different powers
of the strong coupling constant $\alpha_s$ and different kinematic scales and are therefore
technically challenging.
The issue has been addressed over the last 15 years by merging processes with different parton multiplicities before the
parton showering, initially at tree level, and more recently with matrix elements calculated at next-to-leading order (NLO) using multileg matrix-element~(ME) event generators \cite{Hoeche:2012yf,Frederix:2012ps}.

In this paper we present measurements of the differential cross sections for $\cPZ$ boson production in association with jets at $\sqrt{s} = 8\TeV$, in the electron and muon decay channels,  using a data sample corresponding to an integrated luminosity of 19.6\fbinv.
Our measurements are compared with calculations obtained from different multileg ME event generators with leading order (LO) MEs (tree level), NLO MEs and a combination of NLO and LO MEs.  Measurements of the \zjets cross section were previously reported by the CDF and D0 Collaborations in proton-antiproton ($\Pp\Pap$) collisions at a centre-of-mass energy $\sqrt{s} = 1.96\TeV$ \cite{PhysRevLett.100.102001,Abazov:2008ez}. More recent results from proton-proton ($\Pp\Pp$) collisions at $\sqrt{s} = 7\TeV$ were published by the ATLAS \cite{Aad:2013ysa,atlasvjets} and CMS \cite{Chatrchyan:2011ne,Khachatryan:2014zya} Collaborations.

The cross sections are restricted to the phase space where the lepton transverse momenta are greater than $20\GeV$, their absolute pseudorapidities are less than $2.4$, and the dilepton mass is in the interval $91\pm20\GeV$.  The jets are defined using the infrared and collinear safe anti-$k_{\text{t}}$ algorithm applied to all visible particles; the radius parameter is set to 0.5 ~\cite{Cacciari:2008gp}. The
four-momenta of the particles are summed and therefore the jets can be massive. The differential cross sections include only those jets with transverse momentum greater than 30\GeV and further than $R=0.5$ from the leptons in the ($\eta,\phi$)-plane, where $\phi$ is the azimuthal angle. In addition, the absolute jet rapidity is required to be smaller than 2.4.
The jets are referred to as 1$^{\text{st}}$, 2$^{\text{nd}}$, 3$^{\text{rd}}$, etc. according to their transverse momenta, starting with the highest-\pt jet, and denoted as $\text{j}_1$, $\text{j}_2$, $\text{j}_3$, etc.
To further investigate the QCD dynamics towards low Bjorken-x values, multidimensional differential cross sections are measured for
$\cPZ\ +\geq 1$ jet production in an extended phase space with jet rapidities up to 4.7.
The extension of the rapidity coverage from 2.4 to 4.7 is used to tag events from
vector boson fusion (e.g., Higgs production).
Typically, the \zjets events constitute a background for such processes, and a good understanding of their production differential cross section including jets in the forward region is important.

For each jet multiplicity ($N_{\text{jets}}$) a number of measurements are made: the total cross section in the defined phase space, the differential cross sections as functions of the jet transverse momentum scalar sum $\HT$, and the differential cross sections as a function of the individual jet kinematics (transverse momentum \pt, and absolute rapidity $\abs{y}$). For the leading jet a double differential cross section is measured as a function of its absolute rapidity and transverse momentum.
Correlations in the jet kinematics are studied with one-dimensional and multidimensional differential cross section measurements, via 1) the distributions in the azimuthal angles between the $\cPZ$ boson and the leading jet and between the two leading jets, and 2) the rapidity distributions of the $\cPZ$ boson and the leading jet. These two rapidities are used as variables of a three-dimensional differential cross section measurement together with the transverse momentum of the jet. The Lorentz
boost along the beam axis introduces a large correlation amongst the $\cPZ$ boson and the jet rapidities. The two rapidities are combined to form a variable uncorrelated with the event boost along the beam axis, $y_{\text{diff}}=0.5\,|y(\cPZ)-y(\text{j}_i)|$ and a highly boost-dependent variable, $y_{\text{sum}}=0.5\,|y(\cPZ)+y(\text{j}_i)|$. The cross section is measured separately as a function of each of these variables. The distribution of $y_{\text{diff}}$ is mostly
sensitive to the parton scattering, while $y_{\text{sum}}$ is expected to be sensitive to the parton distribution functions (PDFs) of the proton.

The Drell--Yan process, where the $\cPZ$ boson can decay into a pair of neutrinos, is a background to searches for new phenomena, such as dark matter, supersymmetry and other theories beyond the SM that predict the presence of invisible particle(s) in the final state. It is particularly important when the $\cPZ$ boson has a large transverse momentum, leading to a large missing transverse energy. The azimuthal angle between the jets is a good handle to suppress backgrounds
coming from QCD multijet events, while the \HT variable can be used to select events with large jet activity. For such analyses it is important to have a good model of \zjets production and therefore a good understanding of these observables. This motivates the measurement of the distributions of the azimuthal angles between the jets and between the $\cPZ$ boson and the jets for different thresholds applied to the $\cPZ$ boson transverse momentum, the \HT variable, and the jet multiplicity.
 These angles can be measured with high precision, and thus provide an excellent avenue to test the accuracy of SM predictions~\cite{Chatrchyan:2013tna}. The dijet mass is an essential observable in the selection of Higgs boson events produced by vector boson fusion and it is important to model well both this process and its backgrounds. This observable is measured in \zjets events for the two leading jets.

Section~\ref{sec:cms_samples} describes the experimental setup and the data samples used for the measurements, while the object reconstruction and the event selection are presented in Section~\ref{sec:eventSelection}. Section~\ref{UnfoldSec} is dedicated to the subtraction of the background contribution and the correction of the detector response, and Section~\ref{SysSec} to the estimation of the measurement uncertainties. Finally, the results are presented and compared to different theoretical predictions in Section~\ref{sec:Results} and summarised in Section~\ref{sec:summary}.

 \section{The CMS detector, simulation, and data samples}\label{sec:cms_samples}\label{sec:samples}
 The central feature of the CMS apparatus is a superconducting solenoid of 6\unit{m} internal diameter, providing a magnetic field of 3.8\unit{T}. Within the solenoid volume are a silicon pixel and strip tracker covering the range $\abs{\eta}<2.5$ together with a calorimeter covering the  range $\abs{\eta}<3$. The latter consists of a lead tungstate crystal electromagnetic calorimeter (ECAL), and a brass and scintillator hadron calorimeter (HCAL). The HCAL is complemented by an outer calorimeter placed outside the solenoid used to measure the tails of hadron showers. The pseudorapidity coverage is extended up to $\abs{\eta} = 5.2$ by a forward hadron calorimeter built using radiation-hard technology. Gas-ionization detectors exploiting three technologies, drift tubes, cathode strip chambers, and resistive-plate chambers, are embedded in the steel flux-return yoke outside the solenoid and constitute the muon system, used to identify and reconstruct muons over the range $\abs{\eta} < 2.4$. A more detailed description of the CMS detector, together with a definition of the coordinate system used and the relevant kinematic variables, can be found in Ref.~\cite{Chatrchyan:2008zzk}.

Simulated events are used to both subtract the contribution from background processes and to correct for the detector response. The signal and the background (from $\PW\PW$, $\PW\cPZ$, $\cPZ\cPZ$, \ttbar, and single top quark processes) are modelled with the tree-level matrix element event generator \MADGRAPH~v5.1.3.30~\cite{Alwall:2011uj} interfaced with \PYTHIA~6.4.26~\cite{Sjostrand:2006za}. The PDF CTEQ6L1~\cite{Pumplin:2002vw} and the Z2* \PYTHIA~6
tune~\cite{Chatrchyan:2013gfi,Khachatryan:2015pea} are used. For the ME calculations, $\alpha_s$ is set to $0.130$ at the $\cPZ$ boson mass scale. The five processes $\Pp\Pp \to \cPZ\ +\ N_{\text{jets}}\ \text{jets}$,
    $N_{\text{jets}} = 0,\dots,4$, are included in the ME calculations. The $k_t-$MLM~\cite{Alwall:2007fs,Alwall:2008qv} scheme with the merging scale set to $20\GeV$ is used for the matching of the parton showering (PS) with the ME calculations. The same setup is used to estimate the background from $\zjets\to \Pgt^{+}\Pgt^{-} +\ \text{jets}$. The signal sample is normalised to the inclusive cross section calculated at next-to-next-to-leading order (NNLO) with {\FEWZ} 2.0~\cite{Gavin:2010az}
using the CTEQ6M PDF set~\cite{Pumplin:2002vw}.  Samples of $\PW\PW$, $\PW\cPZ$, $\cPZ\cPZ$ events are normalised to the inclusive cross section calculated at NLO using the \MCFM~6.6~\cite{Campbell:2010ff} generator. Finally, an NNLO plus next-to-next-to-leading log (NNLO + NNLL)
calculation~\cite{Czakon:2013goa} is used for the normalisation of the \ttbar sample. When comparing the measurements with the predictions from theory, several other event generators are used for the Drell--Yan process. Those, which are not used for the measurement itself, are described in Section~\ref{sec:Results}.

The detector response is simulated with \GEANTfour~\cite{Allison:2006ve}. The events reconstructed by the detector contain several superimposed proton-proton interactions, including one interaction with a high \pt track that passes the trigger requirements. The majority of interactions, which do not pass trigger requirements, typically produce low energy (soft) particles because of the larger cross section for these soft events. The effect of this superposition of interactions is denoted as pileup. The samples of simulated events are generated with a distribution of the number of proton-proton
interactions per beam bunch crossing close to the one observed in data. The number of pileup interactions, averaging around 20, varies with the beam conditions. The correct description of
pileup is ensured by reweighting the simulated sample to match the measured distribution of pileup interactions.

 \section{Event reconstruction and selection}\label{sec:eventSelection}

Events with at least two leptons (electrons or muons) are selected. The trigger accepts events with two isolated electrons (muons) with a \pt of at least 8  and $17\GeV$. After reconstruction these leptons are restricted to a kinematic and geometric acceptance of $\pt > 20\GeV$ and $\abs{\eta} < 2.4$. We require that the oppositely charged, same-flavor leptons form a pair with an invariant mass within a window of $91\pm20\GeV$. The ECAL barrel-endcap transition
region $1.444 <\abs{\eta}< 1.566$ is excluded for the electrons. The acceptance is extended to the full $\abs{\eta} < 2.4$ region when correcting for the detector response.

Information from all detectors is combined  using the particle-flow (PF) algorithm~\cite{CMS:2009nxa,CMS:2010byl} to produce an event consisting of reconstructed particle candidates. The PF candidates are then used to build jets and calculate lepton isolation. The quadratic sum of transverse momenta of the tracks associated to the reconstructed vertices is used to select the primary vertex (PV) of interest. Because pileup involves typically soft particles, the PV with the highest sum is chosen.

The electrons are reconstructed with the algorithm described in Ref.~\cite{Khachatryan:2015hwa}. Identification criteria based on the electromagnetic shower shape and the energy sharing between ECAL and HCAL are applied. The momentum is estimated by combining the energy measurement in the ECAL with the momentum measurement in the tracker. For each electron candidate, an isolation variable, quantifying the energy flow in the vicinity of its trajectory, is built by summing the transverse momenta of the PF candidates within a cone of size $\DR = \sqrt{\smash[b]{(\Delta\phi)^{2}+(\Delta\eta)^{2}}} = 0.3$, excluding the electron itself and the charged particles not compatible with the primary event vertex. This sum is affected by neutral particles from pileup events, which cannot be rejected with a vertex criterion. An average energy density per unit of $\DR$ is calculated event by event using the method introduced in Ref.~\cite{Cacciari:2007fd} and used to estimate and subtract the neutral particle contribution. The electron is considered isolated if the isolation variable value is less than 15\% of the transverse momentum of the electron.  The electron candidates are required to be consistent with a particle originating from the PV in the event.

Muon candidates are matched to tracks measured in the tracker, and they are required to satisfy the identification and quality criteria described in Ref.~\cite{Chatrchyan:2012xi} that are based on the number of tracker hits and the response of the muon detectors.
 The background from cosmic ray muons, which appear as two back-to-back muons, is reduced by criteria on
the impact
parameter and  by requiring that the muon pairs have an acollinearity larger than 2.5\unit{mrad}. An isolation variable is defined that is similar to that for
electrons, but with a cone size $\DR = 0.4$ and a different approach to the subtraction of the contribution from neutral pileup particles. For the muons this contribution is estimated from the sum of the transverse momenta of the charged particles rejected by the vertex requirement, considered as coming from pileup. This sum is multiplied by a factor of 0.5 to take into account the relative fraction of neutral and charged particles. A muon is considered isolated if the isolation variable value is below 20\% of its transverse momentum.

The efficiencies for the lepton trigger, reconstruction, and identification are measured with the ``tag-and-probe" method~\cite{CMS:2011aa}. The simulation is corrected using the ratios of the efficiencies obtained in the data sample to those obtained in the simulated sample. These scale factors for lepton reconstruction and identification typically range from 0.95 to 1.05 depending on the lepton transverse momentum and rapidity. The overall efficiency of trigger
and event selection is 58\% for the electron channel and 88\% for the muon channel.

The anti-\kt algorithm, with a radius parameter of 0.5, is used to cluster PF candidates to form hadronic jets. The jet momentum is determined as the vectorial sum of all particle momenta.  Charged hadrons identified as coming from a pileup event vertex are rejected from the jet clustering. The remaining contribution from pileup events, which comes from
    neutral hadrons and from charged hadrons whose PV has not been unambiguously identified, is estimated and subtracted event-by-event using a technique based on the jet area method~\cite{Cacciari:2008gp,Cacciari:2008gn}. Jet energy corrections are derived from the simulation, and are confirmed with in situ measurements using the energy balance of dijet and photon+jet events~\cite{Chatrchyan:2011ds}. Jets with a transverse momentum less than $30\unit{GeV}$ or overlapping within $\DR =
    0.5$ with either of the two leptons from the decay of the $\cPZ$ boson are discarded. The single differential cross sections are measured for jet rapidity within $\abs{y} < 2.4$, which is the region with the best jet resolution and pileup rejection. In this measurement, the jet multiplicity refers to the number of jets fulfilling the jet criteria, within the $\abs{y} < 2.4$ boundary for the one-dimensional differential cross sections. For the multidimensional differential cross sections reported in Section~\ref{subsec:multi} the region for the jet rapidity is extended to $\abs{y}<4.7$.
\section{Background subtraction and correction for the detector response}\label{UnfoldSec}

\begin{figure}[htbp!]
\centering
{\includegraphics[width=0.48\textwidth]{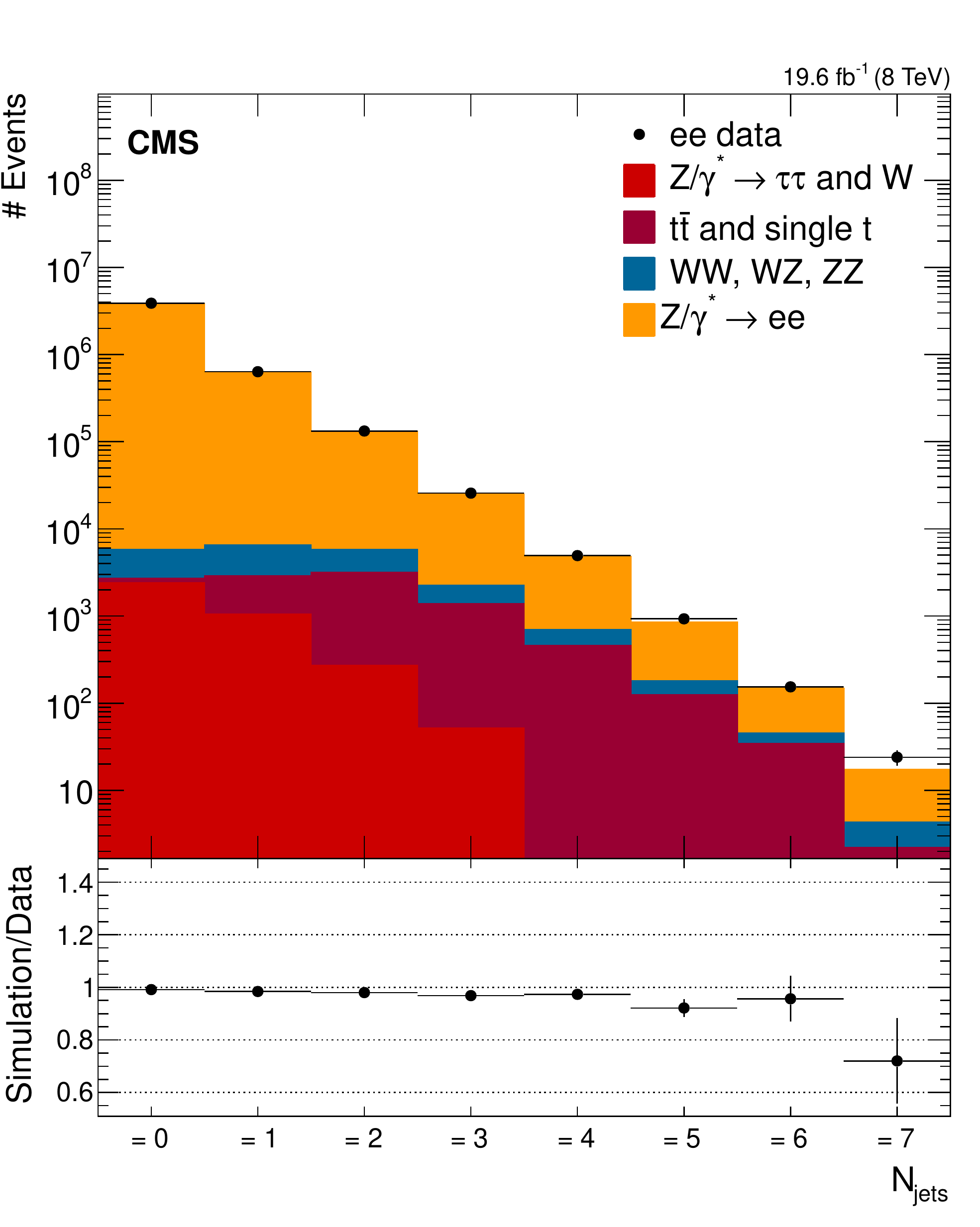}}
{\includegraphics[width=0.48\textwidth]{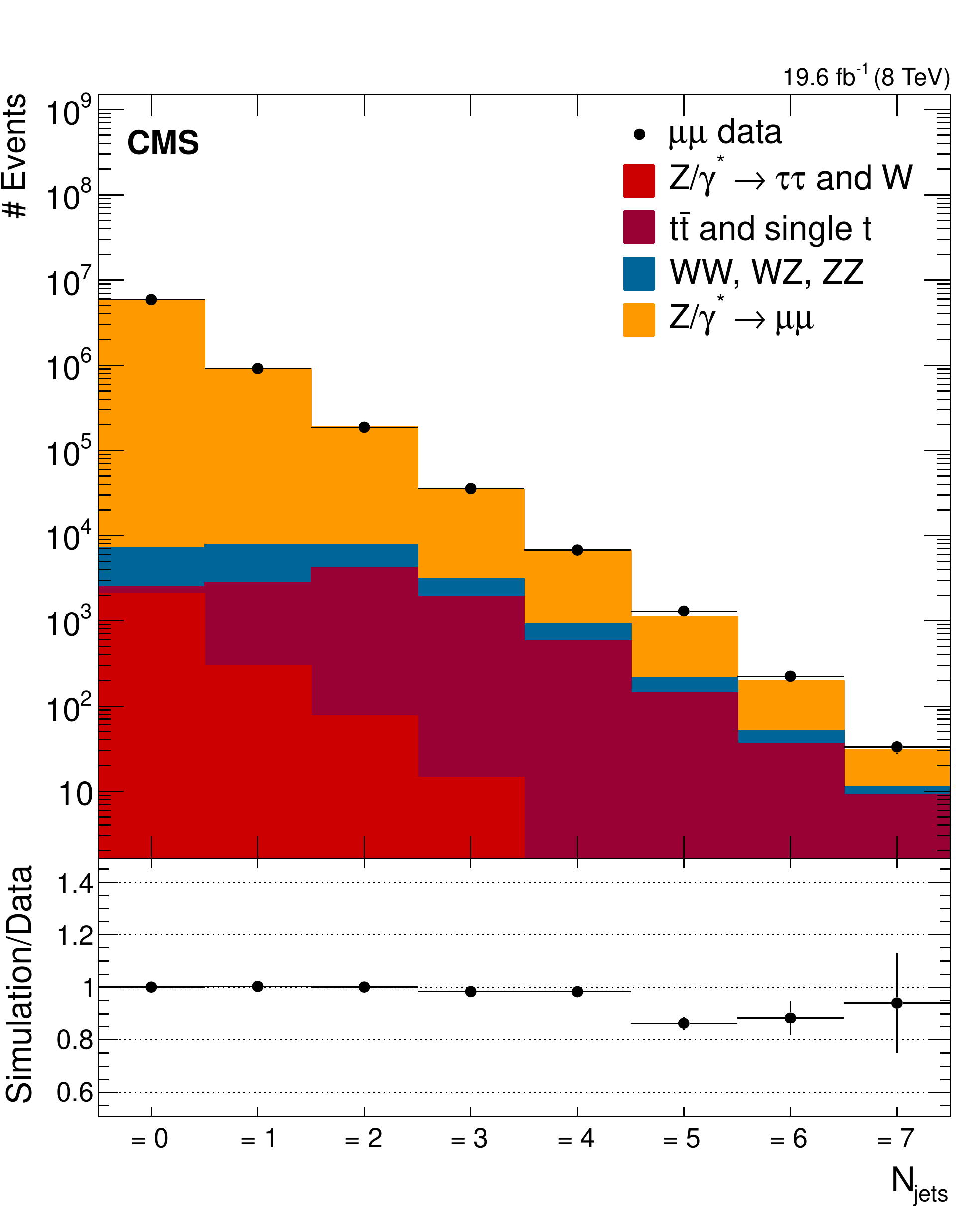}}
\caption{Exclusive jet multiplicity for (left) the electron and (right) the muon channels at detector level in the jet rapidity region $\abs{y} < 2.4$. The data points are shown with statistical error bars. Beneath each plot the ratio of the number of events predicted by the simulation to the measured values is displayed, together with the statistical uncertainties in simulation and data added in quadrature.}
\label{fig:RECO_jetMPexc}
\end{figure}

 In Fig.~\ref{fig:RECO_jetMPexc} the event yield in the electron and muon channels is compared to the
simulation. The agreement between simulation and data is excellent up to four jets. Since the \zjets simulation does not include more than four partons in the ME calculations, we expect a less accurate prediction of the signal for jet multiplicities above four.  Background contamination is below 1\% for a jet multiplicity of one and increases with the jet multiplicity. The background represents 2\% of the event yield for a jet multiplicity of 2 and 20\% for a jet multiplicity of 5.

The background contribution is estimated from the samples of simulated events described in Section~\ref{sec:samples} and subtracted bin-by-bin from the data. The $\tt$ simulations are validated using a $\mu^\pm\Pe^\mp$ data control sample, as explained in Section~\ref{SysSec}.
The background contribution from multijet events where the jets are misidentified as leptons is checked with a lepton control sample using two leptons with the same flavor and charge and found to be negligible.

 Unfolding the detector response corrects the signal distribution for the migration of events between closely separated bins and across boundaries of the fiducial region.
The unfolding procedure also includes a correction for the efficiency of the trigger, and the lepton reconstruction and identification. The unfolding procedure is applied separately to each measured differential cross section. In a first step, the data distribution is corrected to remove the background
contribution and the contribution from signal process events outside of the defined phase space. Then, the iterative D'Agostini method~\cite{D'Agostini:1994zf}, as implemented in the statistical analysis toolkit \textsc{RooUnfold}~\cite{Adye:2011gm}, is used to correct for bin-to-bin migration and for efficiency.
Using the simulation the method generates a response matrix that relates the probability that an event in  bin $i$  of the differential cross section is reconstructed in bin $j$.
 These probabilities include the case of bin-$i$ events that do not pass the selection criteria on the reconstructed event or fall outside the distribution boundaries. For the three-dimensional differential cross section, the method is applied within each $(y(\text{j}_1), y(\cPZ))$ bin, where the unfolding is performed with respect to the $\pt(\text{j}_1)$ observable. Similarly, the unfolding of the double differential cross sections is performed with respect to the most sensitive
variable: $\pt(\text{j}_1)$ for $\rd^2\sigma/\rd\pt(\text{j}_1)\rd y(\text{j}_1)$ and $y(\text{j}_1)$ for $\rd^2\sigma/\rd y(\text{j}_1)\rd y(\Z)$.

The response matrices are built from reconstructed and generated quantities using the \MADGRAPH~5 + \PYTHIA~6 \zjets simulation sample. The generated values refer to the leptons from the decay of the $\cPZ$ boson and to the jets built from the stable particles using the same algorithm as for the measurements. The momenta of all the photons whose $\Delta R$ distance to the lepton axis is smaller than 0.1 are added to the lepton momentum to account for the effects of final-state radiation, and the lepton
is said to be ``dressed''. Although this process does not recover all the final-state radiation; it removes most differences between electrons and muons, and the dilepton mass spectra are identical for the two decay channels after this procedure, as checked with the \MADGRAPH~5 + \PYTHIA~6 simulation. The $\cPZ$ boson is reconstructed from the dressed lepton momentum vectors.

The phase space for the cross section measurement is restricted to $\pt > 20\unit{GeV}$ and $\abs{\eta} < 2.4$ for both dressed
charged leptons, a dilepton mass within $91\pm20\GeV$, and the jet kinematics constrained to $\pt > 30\unit{GeV}$ and $\abs{y} < 2.4$. For the extended multidimensional cross sections the phase space is extended to $\abs{y} < 4.7$. The measured cross section values include the $\cPZ$ branching fraction to a single lepton flavour.

The rejection of jets originating from pileup is more difficult outside the tracker geometric acceptance, since the vertex constraint cannot be used to reject the charged particles coming from pileup. Consequently, despite the jet pileup rejection criterion, a contamination of jets from pileup remains and needs to be subtracted. This region beyond the tracker acceptance, $2.5<\abs{y}<4.7$, is used only for the $\cPZ\ +\ge 1\ \text{jet}$ multidimensional differential cross section measurements, where only
the leading jet is relevant. The fraction of events in which a jet comes from pileup, denoted as $f_{\text{PU}}$, is estimated using a control sample of a $\cPZ$ boson associated with one jet, obtained by requiring one jet with \pt above $30\GeV$ and a veto condition of no other jets with $\pt > 12\GeV$  and above 20\% of the $\cPZ$ boson transverse momentum. Since a $\cPZ$ boson and a jet coming from two different $\Pp\Pp$ collisions are independent, the distribution of $\Delta\phi(\cPZ,\text{j})$ is expected to be flat, which is confirmed by the simulation.
For the $\cPZ$ boson and the jet from a $\Pp\Pp\to\cPZ+1\ \text{jet}$ event the distribution is expected to peak at $\pi$. The constraint on additional jets enforces the \pt balance between the $\cPZ$ boson and the jet, reducing the contribution to low values of $\Delta\phi(\cPZ,\text{j})$. The simulation shows that this contribution is negligible in the region $\Delta\phi(\cPZ,\text{j}) < 1$. Therefore, $f_\text{PU}$ is estimated from the fraction of events in
that region. The value of $f_{\text{PU}}$ is 30\%  in the most forward part ($3.2 < \abs{y} < 4.7$) and lowest
\pt measurement bin ($30\GeV$ to $40\GeV$). The fraction of pileup events estimated from the pileup control sample is used to correct the signal data sample. The same method is applied to the simulation. The ratio of the value of $f_{\text{PU}}$ obtained from the simulation to that measured in data decreases monotonically as a function of \pt. In the \pt bin 30--50\GeV it ranges up to 1.25 (1.35) for $\abs{y(j_1)}$ between 2.5 and 3.2 (3.2 and 4.7). Beyond $\pt=50\GeV$ the discrepancy is negligible and the results are identical with or without the pileup subtraction.

Since the $\abs{y(\text{j})} < 2.4$ region contains the bulk of the events and including the $\abs{y(\text{j})}>2.4$ region does not improve the precision of the measurement, most of the differential cross section measurements are limited to $\abs{y(\text{j})}<2.4$. The subtraction of pileup contributions is not needed when confining measurements to this region.

\section{Systematic uncertainties}\label{SysSec}

The systematic uncertainty in the background-subtracted data distributions is estimated by varying independently each of the contributing factors before the unfolding, and computing the difference induced by the variation in the unfolded distributions. The observed difference between the positive and negative uncertainties is small and the two are therefore averaged. The different sources of uncertainties are
independent and are added in quadrature. The unfolded histogram can be written as a linear combination of the bin contents of the background-subtracted data histogram~\cite{D'Agostini:1994zf}. This linear combination is used to propagate analytically the statistical uncertainties to the unfolded results and calculate the full covariance matrix for each distribution, separately for each $\cPZ$ boson decay channel. The dominant source of systematic uncertainties is the jet energy correction. The
various contributions are listed in Table~\ref{tab:combZNGoodJets_Zexc}.

The jet energy correction uncertainty (JEC in the table) is calculated by varying this correction by one standard deviation. This uncertainty is \pt- and $\eta$-dependent and varies from 1.5\% up to 5\% for $\abs{\eta} < 2.5$ and from 7\% to 30\% for $\abs{\eta} > 2.5$. The uncertainty in the measured cross section is between 5.3\% and 28\% depending on the jet multiplicity. The jet energy resolution (JER) uncertainty is estimated for data and simulation in Ref.~\cite{Khachatryan:2016kdb}.
The resulting uncertainty in the measurement is below 1\% for all the multiplicities.

 \begin{table}[htb]
\centering
\topcaption{Cross section results obtained from the combination of the muon and electron channels as a function of the exclusive jet multiplicity and details of the systematic uncertainties. The column denoted \textit{Tot unc} contains the total uncertainty; the column denoted \textit{Stat} contains the statistical uncertainty; the remaining columns contain the systematic uncertainties.}
\cmsTable{
\begin{tabular}{c|cc|ccccccccc}
 $N_{\text{jets}}$ & $\frac{\rd\sigma}{\rd N_{\text{jets}}}$ & {Tot unc } & {Stat } & {JEC } & {JER } & {Bkg } & {PU } & {Unf stat }  & {Unf sys } & {Lumi } & {Eff } \\
 &[pb]&[\%]&[\%]&[\%]&[\%]&[\%]&[\%]&[\%]&[\%]&[\%]&[\%]\\\hline
$=$0 & 423     & 3.7 & 0.034 & 1.2 & 0.06 & 0.002 & 0.71 & 0.05 & 1.2 & 2.6 & 1.8 \\
$=$1 & 59.9    & 6.3 & 0.11 & 5.3 & 0.23 & 0.042 &  0.26 & 0.075 & 1.4 & 2.6 & 1.8 \\
$=$2 & 12.6    & 9.2 & 0.25 & 8.4 & 0.22 & 0.33  &  0.35 & 0.12  & 1.7 & 2.7 & 1.9 \\
$=$3 & 2.46    & 12 & 0.6 & 11 & 0.22 & 0.76  &  0.42 & 0.22  & 2.7 & 2.9 & 2.0 \\
$=$4 & 0.471   & 16 & 1.4  & 15 & 0.16 & 1.3   &  0.57 & 0.43  & 3.5 & 3.1 & 2.1 \\
$=$5 & 0.0901  & 20 & 3.4  & 19 & 0.28 & 1.9   &  0.75 & 1.0   & 4.6 & 3.2 & 2.3 \\
 $=$6 & 0.0143  & 33 & 9.3  & 28 & 0.72 & 3.3   &  1.9  & 2.4   & 5.5 & 3.7 & 2.6 \\
$=$7 & 0.00230 & 34 & 22  & 23 & 0.61 & 4.3   &  5.6  & 6.3   & 6.4 & 3.9 & 2.8 \\
\end{tabular}}
\label{tab:combZNGoodJets_Zexc}
\end{table}

Other significant background contributions come from \ttbar, diboson, and $\cPZ\to\tau^{+}\tau^{-}$ processes. The related uncertainty (Bkg) is estimated by varying the cross section for each of the background processes (\ttbar, $\cPZ\cPZ$, $\PW\cPZ$, and $\PW\PW$) independently by 10\% for \ttbar and 6\% for diboson processes. The normalisation variation for the \ttbar events is chosen to cover the maximum observed
difference between the simulation and the data in the jet multiplicity, transverse momentum, and rapidity distributions when selecting events with two leptons of oppositely charged, different flavours ($\mu^\pm\Pe^\mp$). The uncertainty in diboson cross sections covers theoretical and PDF contributions. The resulting uncertainty in the measurement increases with the jet multiplicity and reaches 4.3\%.

Another source of uncertainty is the modelling of the pileup (PU).  The number of interactions per bunch crossing in simulated samples is varied by 5\%. This covers effects related to the modeling of simulated minimum bias events of 3\%, the estimate of the number of interactions per bunch crossing in data based on luminosity measurements of 2.6\%, and the experimental uncertainties entering inelastic cross section measurements of 2.9\%. The resulting uncertainties range from 0.26\% to 5.6\% depending on the jet multiplicity. The uncertainty from the pileup subtraction performed in the forward region, $\abs{y}>2.5$, is estimated by varying up and down the pileup fraction $f_{\text{PU}}$ described in Section~\ref{UnfoldSec} by half the difference from the value obtained in the
simulation. In the region covered by the tracker and where no correction is applied, it is verified that the jet multiplicity does not depend on the number of vertices reconstructed in the event. This indicates that the jets from pileup events have a negligible impact on the measurement.

The unfolding procedure has an uncertainty due to its dependence on the simulation used to estimate the response matrix (Unf sys) and to the finite size of the simulation sample (Unf stat). The first contribution is estimated using an alternative event generator, \SHERPA~1.4~\cite{Gleisberg:2008ta}, and taking the difference between the two results to represent the uncertainty. The distribution obtained with the alternative generator differs sufficiently from the nominal one to cover the differences
with the data. The statistical uncertainty in the response matrix is analytically propagated to the unfolded result~\cite{D'Agostini:1994zf}. When added in quadrature and depending on the kinematic variable and jet multiplicity, the total unfolding uncertainty varies up to 10\%.

The uncertainty in the efficiency of the lepton reconstruction, identification, and isolation is propagated to the measurement by varying the total data-to-simulation scale factor by one standard deviation. It amounts to 2.5\% and 2.6\% in the dimuon and dielectron channels, respectively.

The uncertainty in the integrated luminosity amounts to 2.6\%~\cite{CMS:2013gfa}. Since the background event yield normalisation also depends on the integrated luminosity, the effect of the above uncertainty on the background yield (Lumi) can be larger and amounts to 3.9\% in the bins with low signal purity.

\section{Results}\label{sec:Results}

The measurements from the electron and muon channels are consistent and are combined using a weighted average. For each bin of the measured differential cross sections, the results of each of the two measurements are weighted by the inverse of the squared total uncertainty. The covariance matrix of the combination, the diagonal elements of which are used to extract the measurement uncertainties,  is computed assuming full correlation between the two channels for all the uncertainty sources except for statistical uncertainties and those associated with lepton reconstruction and identification, which are taken to be uncorrelated. The measured differential cross sections are compared to the results obtained from three different calculations as described below.

\subsection{Theoretical predictions}

 The measurements are compared to a tree level calculation and two multileg NLO calculations. The first prediction is computed with \MADGRAPH~5~\cite{Alwall:2011uj} interfaced with \PYTHIA~6 (denoted as MG5 + PY6 in the figure legends), for parton showering and hadronisation, with the configuration described in Section~\ref{sec:samples}. The total cross section is normalised to the NNLO cross section computed with {\FEWZ} 2.0~\cite{Gavin:2010az}. Two multileg NLO predictions including
 parton showers using the MC@NLO~\cite{Frixione:2002ik} method are used. For these two predictions the total cross section is normalised to the one obtained with the respective event generators. The total cross section values used for the normalisation are summarised in Table~\ref{tab:theory_xsec}.

The first multileg NLO prediction with parton shower is computed with \SHERPA~2 (2.0.0)~\cite{Gleisberg:2008ta} and
  \BLACKHAT~\cite{Berger:2008ag,Berger:2010gf} for the one-loop corrections. The matrix elements include the five
  processes $\Pp\Pp \to \cPZ+N_{\text{jets}}\ \text{jets}, N_{\text{jets}}\le4$, with an NLO accuracy for $N_{\text{jets}} \leq 2$ and LO accuracy for $N_{\text{jets}}=3$ or $4$. The CT10 PDF~\cite{Gao:2013xoa} is used for both the ME calculations and showering description. The merging of PS and ME calculations is done with the MEPS@NLO method~\cite{Hoeche:2012yf} and the merging scale is set to $20\unit{GeV}$.

The second multileg NLO prediction is computed with MadGraph5\_aMC@NLO~\cite{Alwall:2014hca} (denoted as \MGaMC in the following) interfaced with \PYTHIA~8 using the CUETP8M1 tune~\cite{Khachatryan:2015pea,Skands:2014pea} for parton showering, underlying events, and hadronisation. The matrix elements include the $\cPZ$ boson production processes with 0, 1, and 2 partons at NLO. The FxFx~\cite{Frederix:2012ps} merging scheme is used with a merging scale parameter set to~$30\GeV$. The NNPDF~3.0 NLO PDF~\cite{Ball:2014uwa} is used for the ME calculations
while the NNPDF~2.3 QCD + QED LO~\cite{Ball:2010de,Ball:2011mu} is used for the backward evolution of the showering.
For the ME calculations, $\alpha_s$ is set to the current PDG world average~\cite{Agashe:2014kda} rounded to $\alpha_s(m_\cPZ)=0.118$. For the showering and underlying events the value of the CUETP8M1 tune, $\alpha_s(m_\cPZ)=0.130$, is used. The larger value is expected to compensate for the missing higher order corrections.
NLO accuracy is achieved for $\Pp\Pp \to \cPZ+N_{\text{jets}}\ \text{jets}, N_{\text{jets}}=0,1,2$ and LO accuracy for $N_{\text{jets}}=3$. For this prediction, theoretical uncertainties are computed and include the contribution from the fixed-order calculation\alphasunc[ and]{,} from the NNPDF~3.0 PDF set\alphasunc{, and from the choice of the strong coupling constant value}. The \alphasunc[two]{three} uncertainties are added in quadrature. The fixed-order calculation uncertainties are estimated by varying the renormalisation and the factorisation scales by factors of $1/2$ and 2. The envelope of the variations of all factor combinations, excluding the
two combinations when one scale is varied by a factor $1/2$ and the other by a factor 2, is taken as the uncertainty.
The reweighting method~\cite{Frederix:2011ss} provided by the \MGaMC generator is used to derive the cross sections with the different renormalisation and factorisation scales and with the different PDF replicas used in the PDF uncertainty determination. For the NLO predictions, weighted samples are used (limited to $\pm$1 weights in the case of aMC@NLO), which can lead to larger statistical fluctuations than expected for unweighted samples in some bins of the histograms presented in this section.

\begin{table}
\centering
    \caption{Values of the $\Pp\Pp\to \ell^+\ell^-$ total inclusive cross section used in the predictions in data-theory comparison plots. The cross section used for the plots together with the one obtained from the generated sample (``native'') and their ratio ($k$) are provided. The cross section values correspond to the dilepton mass windows used for the respective samples and indicated in the table.}
\cmsTable{
\begin{tabular}{l|ccccc}
                                         & Dilepton mass    & Native cross     &             & Used cross       &   \\
Prediction                               & window [\GeVns{}] & section [pb] & Calculation & section [pb] & $k$\\
\hline
\MADGRAPH~5 + \PYTHIA~6, $\le 4$ j LO+PS       & $>$50         & 983              & FEWZ NNLO   & 1177             & 1.197 \\
\SHERPA~2, $\le 2$ j NLO, $3,4$ j LO+PS & $[66, 116]$      & 1059             & native      & 1059             & 1     \\
\MGaMC + \PYTHIA~8, $\le$2 j NLO             & $>$50         & 1160             & native      & 1160             & 1     \\
\end{tabular}}
\label{tab:theory_xsec}

\end{table}

\subsection{Jet multiplicity}

\begin{figure}[htbp!]
\centering
\includegraphics[width=0.48\textwidth]{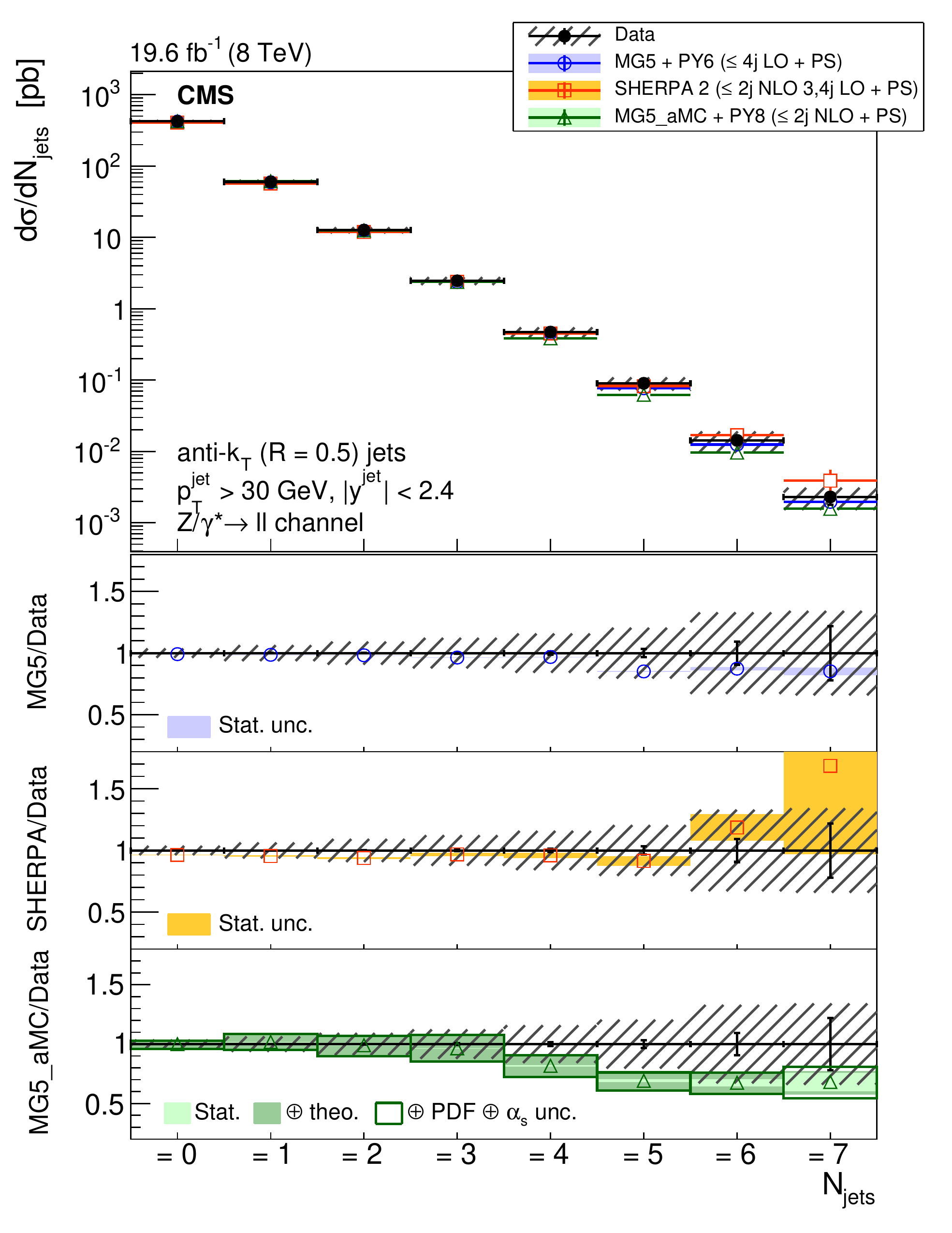}
\includegraphics[width=0.48\textwidth]{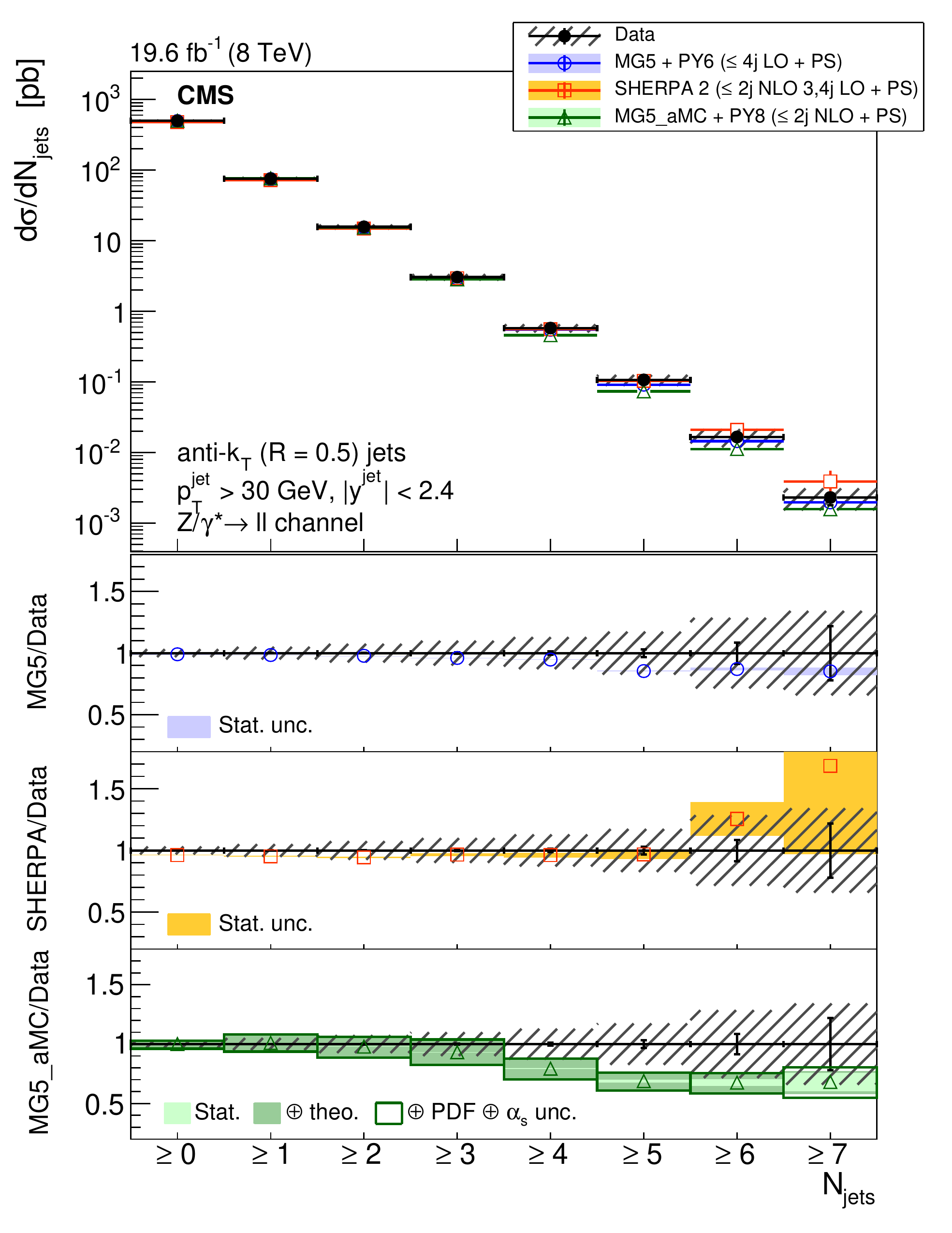}
\caption{ The cross section for \zlljets production measured as a function of the (left) exclusive and (right) inclusive jet multiplicity distributions compared to the predictions calculated with \MADGRAPH~5 + \PYTHIA~6, \SHERPA~2, and \MGaMC + \PYTHIA~8. \plotstdcapt}
\label{fig:CombXSec_jetMPexc}
\end{figure}

The cross sections for jet multiplicities from 0 to 7, and the comparisons with various predictions are presented in this section.
 Figure~\ref{fig:CombXSec_jetMPexc} shows the cross section for both inclusive and exclusive jet multiplicities and the numbers are compared with the prediction obtained with \MGaMC+ \PYTHIA~8 in Table~\ref{tab:nj_unc} for the exclusive case. The agreement with the predictions is very good for jet multiplicities up to the maximum number of final-state partons included in the ME calculations, namely three for \MGaMC + \PYTHIA~8 and four for both \MADGRAPH~5 + \PYTHIA~6 and \SHERPA~2. The level of precision of the measurement does not allow us to probe the improvement expected from the additional NLO terms. The
cross section is reduced by a factor of five for each additional jet.

The predictions already agree well at tree level (\MADGRAPH~5 + \PYTHIA~6) renormalised to the NNLO inclusive cross section. For $N_{\text{jets}}= 4$, the \MGaMC + \PYTHIA~8 calculation, which does not include this jet multiplicity in the matrix elements, predicts a different cross section from those that do. The predictions that include four jets in the matrix elements are in better agreement with the data, but the difference between the predictions is limited to
roughly one standard deviation of the measurement uncertainty.
The large uncertainty is due to the sensitivity of the jet \pt threshold acceptance to the jet energy scale.
 The \SHERPA~2 prediction for $N_{\text{jets}}=5$ is closer to the measurement than \MADGRAPH~5 + \PYTHIA~6, while neither of these includes this multiplicity in the ME calculations. The theoretical uncertainty shown in the figure for the \MGaMC + \PYTHIA~8 prediction uses the standard method described in the previous subsection. In the case of the exclusive jet
multiplicity, the presence of large logarithms in the perturbative calculation can lead to an underestimate of this uncertainty, so the Steward and Tackmann prescription (ST) provides a better estimate~\cite{Stewart:2011cf}. The uncertainties calculated with both prescriptions are provided in Table~\ref{tab:nj_unc}. For the calculations considered here, the increase of the ST uncertainty with respect to the standard one is moderate. This is consistent with the observation that the agreement with the measurement and the coverage of the difference by the theoretical uncertainty in Fig.~\ref{fig:CombXSec_jetMPexc} is similar for the inclusive and exclusive jet multiplicities.

\begin{table}[htb]
\centering
    \topcaption{Measured ($\sigma_{\text{data}}$) and calculated cross sections of the production of $\Z+N_{\text{jets}}\text{ jets}$ events. The cross section calculated with \MGaMC is given in the third column together with the total uncertainty that covers the theoretical (standard method), PDF, $\alpha_s$, and statistical uncertainties. The theoretical uncertainty obtained with the standard and ST methods are compared in the two last columns. The uncertainty on the measurement is separated in systematic and statistical components when the latter is not negligible.}
\cmsTable{
\def\arraystretch{1.5}
\begin{tabular}{c|c@{$\;\pm$}l@{$\;$}l|c@{$$}l|c|c}
  $N_{\text{jets}}$ & \multicolumn{3}{c|}{$\sigma_{\text{data}}$\,[pb] } & \multicolumn{2}{c|}{$\sigma_{\MGaMC}$\,[pb]} & \shortstack[c]{Standard\\theo.\\ uncert. } & \shortstack[c]{ST\\ theo.\\ uncert. } \\
    \hline
$=$0 &     423    & 16	           &                & 423    &   $^{+13}_{-17}$		 & $^{+10}_{-15}$	   &  $^{+12}_{-18}$		    \\
$=$1 &     59.9   & 3.8\syst     & $\pm$0.1\stat &  61.0     &   $^{+4.1}_{-4.0}$	 & $^{+3.9}_{-3.8}$	   &  $^{+5.3}_{-5.4}$	    \\
$=$2 &     12.6   & 1.2            &                   &  12.5     &   $^{+1.0}_{-1.2}$	 & $^{+0.97}_{-1.1}$	   &  $^{+1.3}_{-1.4}$	    \\
$=$3 &     2.46   & 0.29\syst    & $\pm$0.02\stat &  2.37     &   $^{+0.28}_{-0.27}$	 & $^{+0.27}_{-0.27}$	   &  $^{+0.32}_{-0.32}$	    \\
$=$4 &     0.471  & 0.075\syst   & $\pm$0.007\stat &  0.385    &   $^{+0.042}_{-0.044}$	 & $^{+0.041}_{-0.044}$     &$^{+0.049}_{-0.053}$	    \\
$=$5 &     0.0901 &  0.018\syst  & $\pm$0.003\stat  & 0.0622   &   $^{+0.0063}_{-0.0072}$	 & $^{+0.0062}_{-0.0070}$   &$^{+0.0073}_{-0.0084}$	    \\
$=$6 &     0.0143 &  0.0045\syst  & $\pm$0.0013\stat &  0.0096  &   $^{+0.0011}_{-0.0013}$	 & $^{+0.001}_{-0.0011}$    &$^{+0.0011}_{-0.0013}$	    \\
$=$7 &     0.00230 &  0.00060\syst & $\pm$0.00051\stat &  0.00157  &   $^{+0.00023}_{-0.00026}$ & $^{+0.00012}_{-0.00017}$ &$^{+0.00013}_{-0.00019}$    \\
\end{tabular}
}
\label{tab:nj_unc}
\end{table}

\subsection{Jet transverse momentum}
\begin{figure}[htbp!]
\centering
\includegraphics[width=0.45\textwidth]{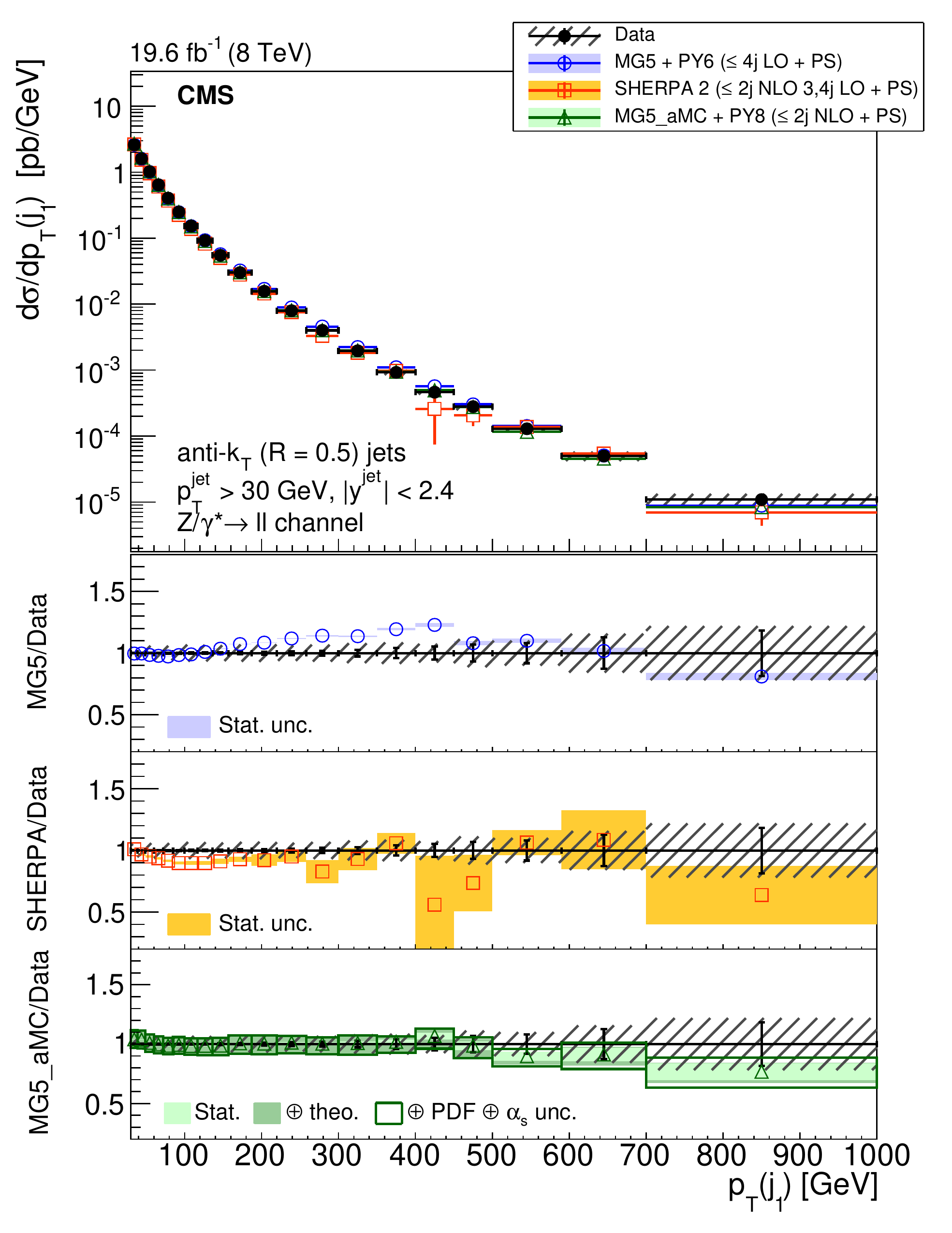}
\includegraphics[width=0.45\textwidth]{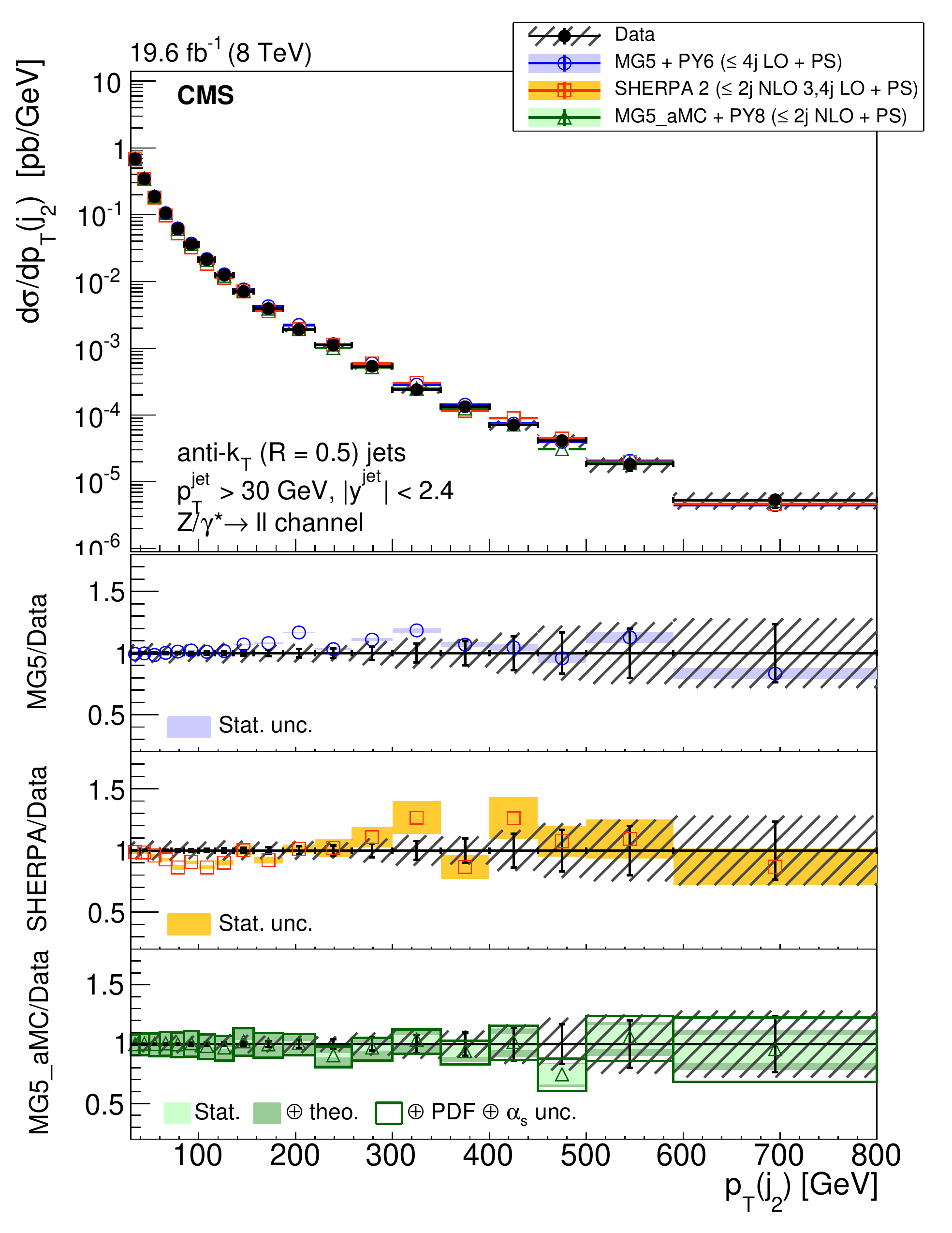}
\caption{The differential cross section for \zlljets production measured as a function of the (left) $1^{\text{st}}$ and (right) $2^{\text{nd}}$ jet \pt compared to the predictions calculated with \MADGRAPH~5 + \PYTHIA~6, \SHERPA~2, and \MGaMC + \PYTHIA~8. \plotstdcapt}
\label{fig:CombXSec_FirstJetPt}
\end{figure}

\begin{figure}[htbp!]
\centering
{\includegraphics[width=0.48\textwidth]{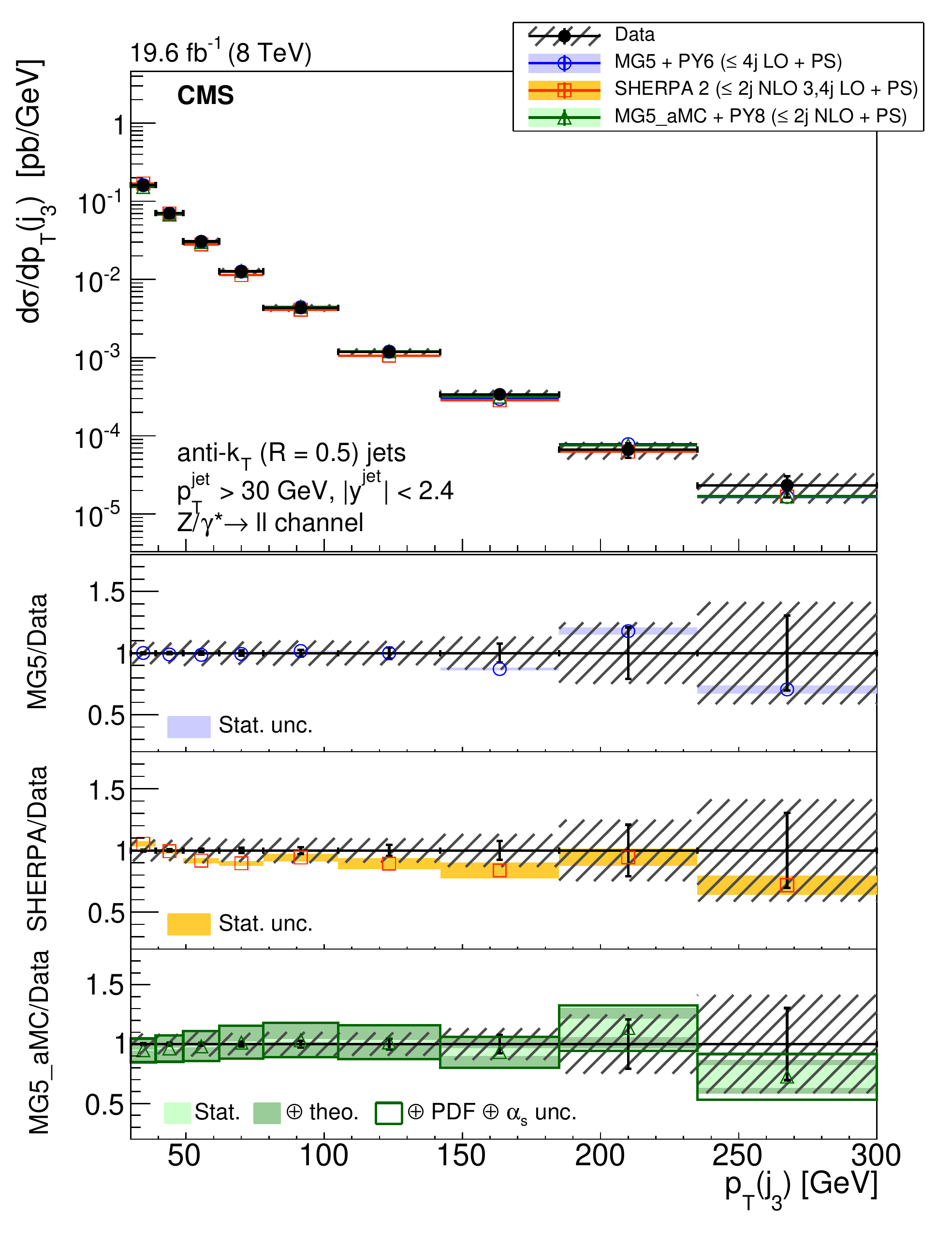}}
{\includegraphics[width=0.48\textwidth]{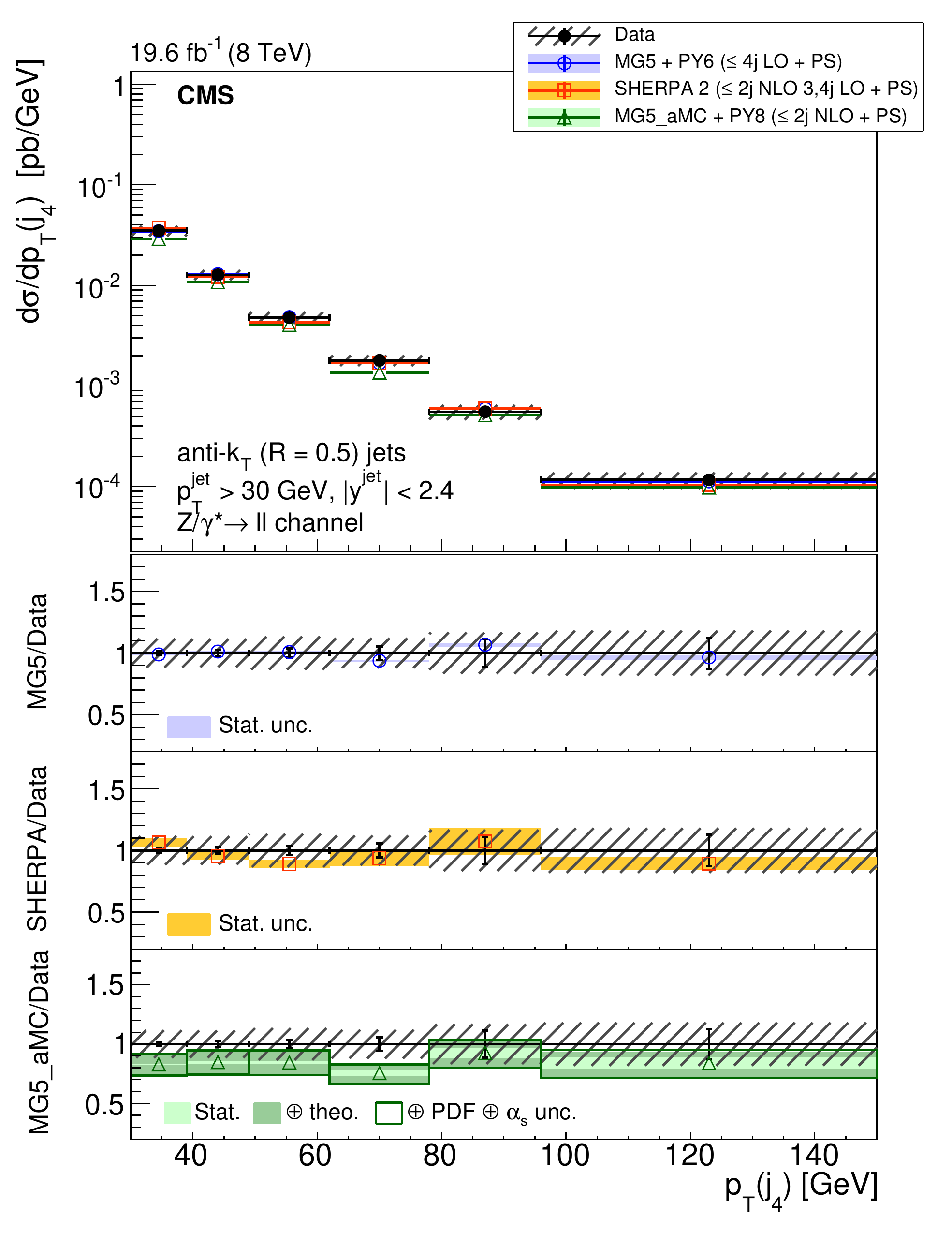}}
\caption{The differential cross section for \zlljets production measured as a function of the (left) $3^{\text{rd}}$ and (right) $4^{\text{th}}$ jet \pt compared to the predictions calculated with \MADGRAPH~5 + \PYTHIA~6, \SHERPA~2, and \MGaMC + \PYTHIA~8. \plotstdcapt}
\label{fig:CombXSec_ThirdJetPt}
\end{figure}

\begin{figure}[htbp!]
\centering
\includegraphics[width=0.48\textwidth]{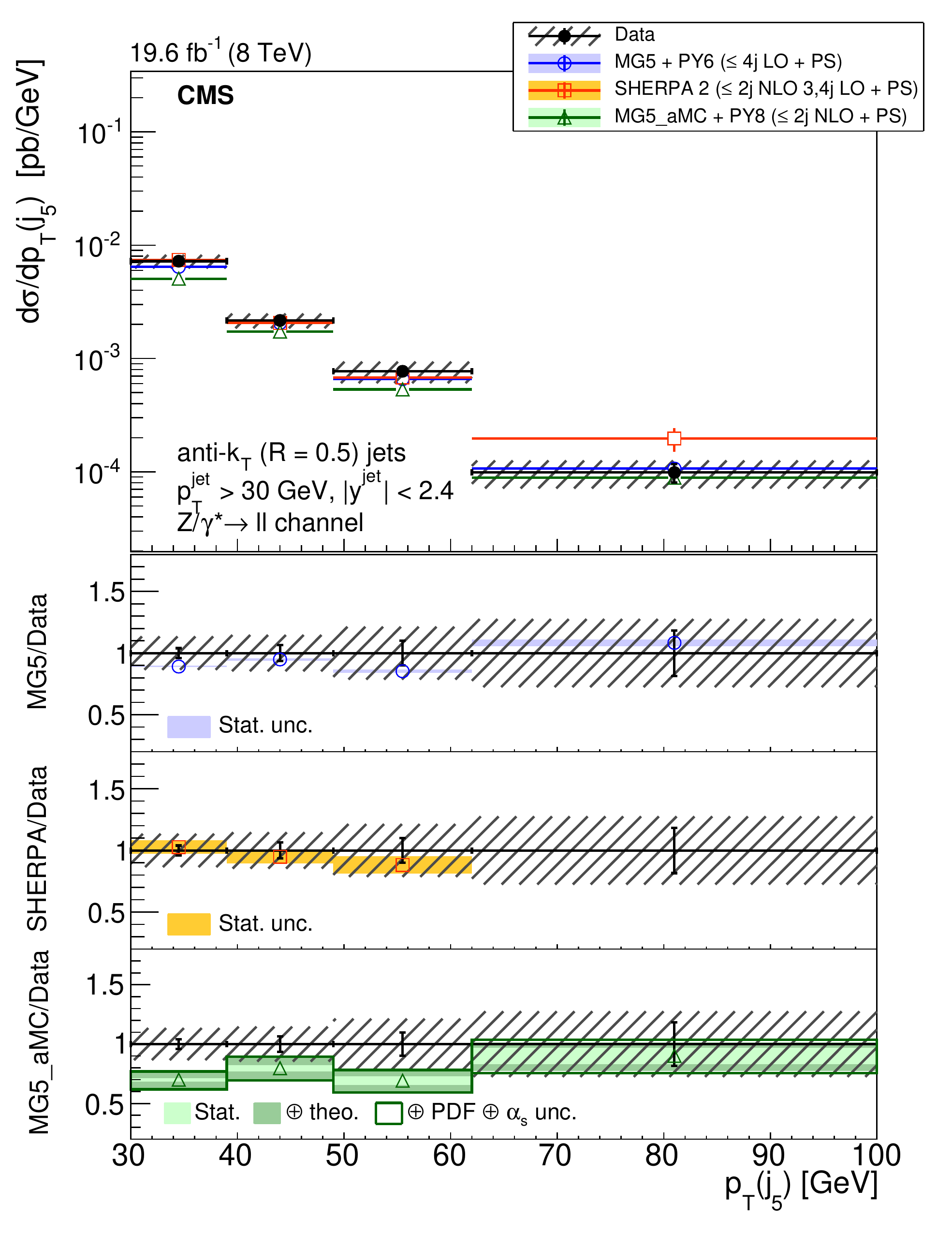}
\caption{The differential cross section for \zlljets production measured as a function of the $5^{\text{th}}$ jet \pt compared to the predictions calculated with \MADGRAPH~5 + \PYTHIA~6, \SHERPA~2, and \MGaMC + \PYTHIA~8. \plotstdcapt}
\label{fig:CombXSec_FifthJetPt}
\end{figure}

Knowledge of the kinematics of SM events with large jet multiplicity is essential for the LHC experiments since these events are backgrounds to searches for new physics that predict decay chains of heavy coloured particles, such as squarks, gluinos, or heavy top quark partners.
The measured differential cross sections as a function of jet \pt for the $1^{\text{st}}$, $2^{\text{nd}}$, $3^{\text{rd}}$, $4^{\text{th}}$, and $5^{\text{th}}$ jets are presented in Figs.~\ref{fig:CombXSec_FirstJetPt}--\ref{fig:CombXSec_FifthJetPt}. The cross sections fall rapidly with increasing \pt. The cross section for the leading jet is measured for \pt values between $30\GeV$ and $1\TeV$ and decreases by more than five orders of magnitude over this range. The
cross section for the fifth jet is measured for \pt values between 30 and 100\GeV and decreases even faster, mainly because of the phase space covered.

For the leading jet, the agreement of the \MADGRAPH~5 + \PYTHIA~6 prediction with the measurement is very good up to ${\approx}$150\GeV. Discrepancies are observed from ${\approx}$150 to ${\approx}$450\GeV.  A similar excess in the ratio with the tree-level calculation was observed at $\sqrt{s}=7\TeV$ in the CMS measurement~\cite{Khachatryan:2014zya}, using predictions from the same generators, as well as in the ATLAS measurement~\cite{Aad:2013ysa}, which used {\sc
Alpgen}~\cite{Mangano:2002ea} interfaced to {\HERWIG}~\cite{Corcella:2000bw} for the predictions. The calculations that include NLO terms for this jet multiplicity do not show this discrepancy. The prediction from \SHERPA~2 shows some disagreement with data in the low transverse momentum region. The second jet shows similar behaviour. Both \MADGRAPH~5 + \PYTHIA~6 and \MGaMC + \PYTHIA~8 are in good agreement with the measurement for the third jet \pt spectrum. The shape predicted by
the calculations from \SHERPA~2 differs from the measurement since the predicted spectrum is harder. For the $4^{\text{th}}$ jet, the three predictions agree well with the measurements. Calculations from \SHERPA~2 and \MGaMC + \PYTHIA~8 predict different spectra. Based on the experimental uncertainties it is difficult to arbitrate between the two, although we expect the one that includes four partons in the matrix elements to be more accurate. The agreement of \SHERPA~2 and \MADGRAPH~5 + \PYTHIA~6
calculations with the measured $5^{\text{th}}$ jet \pt spectrum is similar.

In summary, including many jet multiplicities in the matrix elements provides a good description of the different jet transverse momentum spectra. Including NLO terms improves the agreement with the measured spectra. Nevertheless, some differences are observed between the predictions calculated with \SHERPA~2 and \MGaMC + \PYTHIA~8. The two calculations differ in many ways, other than the fixed order: different PDF choices, different jet merging schemes, and different showering models.
In Ref.~\cite{Khachatryan:2014zya} it was shown that the jet \pt spectra have little dependence on the PDF choice, therefore the difference between the two generator is likely to be due to the different parton showerings or jet merging schemes.

  \subsection{Jet and Z boson rapidity}\label{sec:rap}

\begin{figure}[htbp!]
\centering
\includegraphics[width=0.48\textwidth]{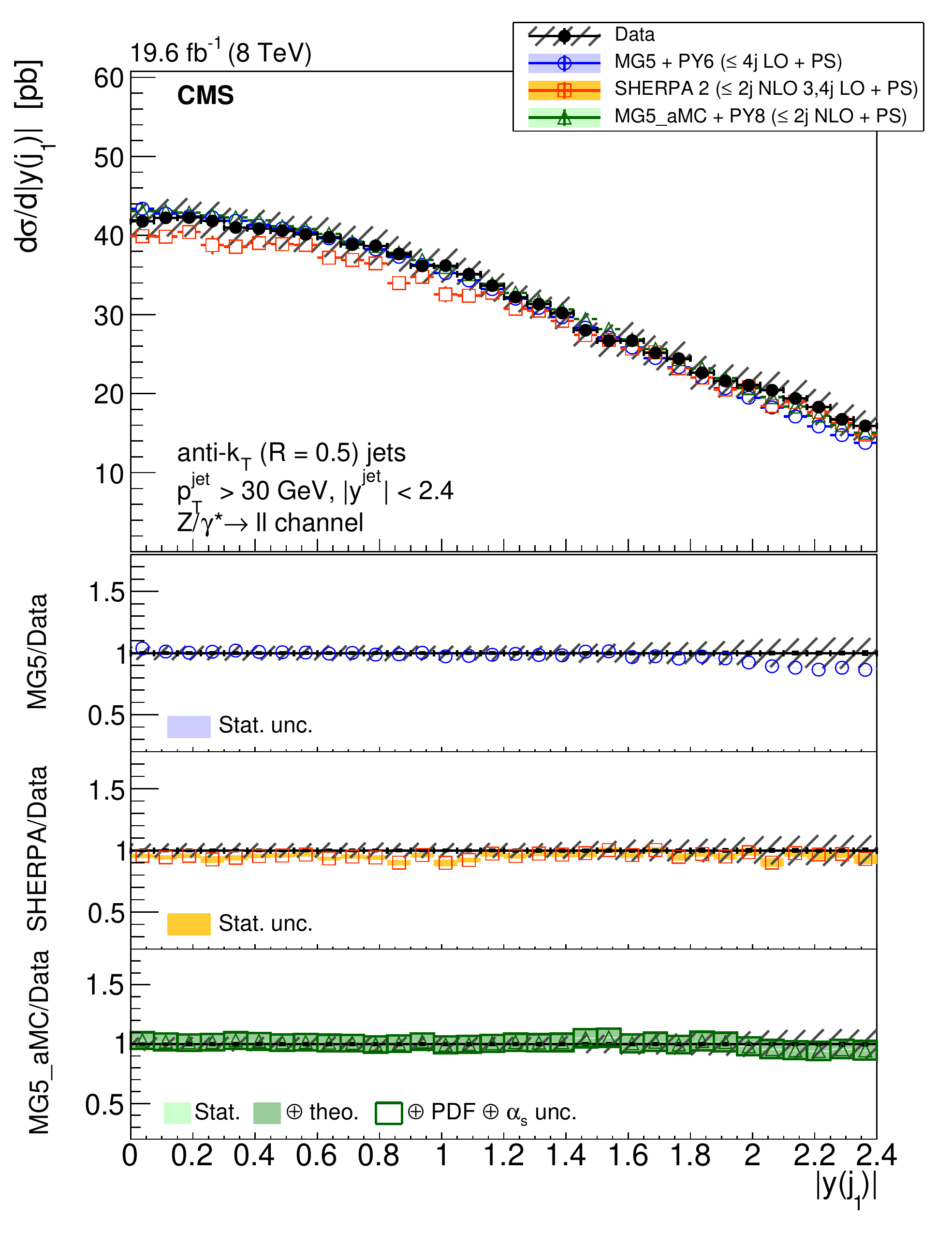}
\includegraphics[width=0.48\textwidth]{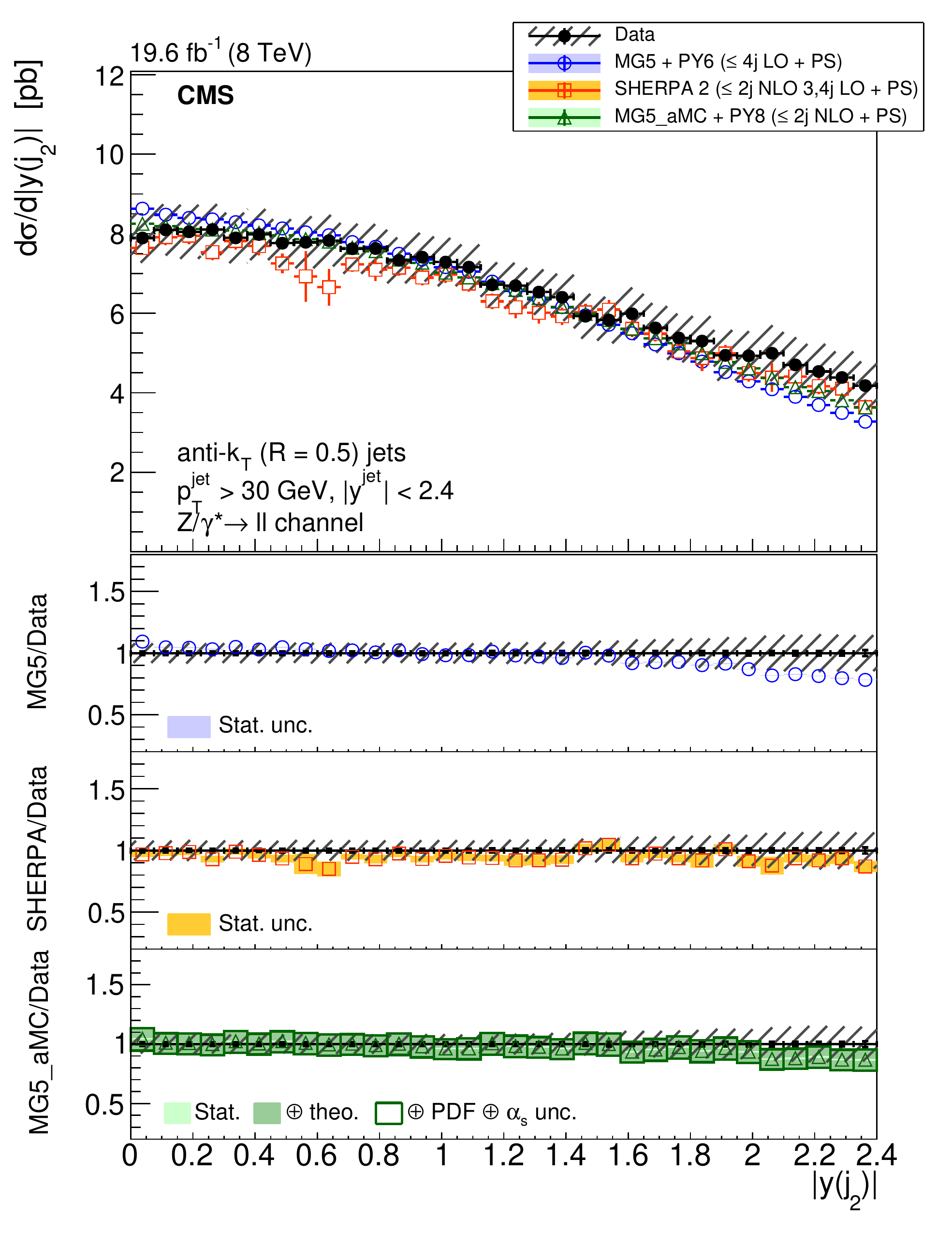}
\caption{The differential cross section for \zlljets production measured as a function of the (left) $1^{\text{st}}$ and (right) $2^{\text{nd}}$ jet $\vert y \vert$ compared to the predictions calculated with \MADGRAPH~5 + \PYTHIA~6, \SHERPA~2, and \MGaMC + \PYTHIA~8. \plotstdcapt}
\label{fig:CombXSec_FirstJetEta}
\end{figure}

\begin{figure}[htbp!]
\centering
{\includegraphics[width=0.48\textwidth]{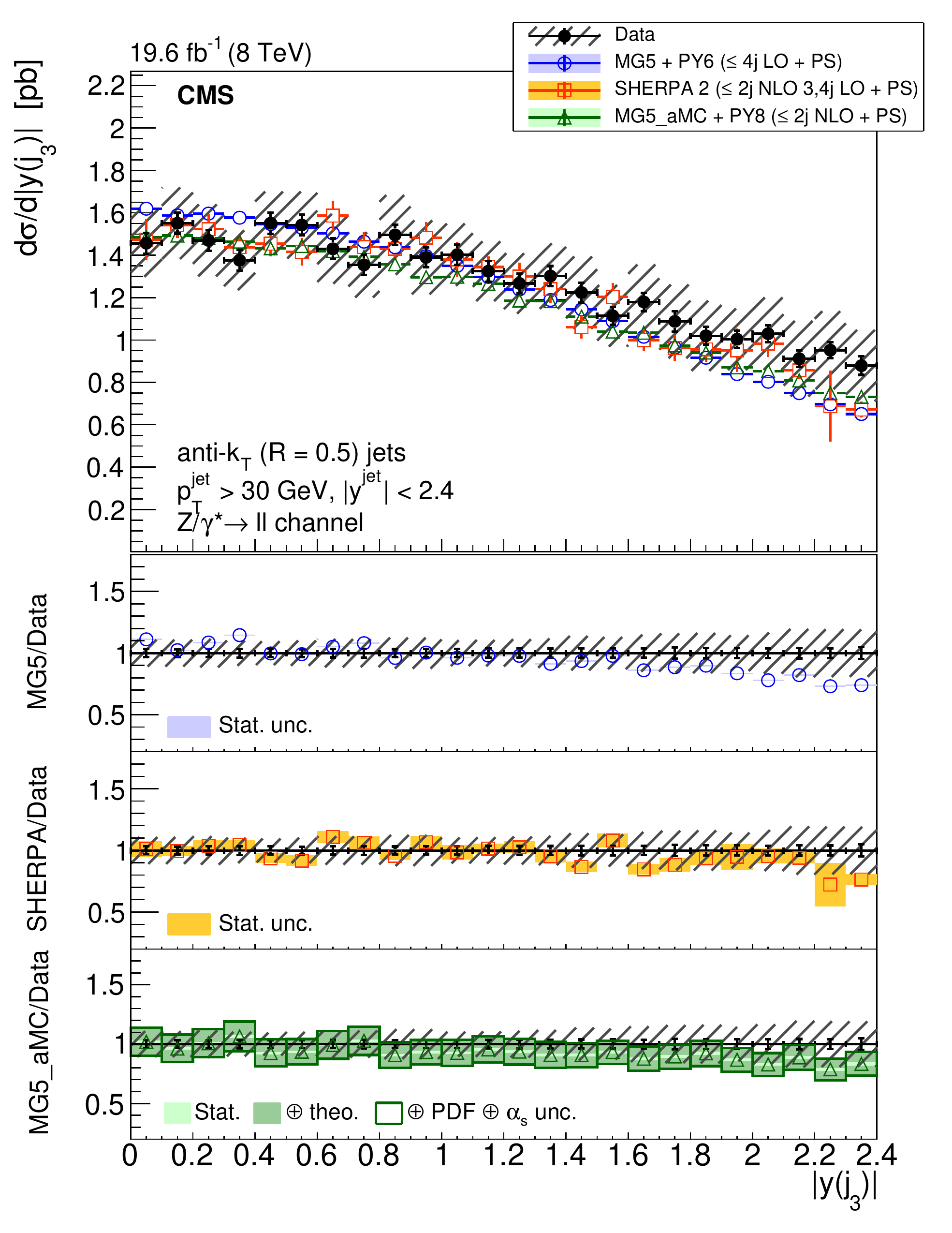}}
{\includegraphics[width=0.48\textwidth]{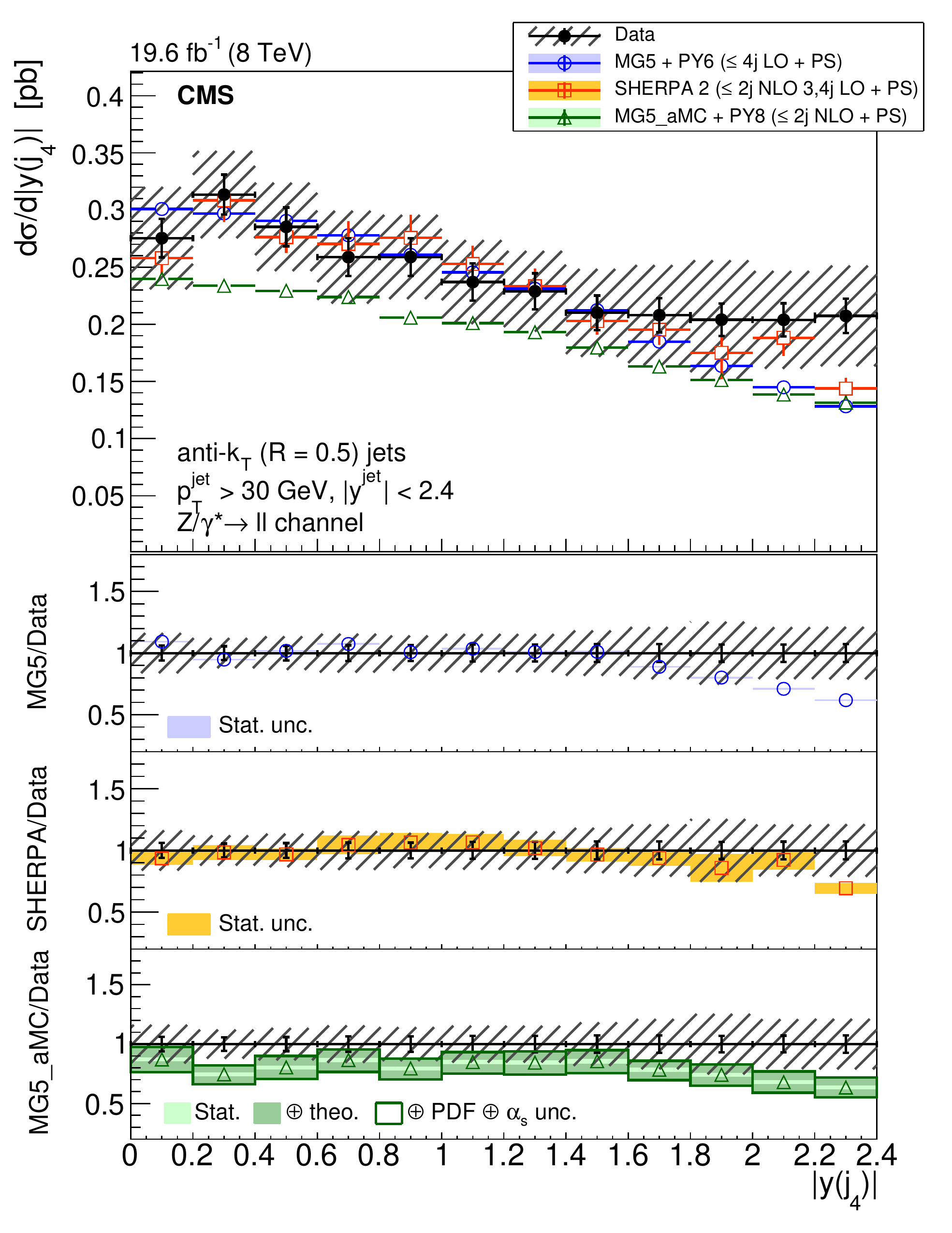}}
\caption{The differential cross section for \zlljets production measured as a function of the (left) $3^{\text{rd}}$ and (right) $4^{\text{th}}$ jet $\abs{y}$ compared to the predictions calculated with \MADGRAPH~5 + \PYTHIA~6, \SHERPA~2, and \MGaMC + \PYTHIA~8. \plotstdcapt}
\label{fig:CombXSec_ThirdJetEta}
\end{figure}

\begin{figure}[!hptb]
\centering
\includegraphics[width=0.48\textwidth]{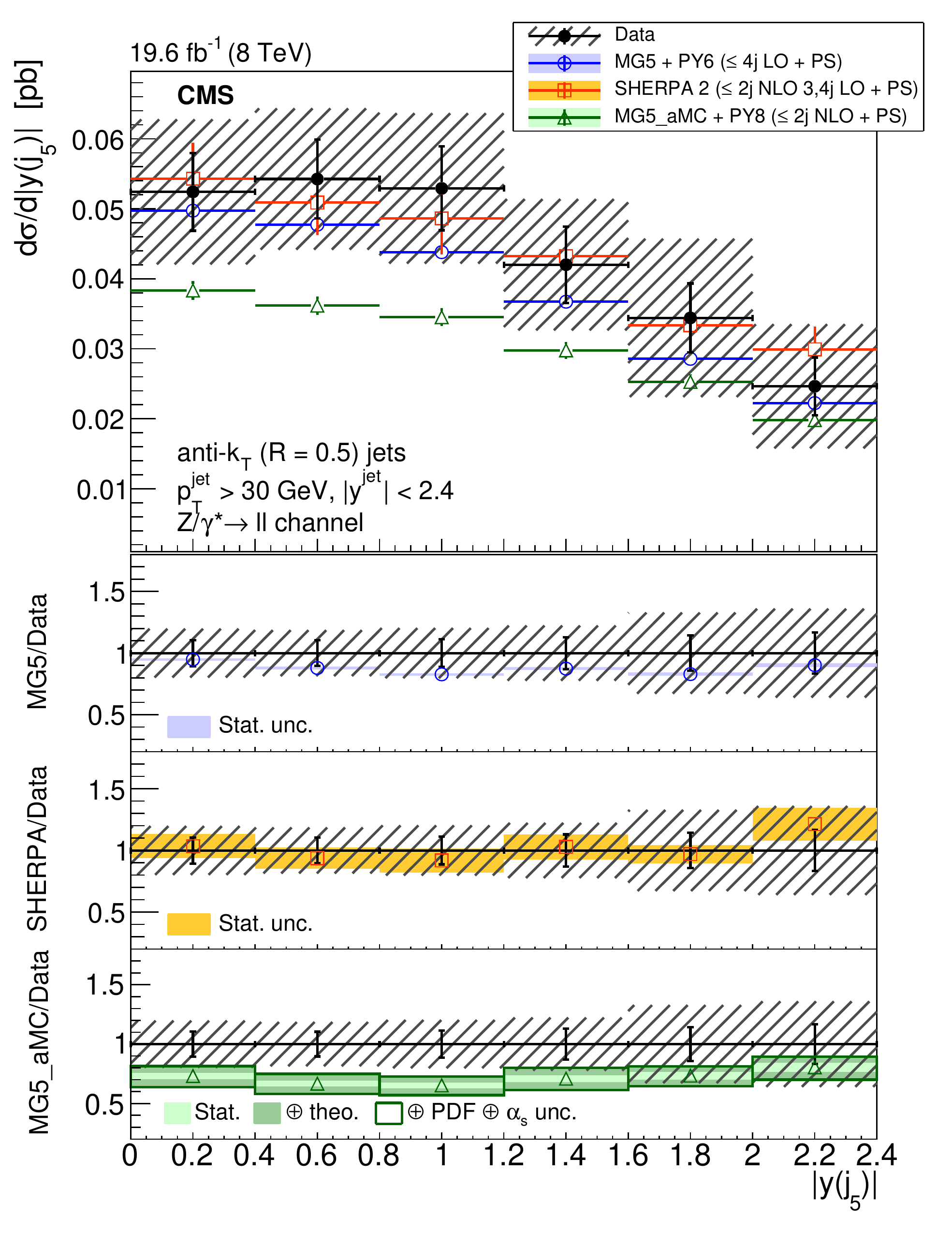}

\caption{The differential cross section for \zlljets production measured as a function of the $5^{\text{th}}$ jet $\vert y \vert$ compared to the predictions calculated with \MADGRAPH~5 + \PYTHIA~6, \SHERPA~2, and \MGaMC + \PYTHIA~8. \plotstdcapt}
\label{fig:CombXSec_FifthJetEta}
\end{figure}

The differential cross sections as a function of the absolute rapidity of the first, second, third, fourth, and fifth jets are presented in Figs.~\ref{fig:CombXSec_FirstJetEta}, \ref{fig:CombXSec_ThirdJetEta}, and \ref{fig:CombXSec_FifthJetEta}, including all events with at least one, two, three, four, and five jets. The differential cross sections in $\abs{y}$ have similar shapes for all jets while they vary by about a factor 2 in the range from 0 to 2.4.

The predictions obtained with \SHERPA~2 provide the best overall description regarding the shape of data distributions.
The predictions of both \MADGRAPH~5 + \PYTHIA~6 and \MGaMC + \PYTHIA~8 have a more central distribution than is measured for jets 1 to 4, although this behaviour is less pronounced for the latter. The difference could be attributed to the different showering methods and the different PDF choices for the three predictions. Given the experimental uncertainties, the shape of the spectrum of the $5^{\text{th}}$ jet rapidity is equally well described by the three calculations.

\begin{figure}[h!t]
\centering
{\includegraphics[width=0.32\textwidth]{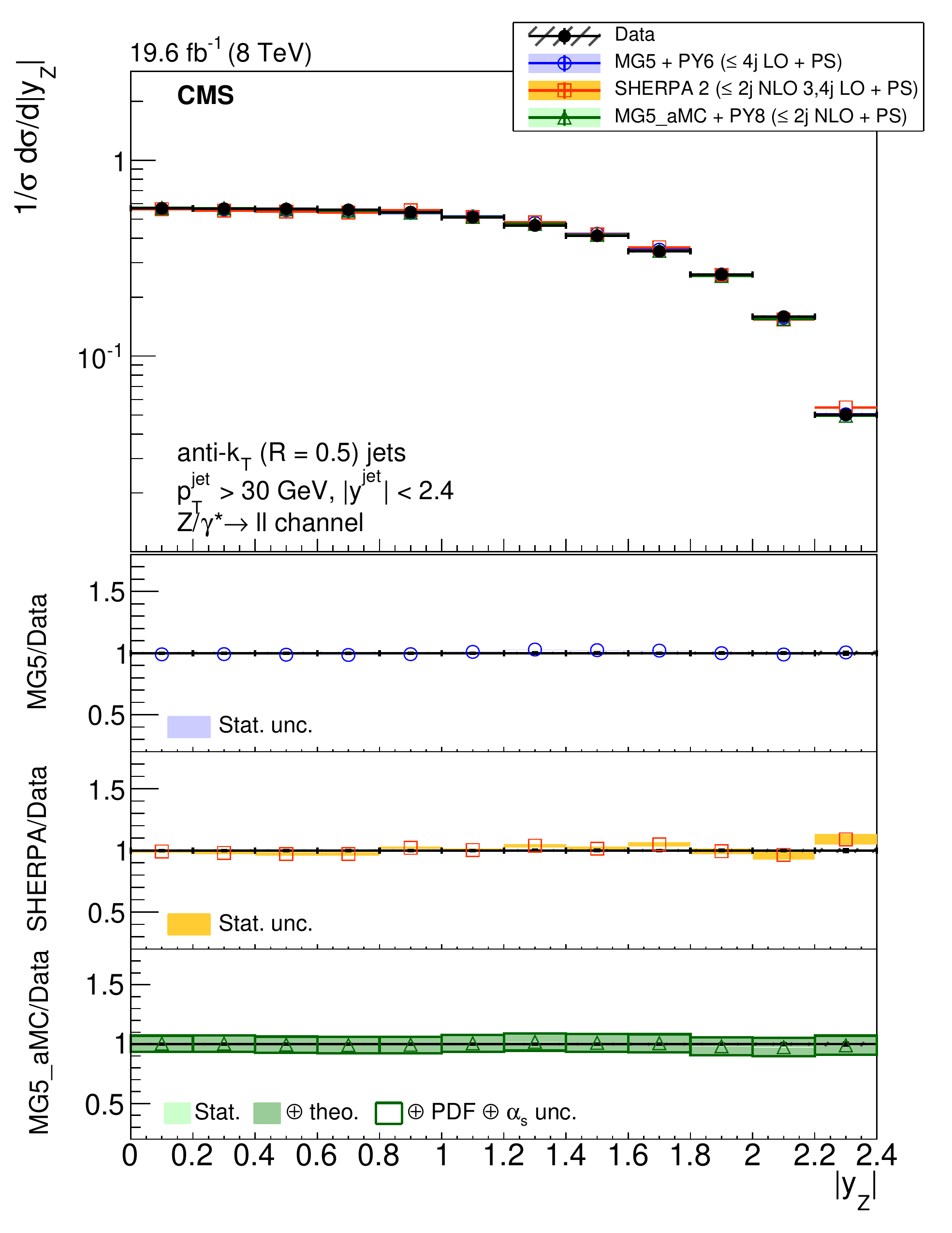}}
{\includegraphics[width=0.32\textwidth]{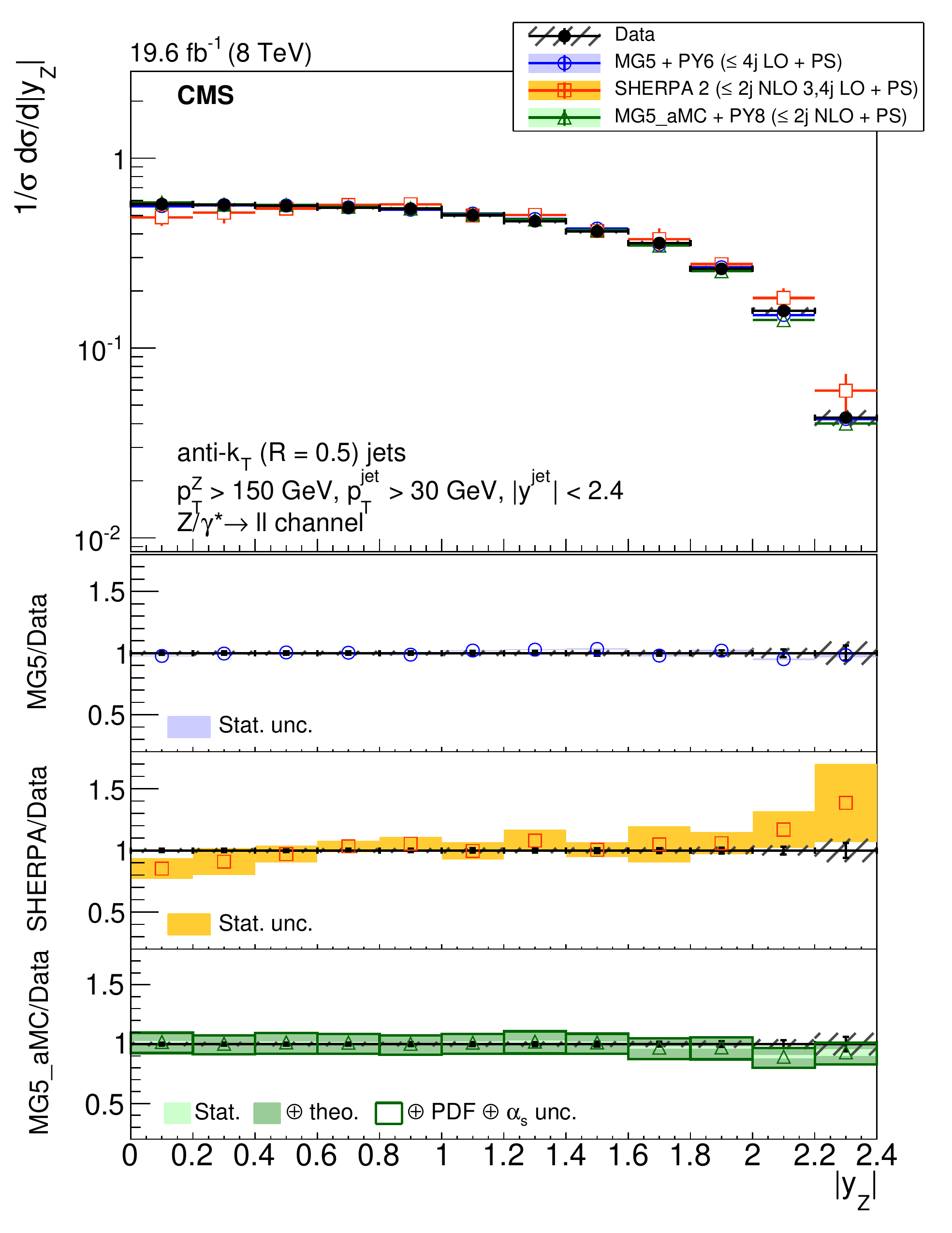}}
{\includegraphics[width=0.32\textwidth]{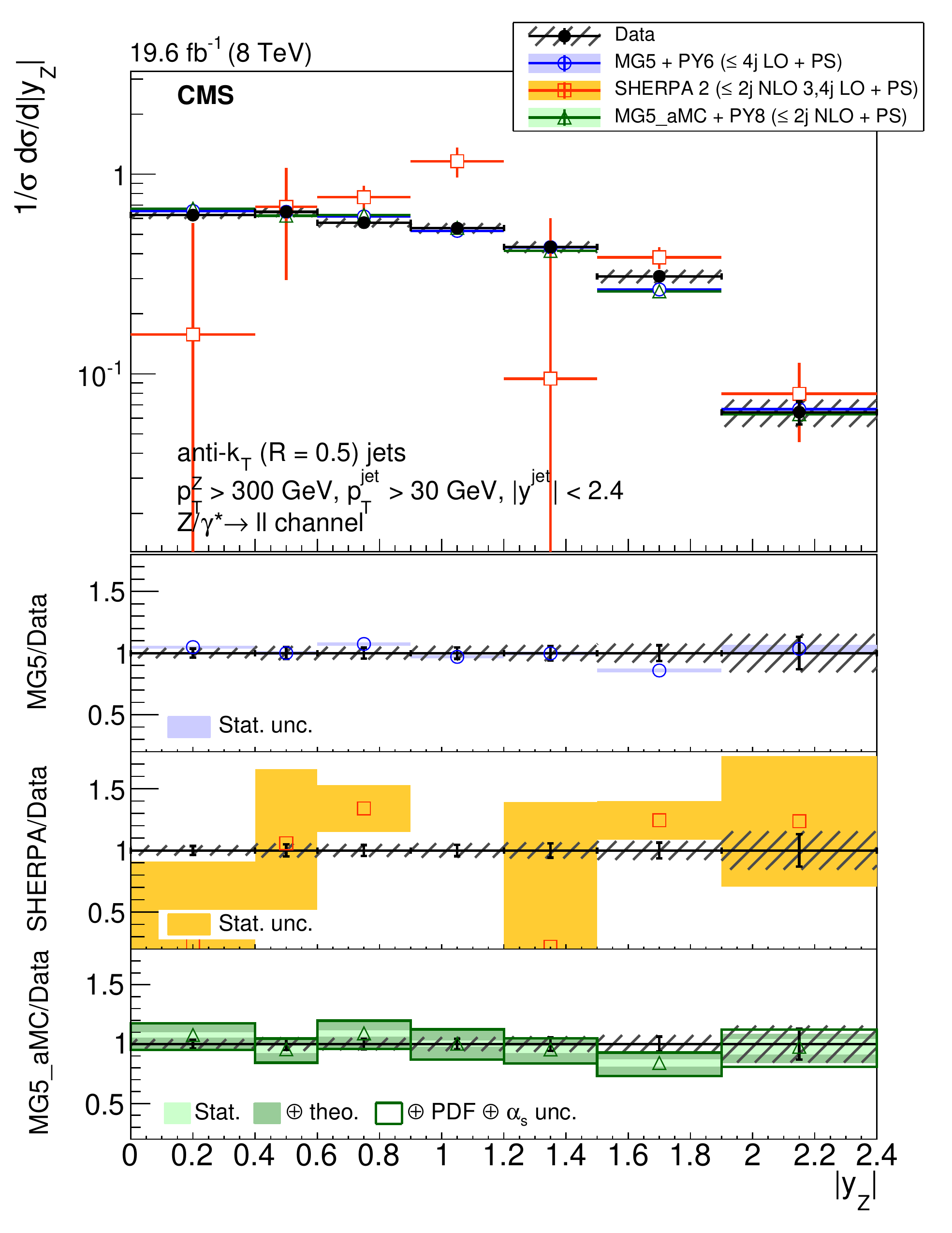}}
\caption{The normalised differential cross section for \zlljets production measured as a function of $\cPZ$ boson rapidity compared to the predictions calculated with \MADGRAPH~5 + \PYTHIA~6, \SHERPA~2, and \MGaMC + \PYTHIA~8. The cross section is measured (left) inclusively with respect to the $\cPZ$ boson \pt, (middle) for $\pt>150\GeV$, and (right) for $\pt>300\GeV$. \plotstdcapt}
 \label{fig:xsecabsyz}
\end{figure}

The $\cPZ$ boson rapidity distribution is presented in Fig.~\ref{fig:xsecabsyz} with no requirement on the $\cPZ$ boson transverse momentum. To minimize the uncertainties the measurement is done for the normalized distributions.
The relative contributions of matrix elements and parton shower depend on the $\cPZ$ transverse momentum. The measurement is also performed with a lower limit of 150 and 300\GeV on the $\cPZ$ boson transverse momentum. Each distribution is normalised to unity. The three calculations are in very good agreement with the measured values. The agreement of the prediction calculated with \SHERPA~2 degrades when applying a threshold on the $\cPZ$ boson \pt, though it is still consistent with data within the statistical uncertainty.

\begin{figure}[!hptb]
\centering
{\includegraphics[width=0.32\textwidth]{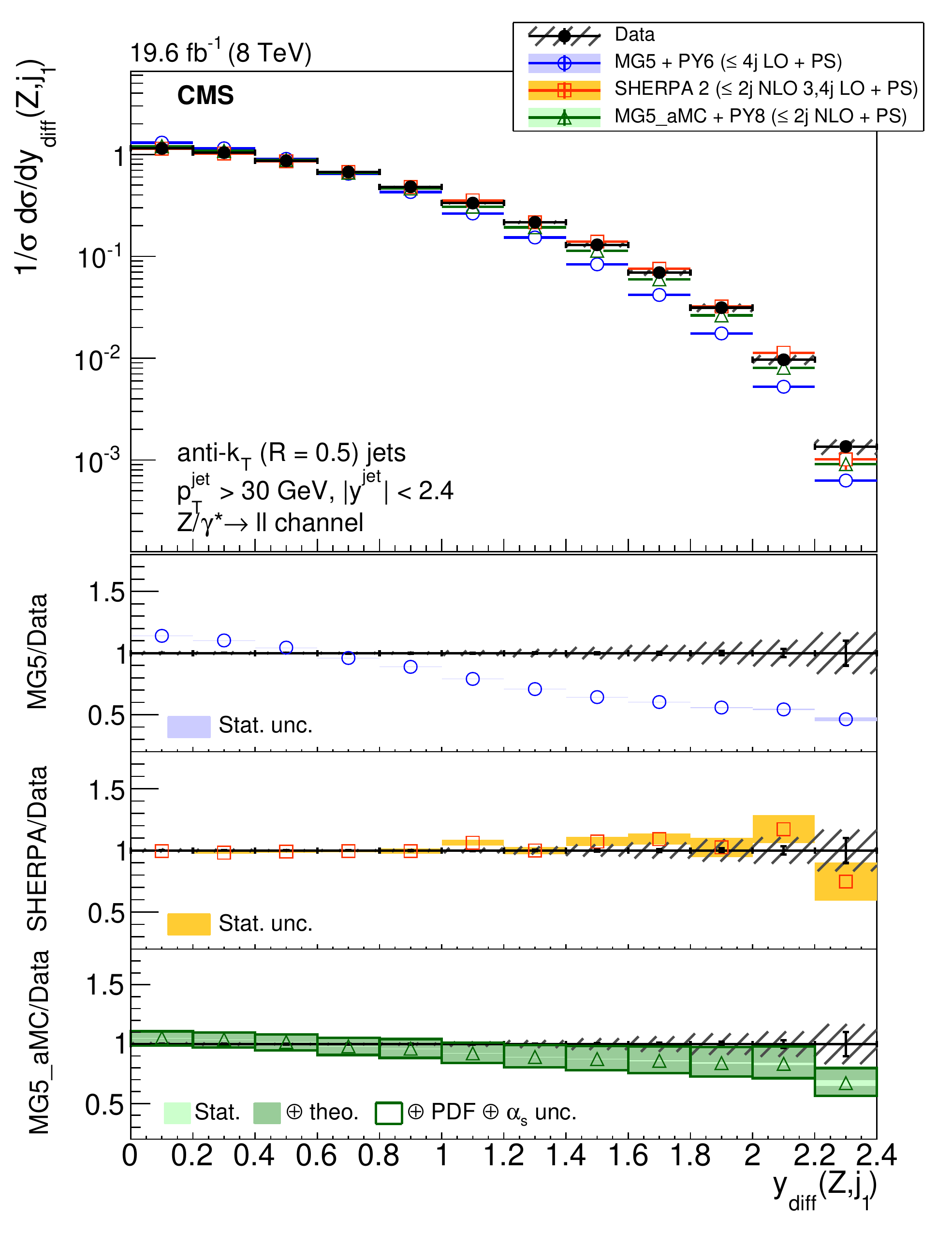}}
{\includegraphics[width=0.32\textwidth]{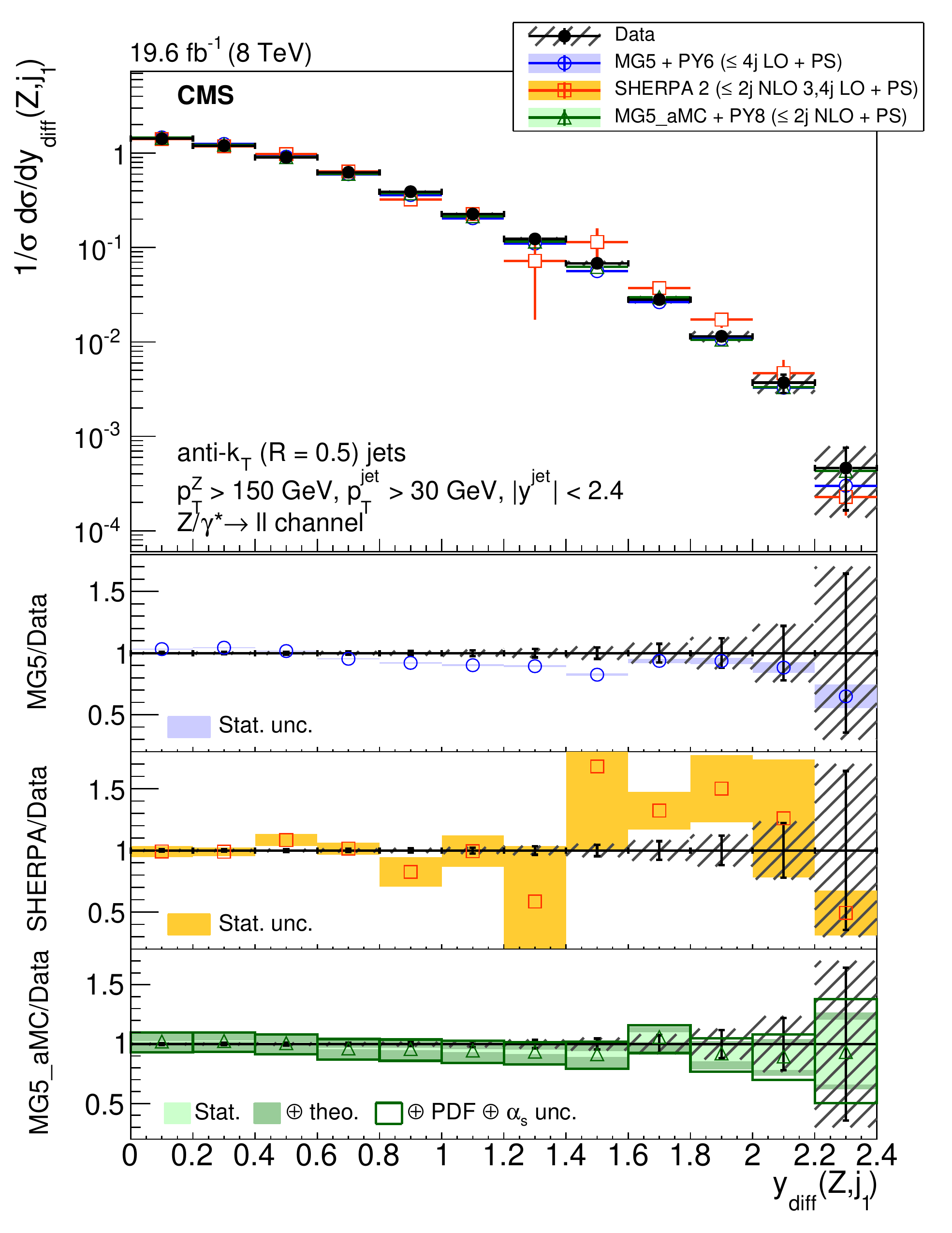}}
{\includegraphics[width=0.32\textwidth]{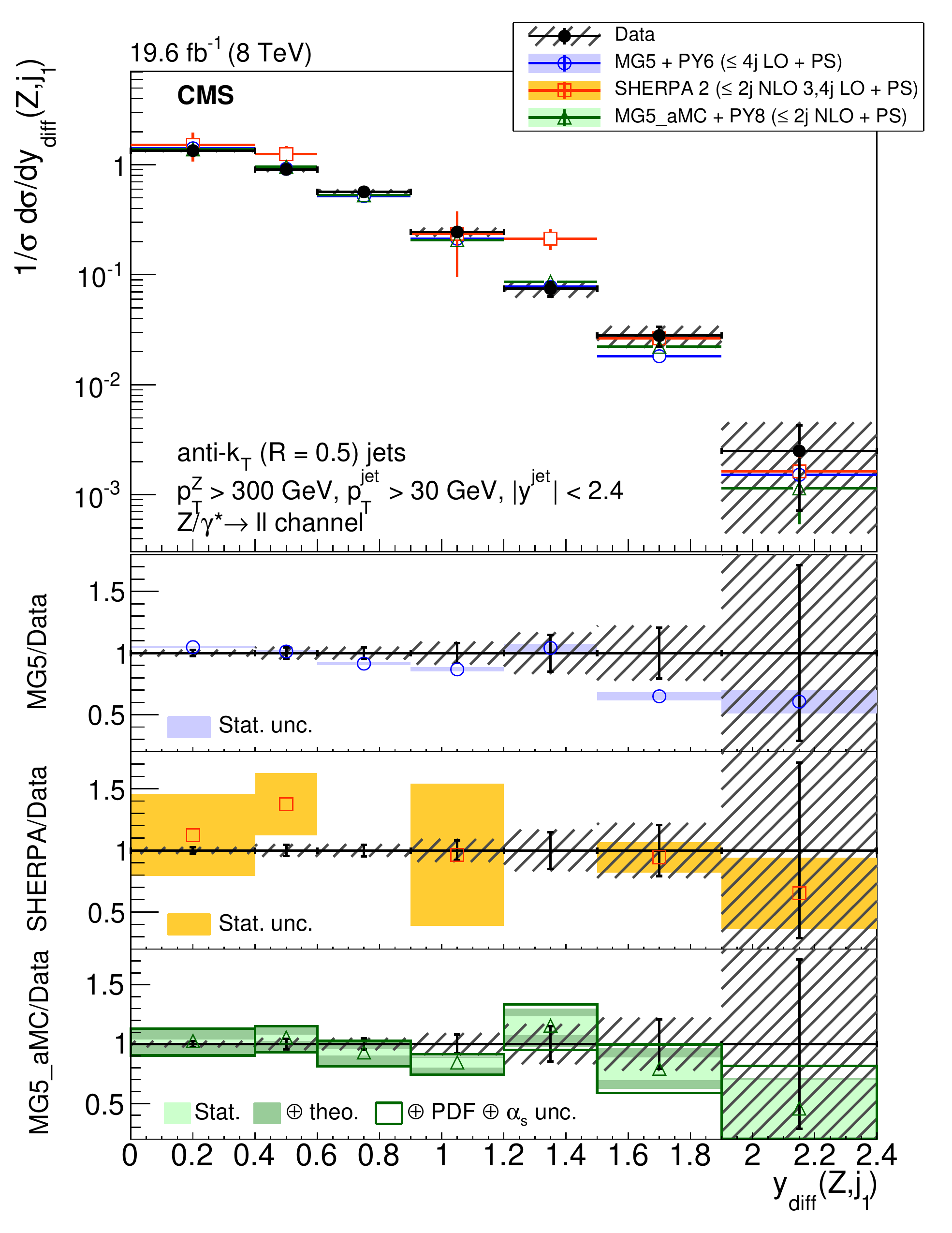}}
\caption{The normalised differential cross section for \zlljets ($N_{\text{jets}} \geq$ 1) production measured as a function of the \ydiff of the $\cPZ$ boson and the leading jet compared to the predictions calculated with \MADGRAPH~5 + \PYTHIA~6, \SHERPA~2, and \MGaMC + \PYTHIA~8. (left) The cross section is measured inclusively with respect to the $\cPZ$ boson \pt and for two different $\pt(\cPZ)$ thresholds. The ratio of the prediction to the measurements is shown for (left) $\pt>0\GeV$, (middle) $\pt>150\GeV$, and (right) $\pt>300\GeV$. \plotstdcapt}
\label{fig:ydiff_Z_j1}
\end{figure}

\begin{figure}[!hptb]
 \centering
{\includegraphics[width=0.32\textwidth]{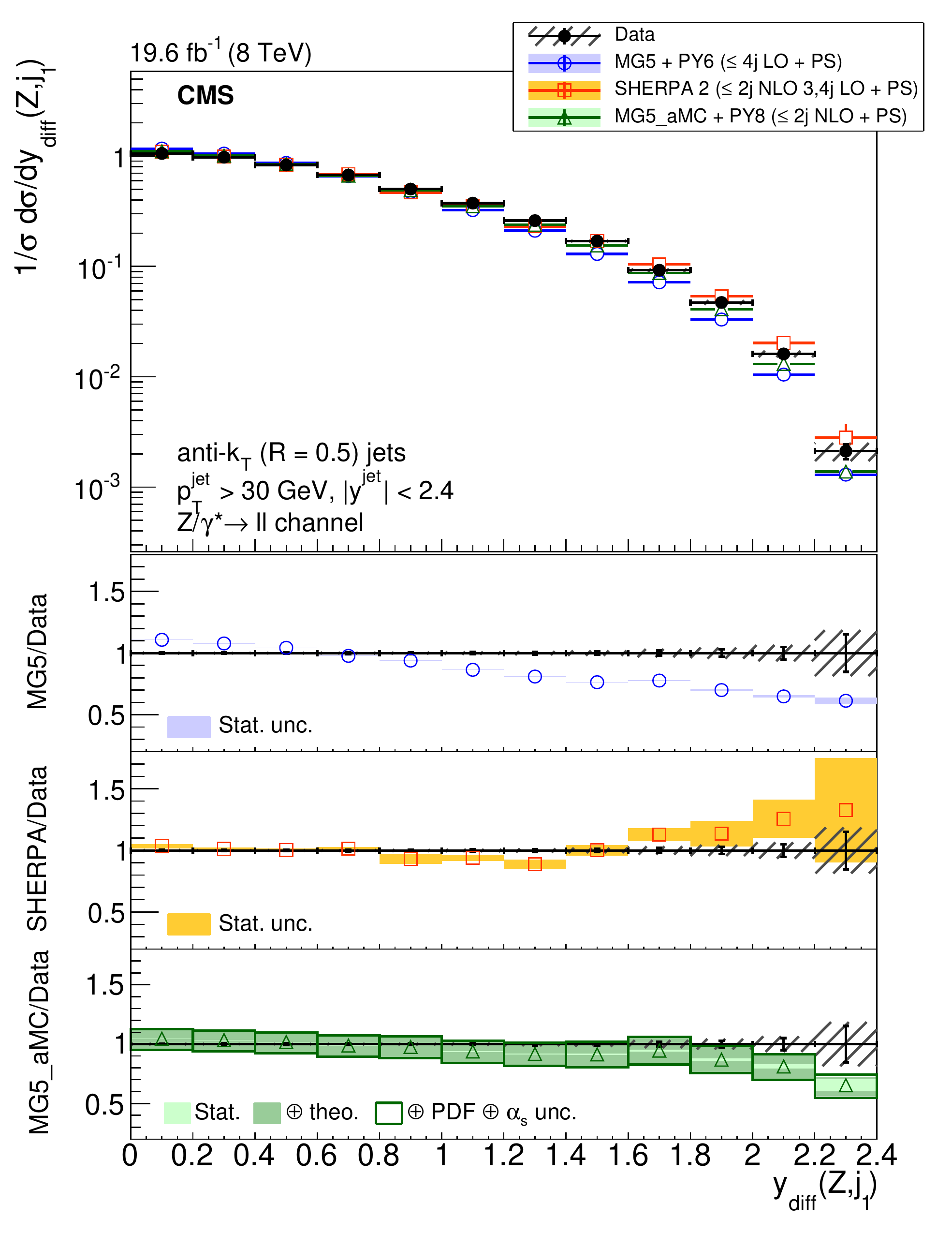}}
{\includegraphics[width=0.32\textwidth]{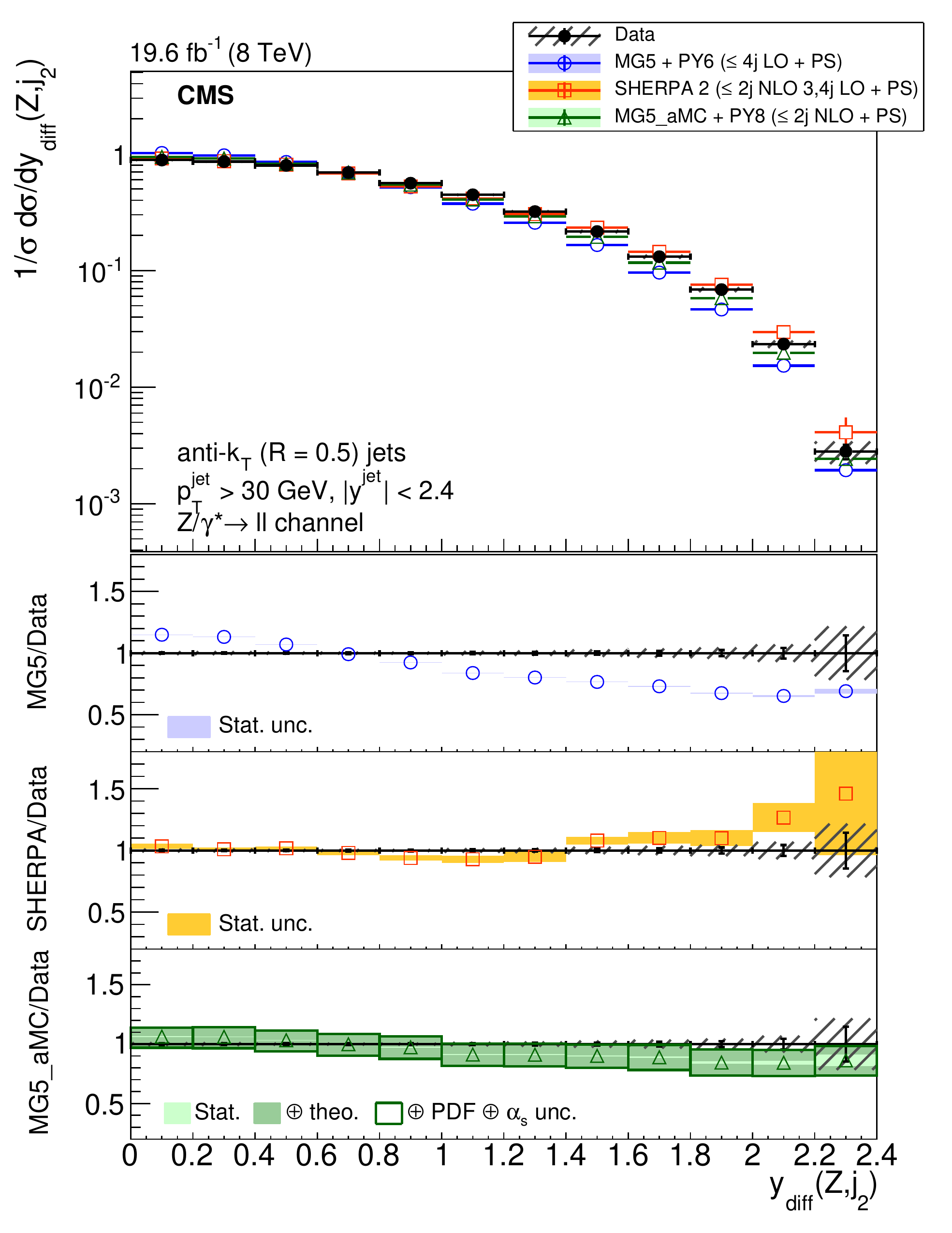}}
{\includegraphics[width=0.32\textwidth]{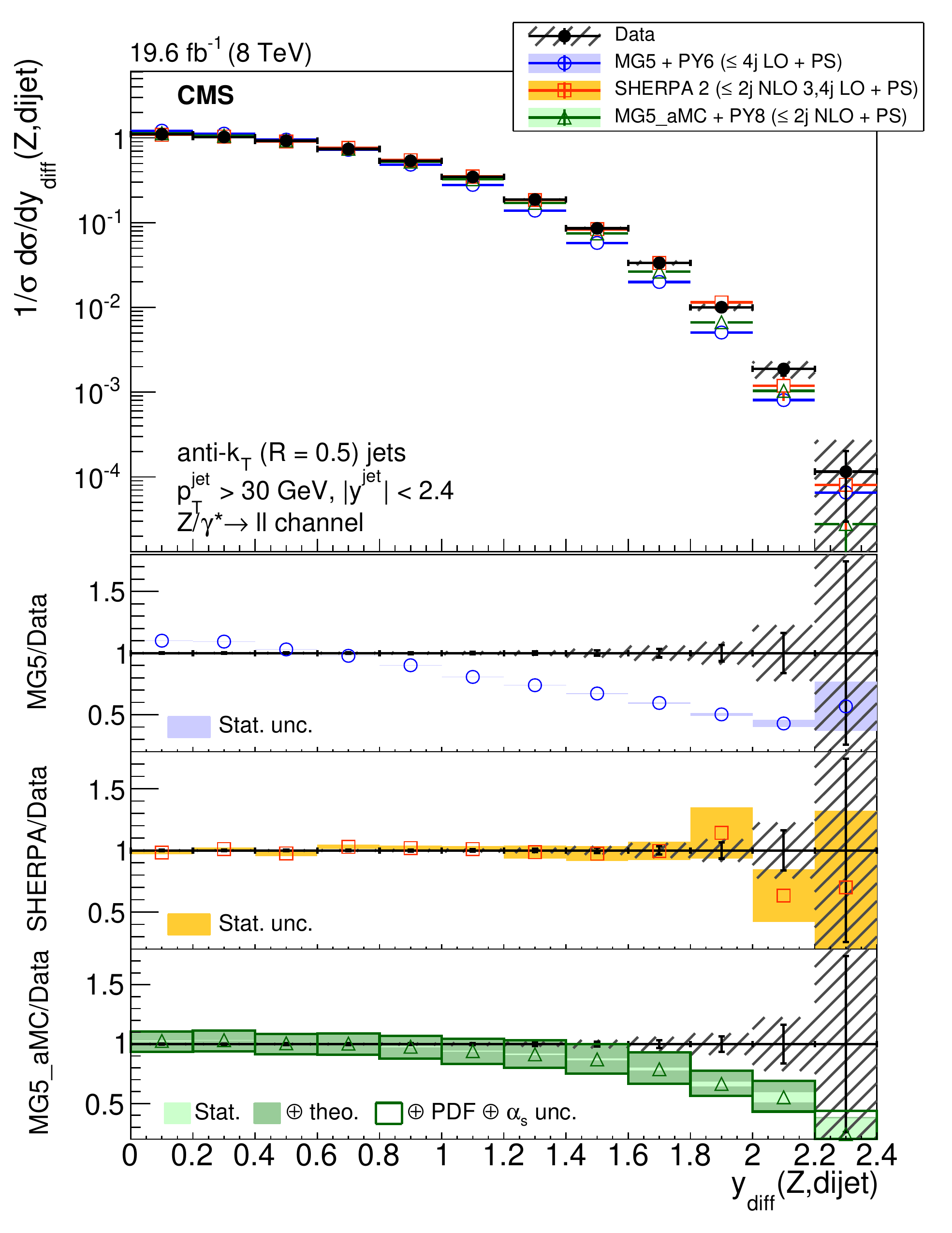}}
\caption{The normalised differential cross section for \zlljets ($N_{\text{jets}} \geq$ 2) production measured as a function of the \ydiff of the $\cPZ$ boson and (left) the leading jet, (middle) the second-leading jet, and (right) the system constituted by these two jets. The measurement is compared to the predictions calculated with \MADGRAPH~5 + \PYTHIA~6, \SHERPA~2, and \MGaMC + \PYTHIA~8. \plotstdcapt}
\label{fig:ydiff_Z_2j}
\end{figure}

\begin{figure}[!hptb]
   \centering
   {\includegraphics[width=0.48\textwidth]{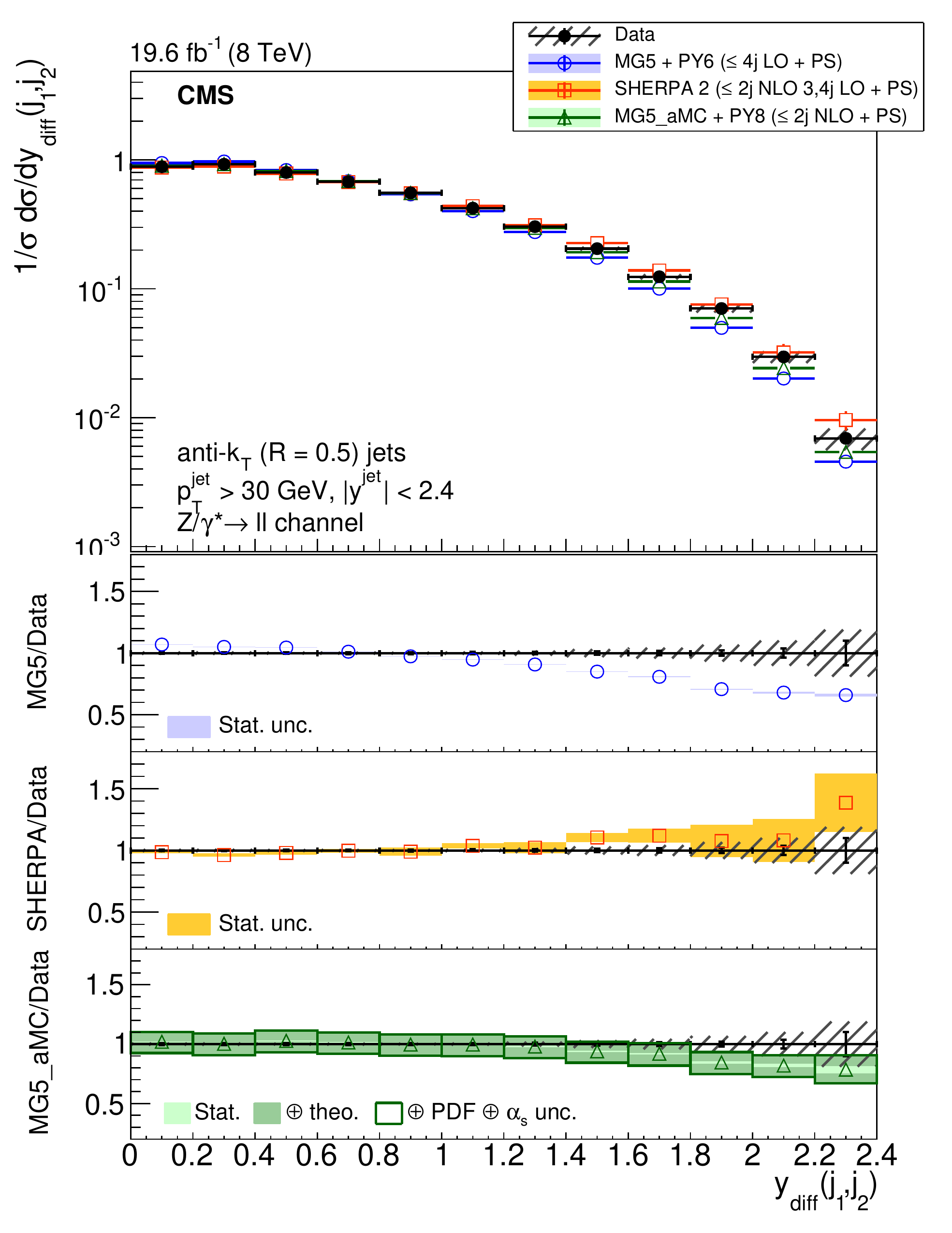}}
   {\includegraphics[width=0.48\textwidth]{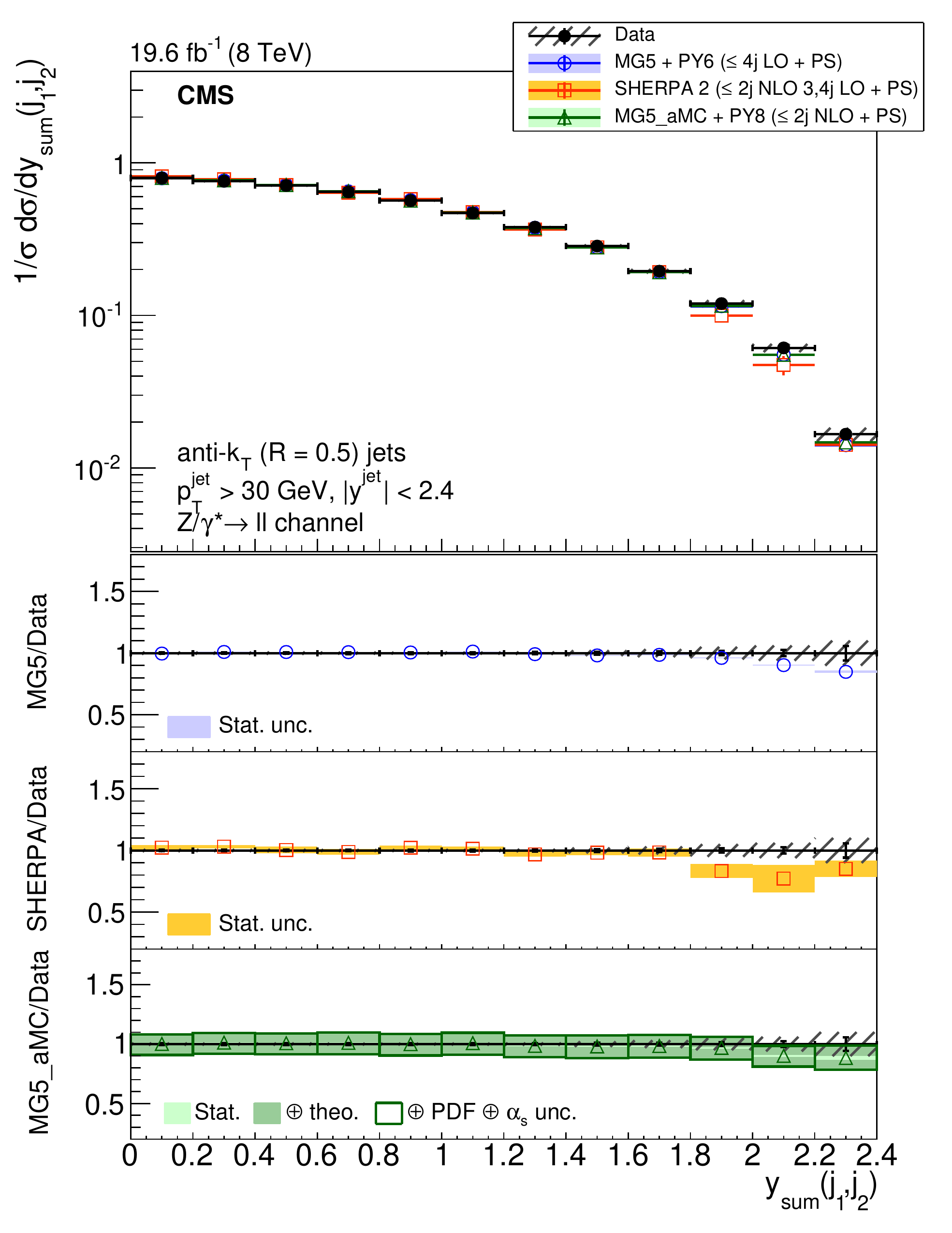}}
   \caption{The normalised differential cross section for \zlljets ($N_{\text{jets}} \geq$ 2) production measured as a function of the (left) \ydiff and (right) \ysum of the two leading jets. The measurement is compared to the predictions calculated with \MADGRAPH~5 + \PYTHIA~6, \SHERPA~2, and \MGaMC + \PYTHIA~8. \plotstdcapt}
\label{fig:ysum_ydiff_j1j2}
\end{figure}

\begin{figure}[!hptb]
 \centering
 {\includegraphics[width=0.32\textwidth]{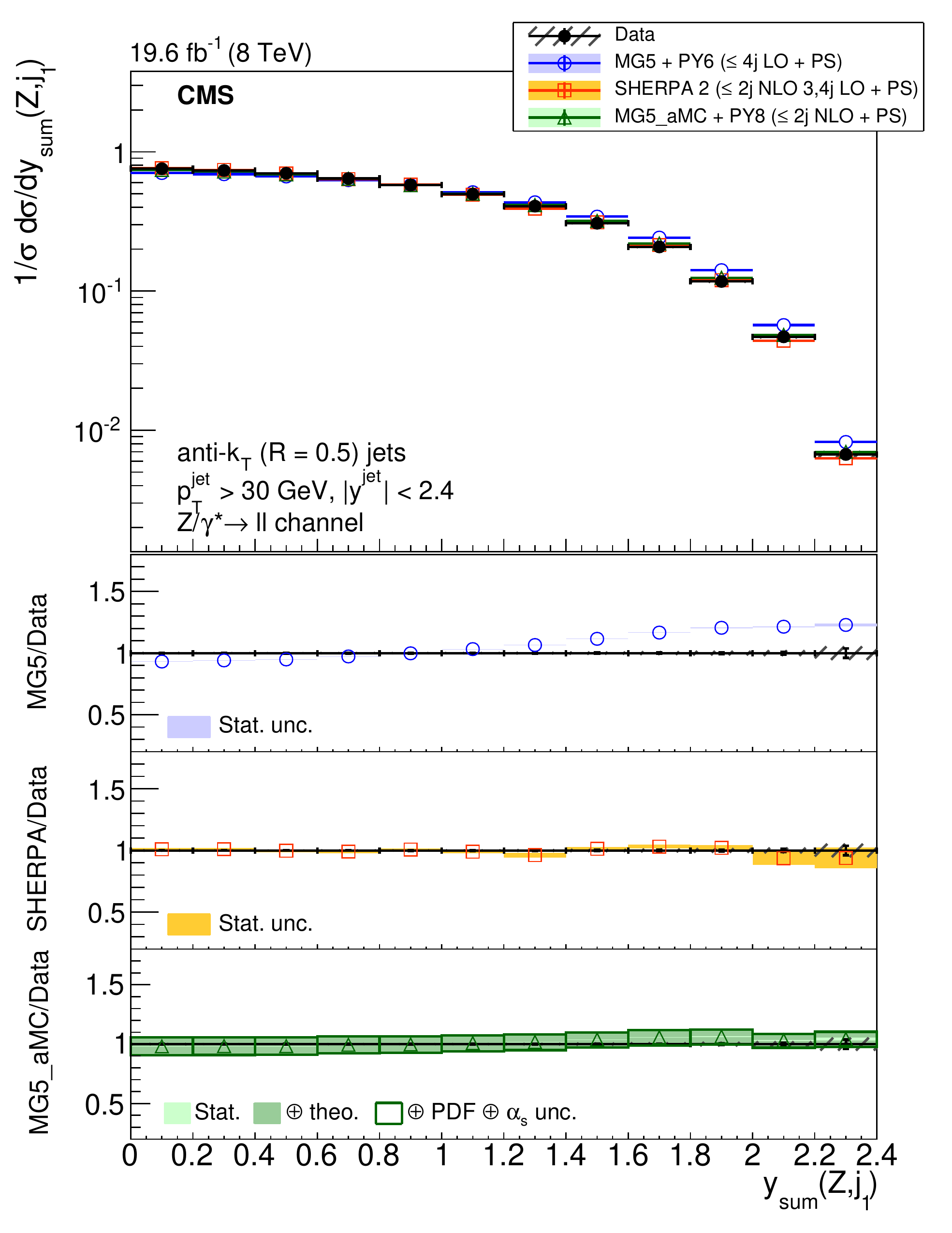}}
 {\includegraphics[width=0.32\textwidth]{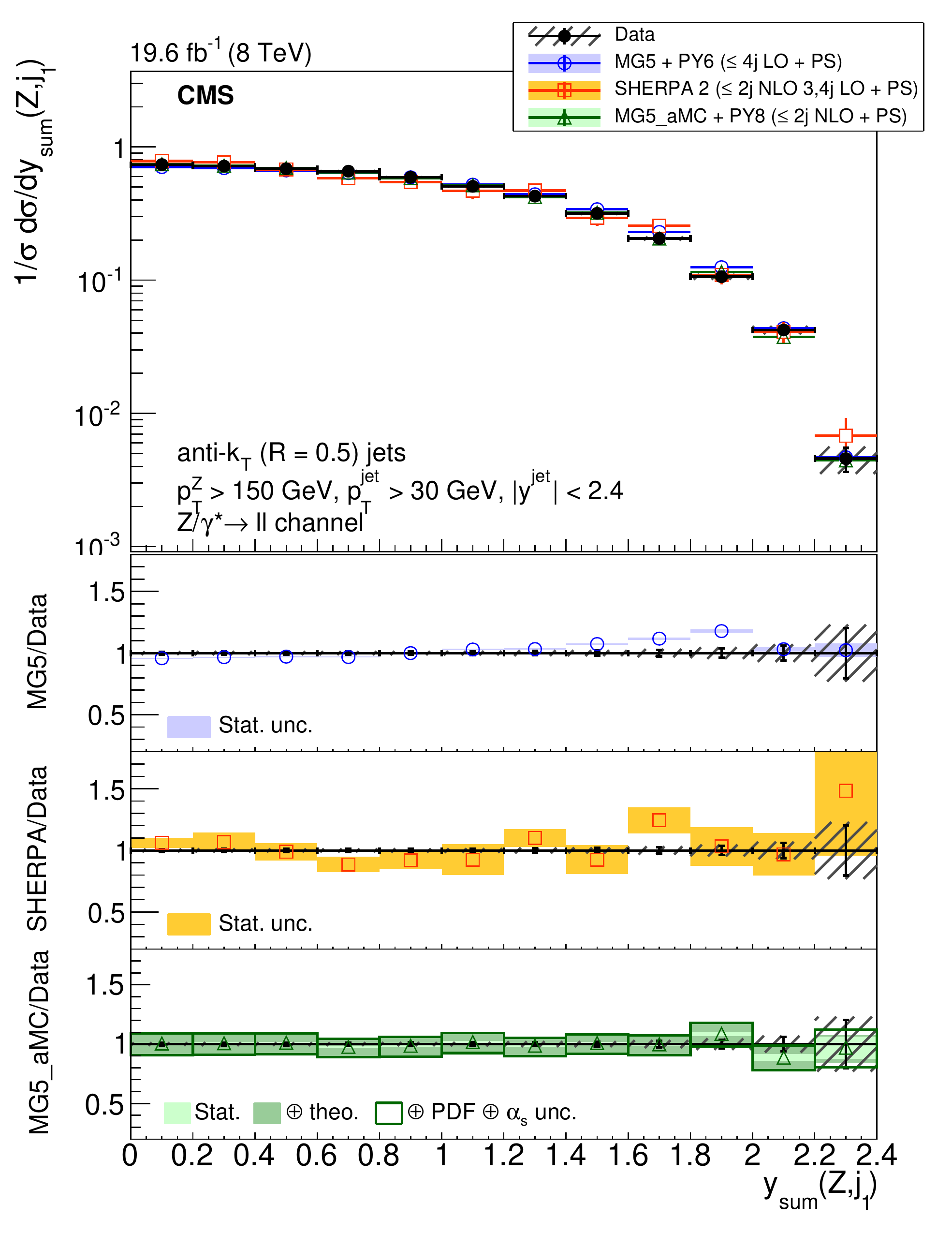}}
 {\includegraphics[width=0.32\textwidth]{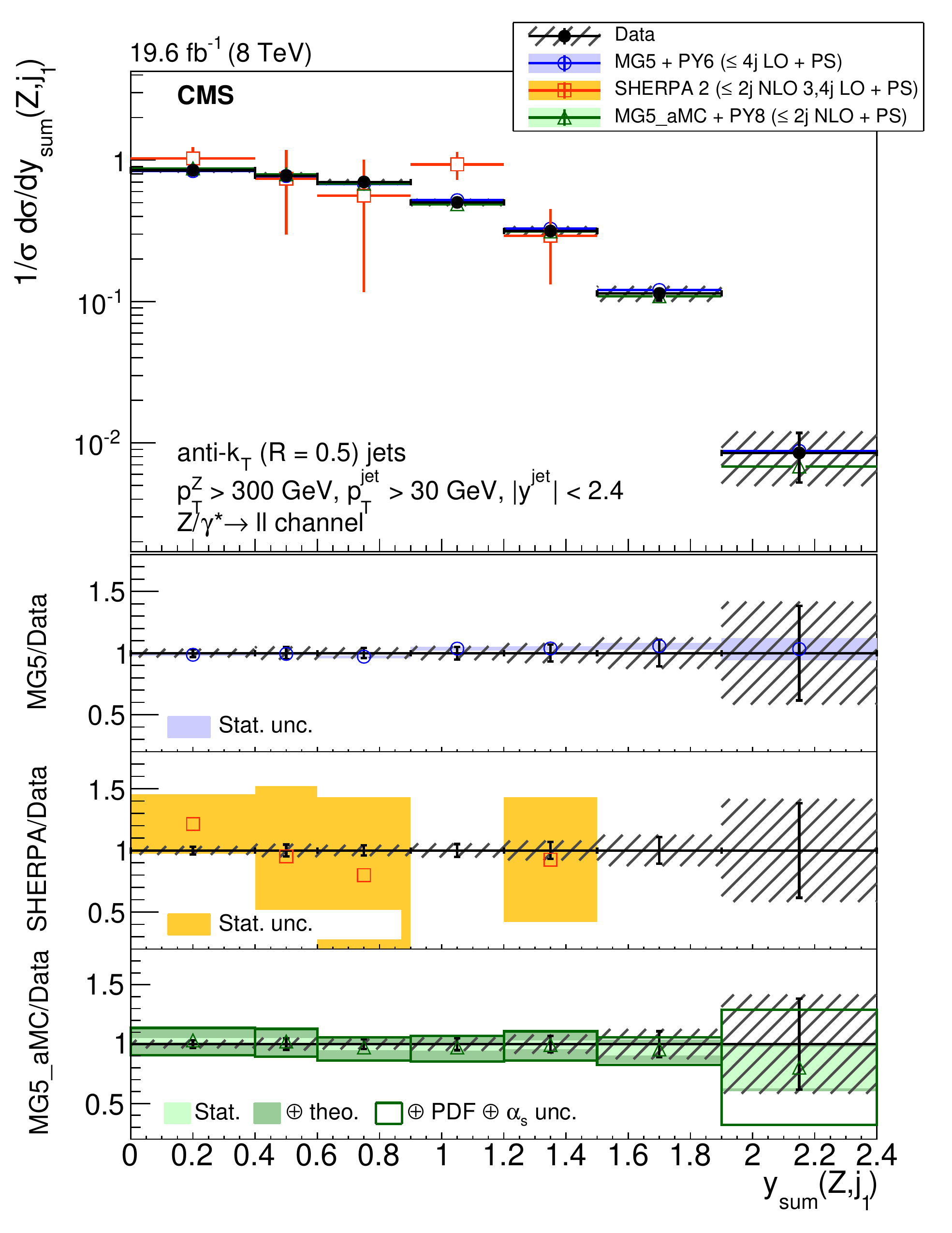}}
  \caption{The normalised differential cross section for \zlljets ($N_{\text{jets}} \geq 1$) production measured as a function of the \ysum of the $\cPZ$ boson and the leading jet compared to the predictions calculated with \MADGRAPH~5 + \PYTHIA~6, \SHERPA~2, and \MGaMC + \PYTHIA~8. The cross section is measured inclusively with respect to the $\cPZ$ boson \pt and for two different $\pt(\cPZ)$ thresholds. The ratio of the prediction to the measurements is shown for (left) $\pt>0\GeV$, (middle) $\pt>150\GeV$, and (right) $\pt>300\GeV$. \plotstdcapt}
\label{fig:ysum_Z_j1}
 \end{figure}

\begin{figure}[!hptb]
   \centering
\includegraphics[width=0.32\textwidth]{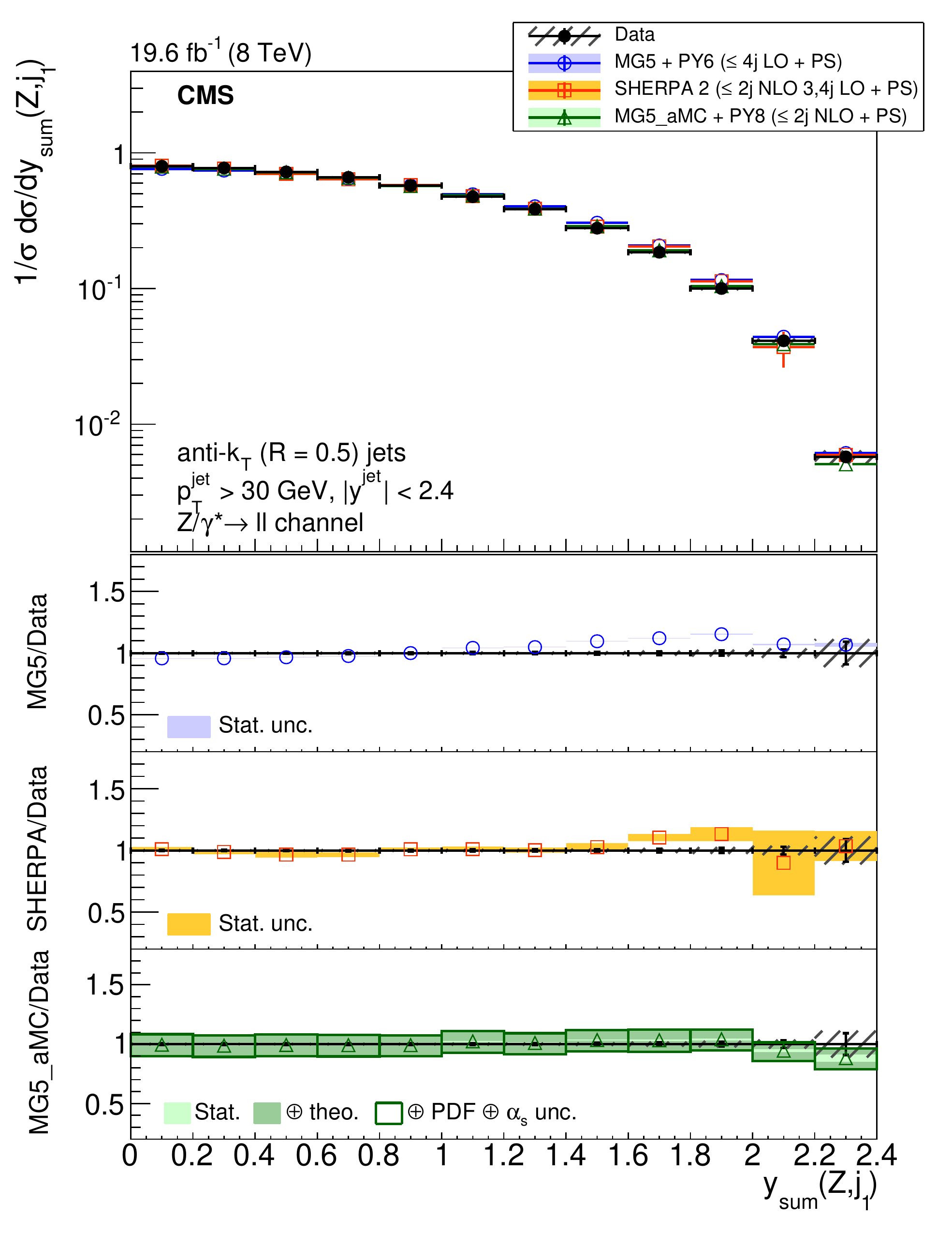}
\includegraphics[width=0.32\textwidth]{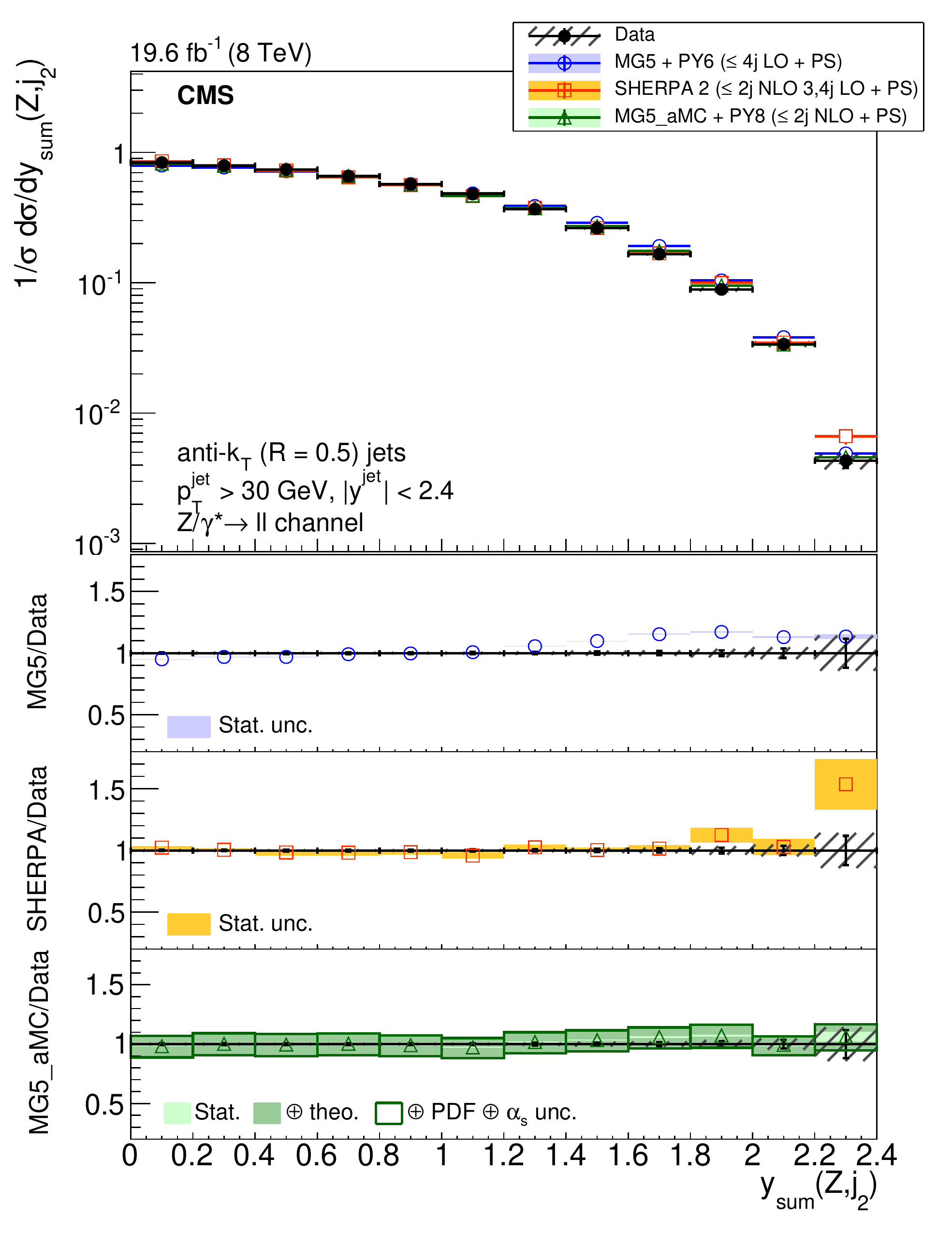}
\includegraphics[width=0.32\textwidth]{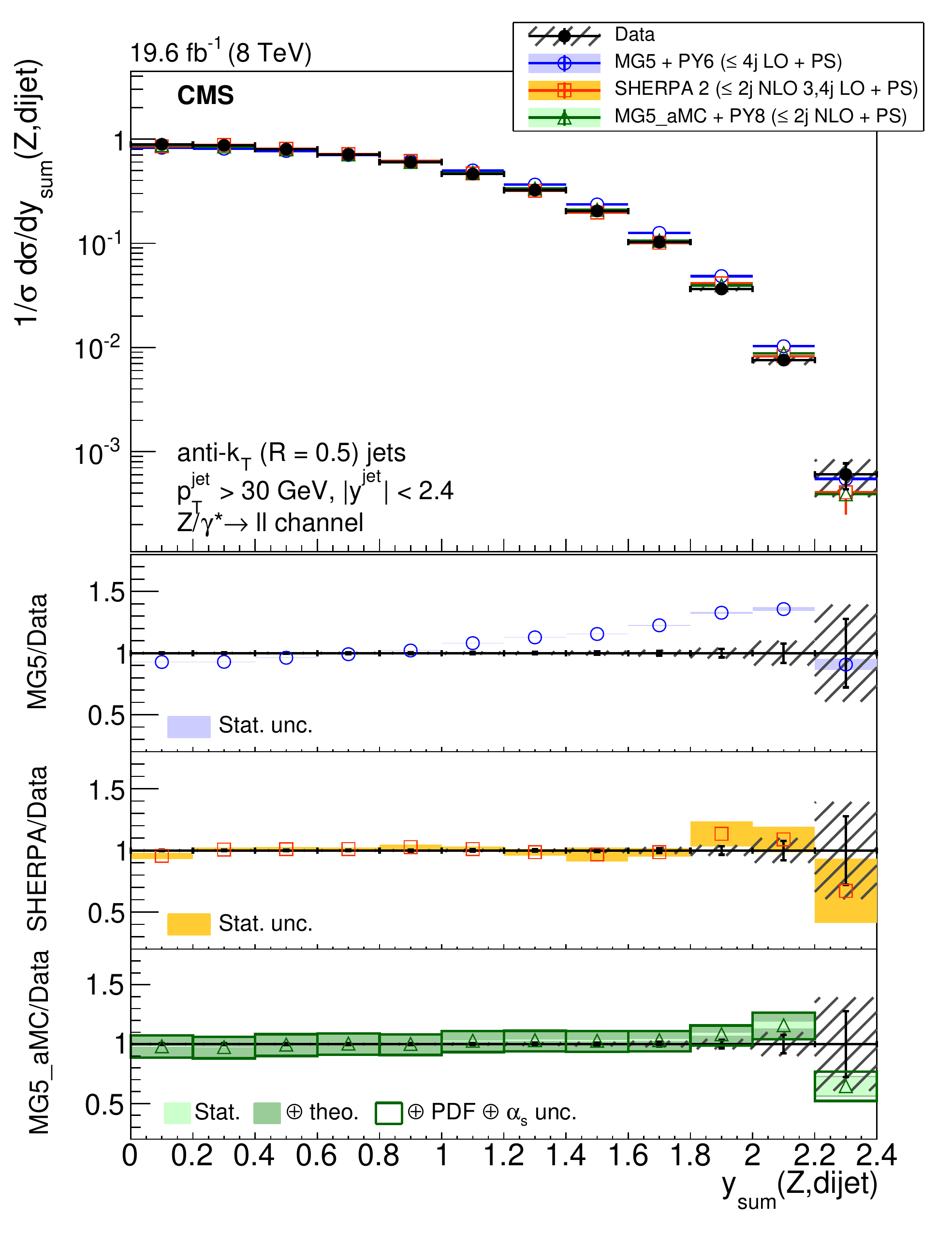}
\caption{The normalised differential cross section for \zlljets ($N_{\text{jets}} \geq$ 2) production measured as a function of the \ysum of the $\cPZ$ boson and (left) the leading jet, (middle) the second-leading jet, and (right) the system constituted by these two jets. The measurement is compared to the predictions calculated with \MADGRAPH~5 + \PYTHIA~6, \SHERPA~2, and \MGaMC + \PYTHIA~8. \plotstdcapt}
\label{fig:ysum_Z_2j}
 \end{figure}

The correlations in rapidity between the different objects ($\cPZ$ boson and jets) are shown in Figs.~\ref{fig:ydiff_Z_j1} to \ref{fig:ysum_Z_2j}. The normalised cross section is presented as a function of the rapidity difference between the $\cPZ$ boson and the leading jet, $\ydiff(\Z,\text{j}_1) = 0.5 |y(\cPZ) - y({\text{j}_1)}|$ in Fig.~\ref{fig:ydiff_Z_j1}. A large discrepancy is observed between the measured cross section and that predicted by \MADGRAPH~5 + \PYTHIA~6. Such an effect was
previously observed at $\sqrt{s} =7\TeV$~\cite{Chatrchyan:2013oda} and is confirmed here with an increased statistical precision and with an extended range in $\ydiff(\cPZ,\text{j}_1)$. The discrepancy is significantly reduced when a threshold is applied to the transverse momentum of the $\cPZ$ boson as shown in the same figure.  This observation supports the attribution of the discrepancy to the matching procedure between the ME and PS, as discussed in
\cite{Chatrchyan:2013oda}.  By contrast, a quite good agreement is found, independently of any threshold on the $\cPZ$ boson transverse momentum, for the NLO predictions of \SHERPA~2, and \MGaMC + \PYTHIA~8. This improvement is expected to come from additional diagrams at NLO with a gluon propagator in the $t$-channel that populate the forward rapidity regions.

The presence of additional jets in the event should reduce the dependence on the ME/PS matching for the first jet since this jet will have a larger \pt on average. Figure~\ref{fig:ydiff_Z_2j} shows the normalised cross section for Z production with at least two jets as a function of the rapidity difference between the $\cPZ$ boson and the leading jet, $\ydiff(\cPZ,\text{j}_1)$, between the $\cPZ$ boson and the second-leading jet, $\ydiff(\cPZ,\text{j}_2)$, and between the $\cPZ$
boson and the system formed by the two leading jets, $\ydiff(\cPZ,\text{dijet})$.
The discrepancies between the measured cross sections and the \MADGRAPH~5 + \PYTHIA~6 predictions are present in all three cases, but they are less pronounced than in the one-jet case (Fig.~\ref{fig:ydiff_Z_2j}a compared to Fig.~\ref{fig:ydiff_Z_j1}a). The NLO predictions from \SHERPA~2 and \MGaMC + \PYTHIA~8 reproduce the measured dependencies much better than \MADGRAPH~5 + \PYTHIA~6 does.

The rapidity correlation of the two leading jets, independently of the $\cPZ$ boson rapidity, is displayed in Fig.~\ref{fig:ysum_ydiff_j1j2}, showing the rapidity sum and rapidity difference between the two jets. There is a good agreement between the measured cross section and the three predictions for the rapidity sum dependence. The rapidity difference presents a discrepancy with \MADGRAPH~5 + \PYTHIA~6 at large values, while the NLO predictions of \SHERPA~2, and \MGaMC + \PYTHIA~8 are in good agreement with the data.

The rapidity sum for the system of the $\cPZ$ boson and the leading jet is studied with different thresholds applied to the transverse momentum of the $\cPZ$ boson. Figure~\ref{fig:ysum_Z_j1} shows the normalised cross section as a function of the rapidity sum of the $\cPZ$ boson and the leading jet, $\ysum(\Z,\text{j}_1) = 0.5 |y(\cPZ)+ y(\text{j}_1)|$ for $\cPZ$ boson transverse momentum above 0, 150, and 300\GeV. The observed discrepancy between the measured cross section and that
predicted by \MADGRAPH~5 + \PYTHIA~6 is similar to the effect that has been found at 7\TeV \cite{Chatrchyan:2013oda}, and is confirmed here with increased statistical precision.  The discrepancy almost vanishes when the transverse momentum of the $\cPZ$ boson is required to be larger than 150\GeV. The NLO predictions of \SHERPA~2, and \MGaMC + \PYTHIA~8 are in good agreement with the measured cross section independently of the $\cPZ$ boson transverse momentum. This improvement with respect to \MADGRAPH~5 + \PYTHIA~6 can be attributed to either the different PDF choice, or to the NLO terms.

For dijet events, Fig.~\ref{fig:ysum_Z_2j} shows cross sections as a function of rapidity sums, for the Z boson and the leading jet, for the Z boson and the second-leading jet, and for the Z boson and the dijet system of the two leading jets. Comparison between the measured cross sections and the \MADGRAPH~5 + \PYTHIA~6 predictions exhibit a small disagreement for a rapidity sum above 1 for each jet, and the discrepancies increase when the dijet system is considered. Comparison with NLO predictions from
\SHERPA~2, and from \MGaMC + \PYTHIA~8 shows a very good agreement.

The rapidity correlation study confirms the observations made at $\sqrt{s}=7\TeV$, and shows that the behaviour with respect to the tree-level prediction is similar for the correlation with the second jet and enhanced when considering the dijet system consisting of the two leading jets. The study demonstrates that the two NLO predictions improve the agreement with the measurements, especially for the rapidity difference observables.

\subsection{Differential cross section in jet \texorpdfstring{$\HT$}{HT}}

\begin{figure}[htbp!]
\centering
{\includegraphics[width=0.48\textwidth]{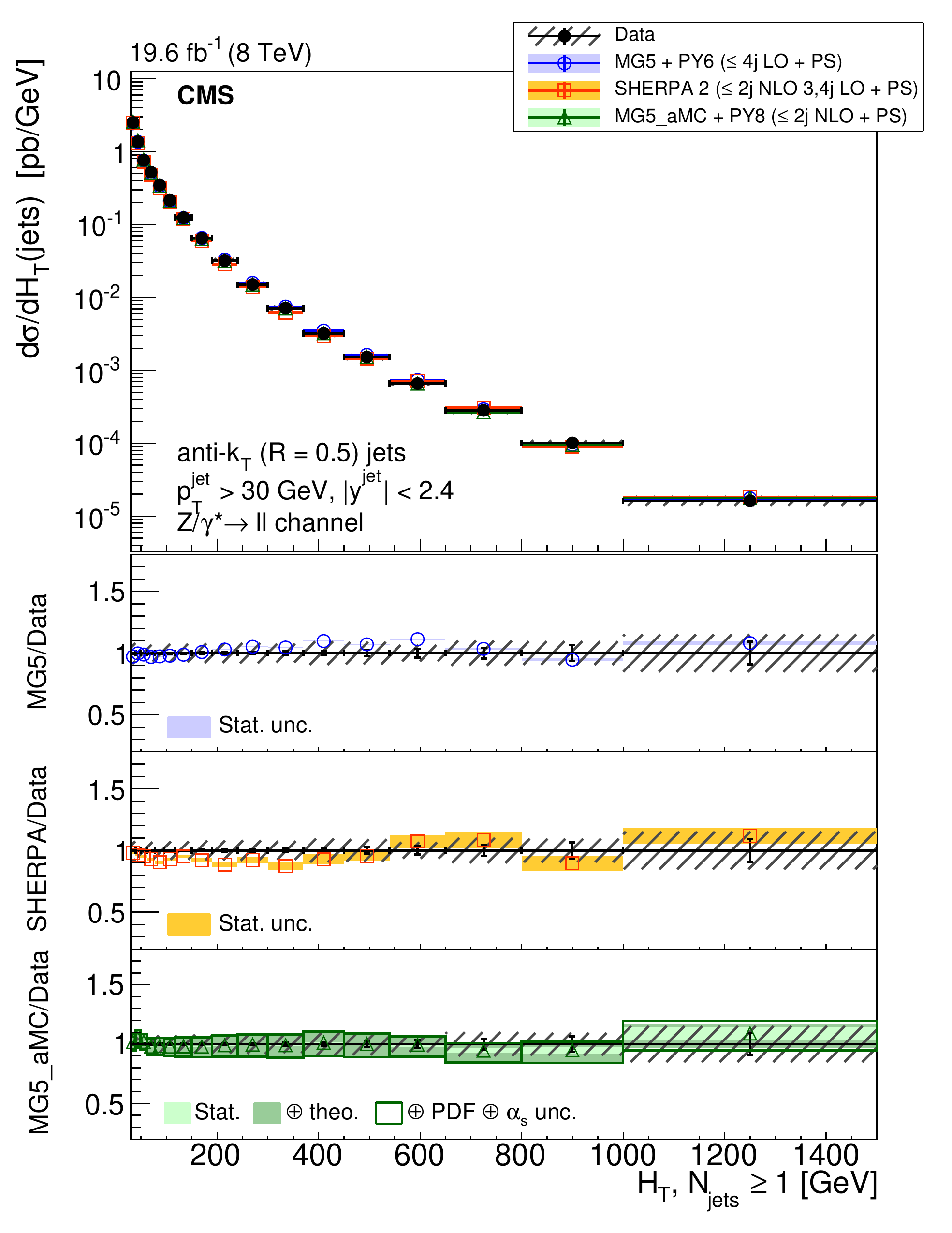}}
{\includegraphics[width=0.48\textwidth]{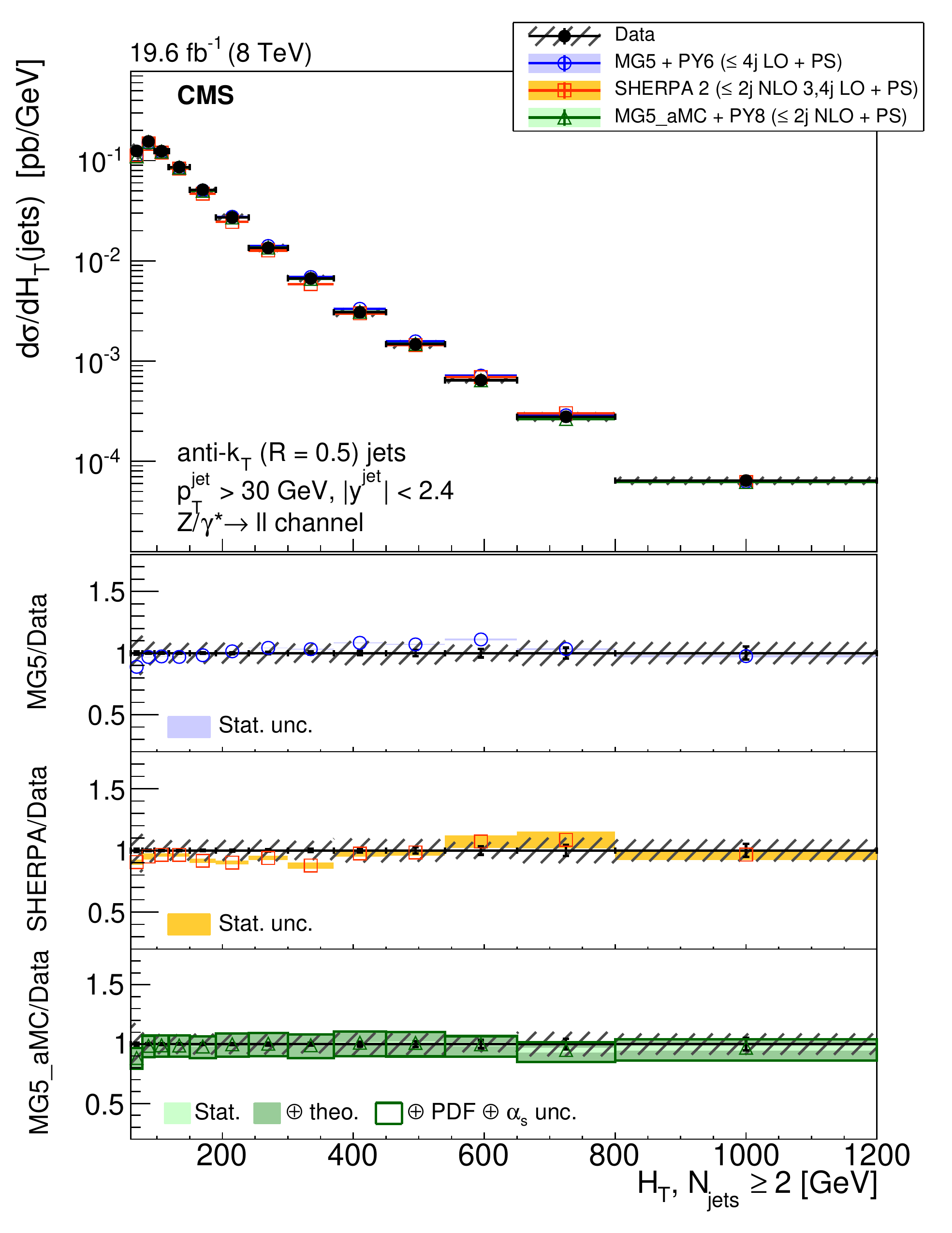}}
\caption{The differential cross section for \zlljets production measured as a function of \HT for (left) $N_{\text{jets}} \geq 1$ and (right) $N_{\text{jets}} \geq 2$ compared to the predictions calculated with \MADGRAPH~5 + \PYTHIA~6, \SHERPA~2, and \MGaMC + \PYTHIA~8. \plotstdcapt}
\label{fig:CombXSec_JetsHT_1and2}
\end{figure}

\begin{figure}[htbp!]
\centering
{\includegraphics[width=0.48\textwidth]{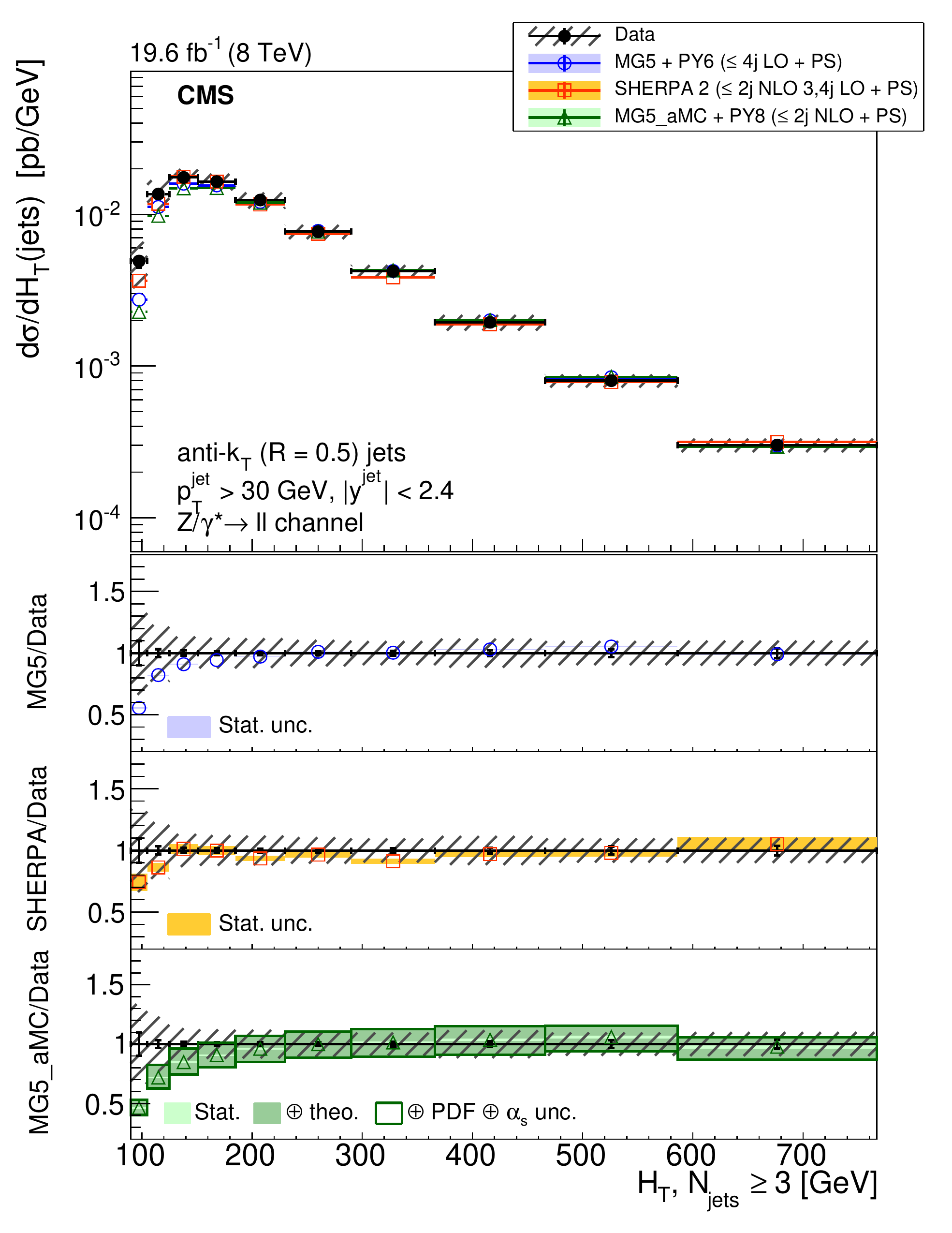}}
{\includegraphics[width=0.48\textwidth]{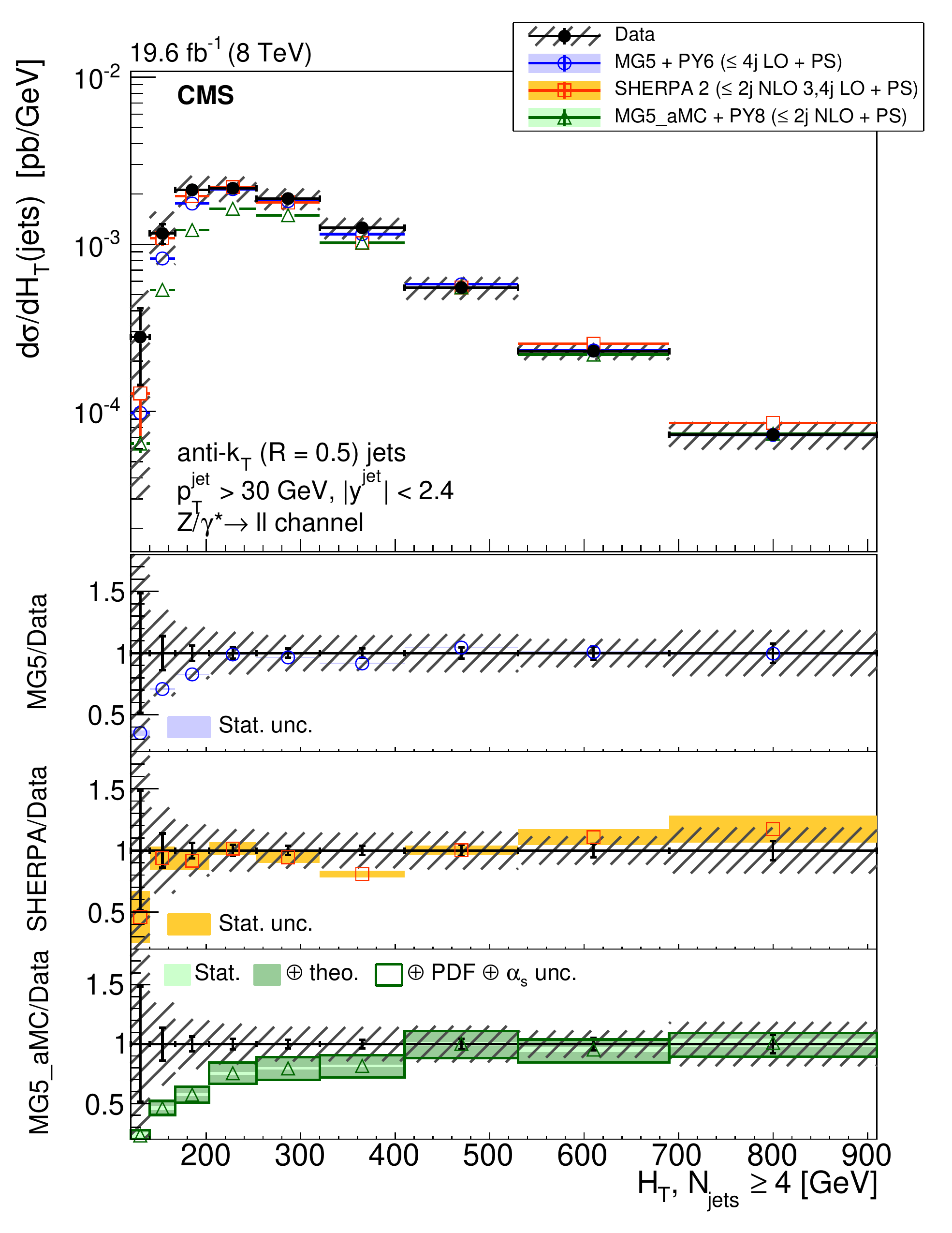}}
\caption{The differential cross section for \zlljets production measured as a function of \HT for (left) $N_{\text{jets}} \geq 3$ and (right) $N_{\text{jets}} \geq 4$ compared to the predictions calculated with \MADGRAPH~5 + \PYTHIA~6, \SHERPA~2, and \MGaMC + \PYTHIA~8. \plotstdcapt}
\label{fig:CombXSec_JetsHT_3and4}
\end{figure}

\begin{figure}[htbp!]
\centering
\includegraphics[width=0.48\textwidth]{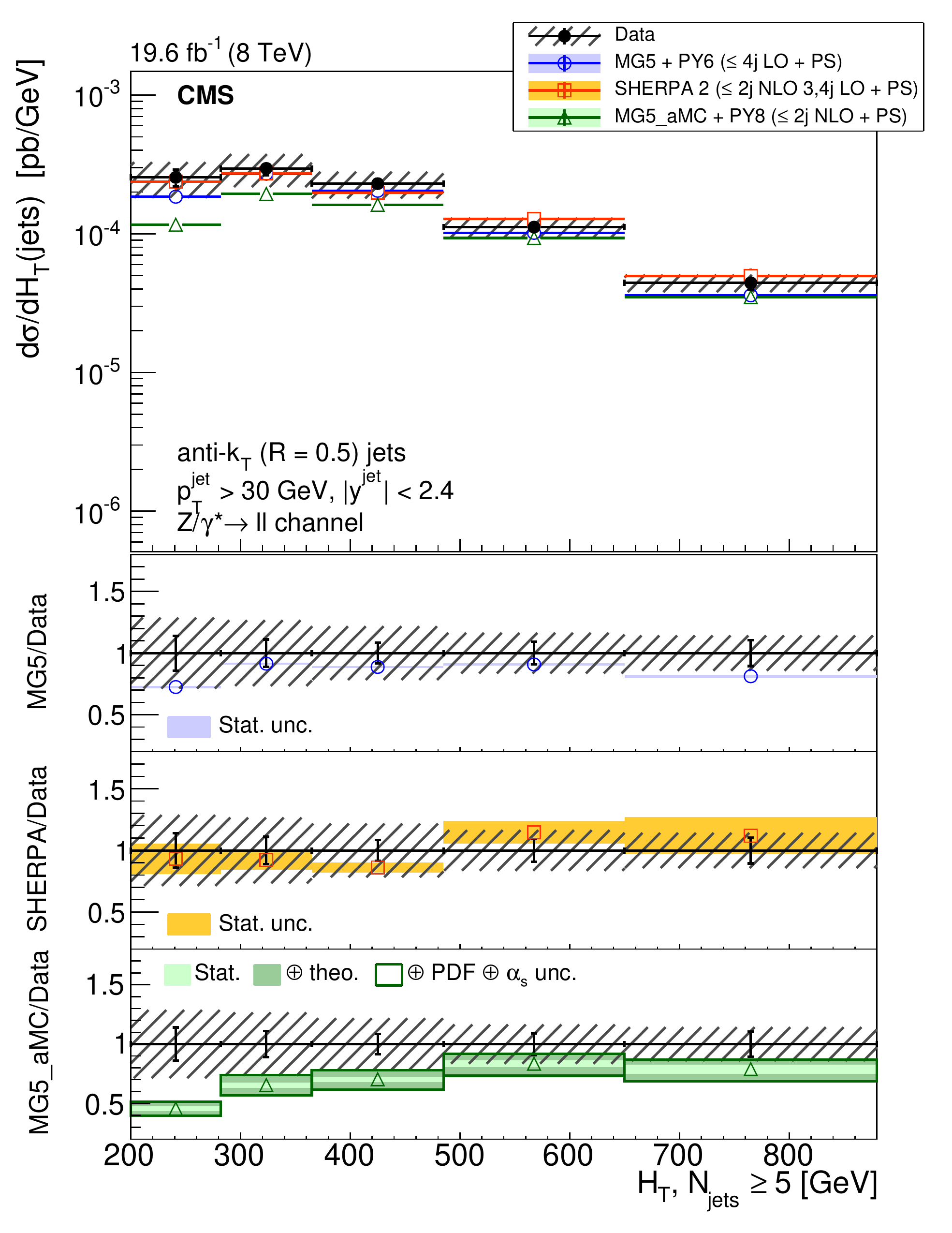}
\caption{The differential cross section for \zlljets production measured as a function of \HT for $N_{\text{jets}} \geq 5$ compared to the predictions calculated with \MADGRAPH~5 + \PYTHIA~6, \SHERPA~2, and \MGaMC + \PYTHIA~8. \plotstdcapt}
\label{fig:CombXSec_JetsHT_5}
\end{figure}

The hadronic activity of an event can be probed with the scalar sum of the transverse momenta of the jets, $\HT$. Measuring hadronic activity is important in searches for signatures with high jet activity or, by contrast, when wishing to veto such activity, for instance in the central region when looking for vector boson fusion induced processes. In this section we present measurements of the spectra for this variable in \zjets events. The differential cross sections are shown in
Figs.~\ref{fig:CombXSec_JetsHT_1and2}--\ref{fig:CombXSec_JetsHT_5} for the different inclusive jet multiplicities.

The predictions of the generators agree well with the measurements within the experimental uncertainties. For events with three or more jets (Figs.~\ref{fig:CombXSec_JetsHT_3and4}), all three simulations predict a distribution that falls more steeply at low values of $\HT$. For the normalised distributions the total uncertainties in the measurements for $\cPZ+\ge3\ \text{jets}$ reduce for the three first bins to 21\%, 10\%, and 3.3\%, respectively. This indicates that the
difference in the shape is significant for \MADGRAPH~5 + \PYTHIA~6 and \MGaMC + \PYTHIA~8 predictions.

\subsection{Azimuthal angles}

\begin{figure}[h!t]
{\centering
 {\includegraphics[width=0.32\textwidth]{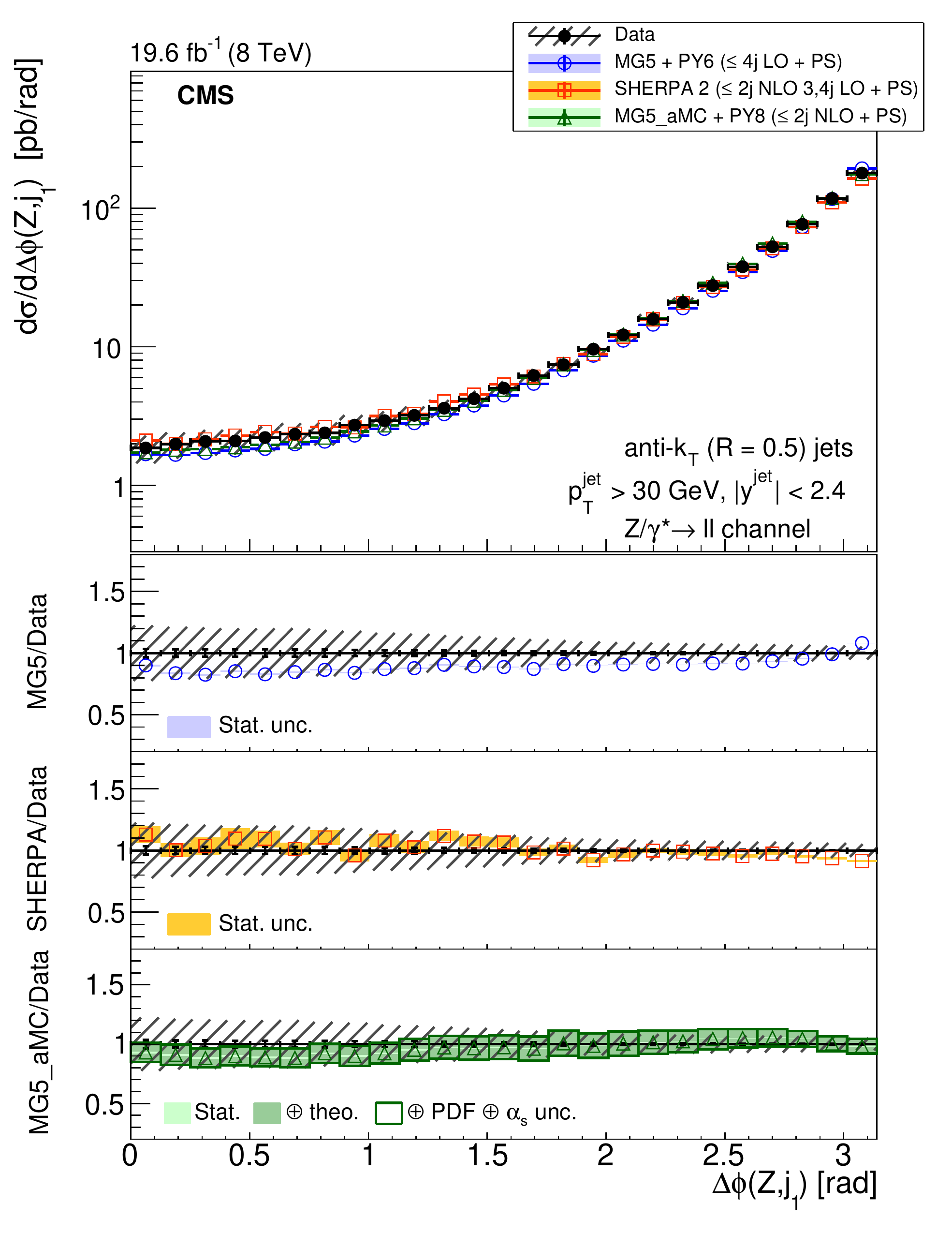}}
 {\includegraphics[width=0.32\textwidth]{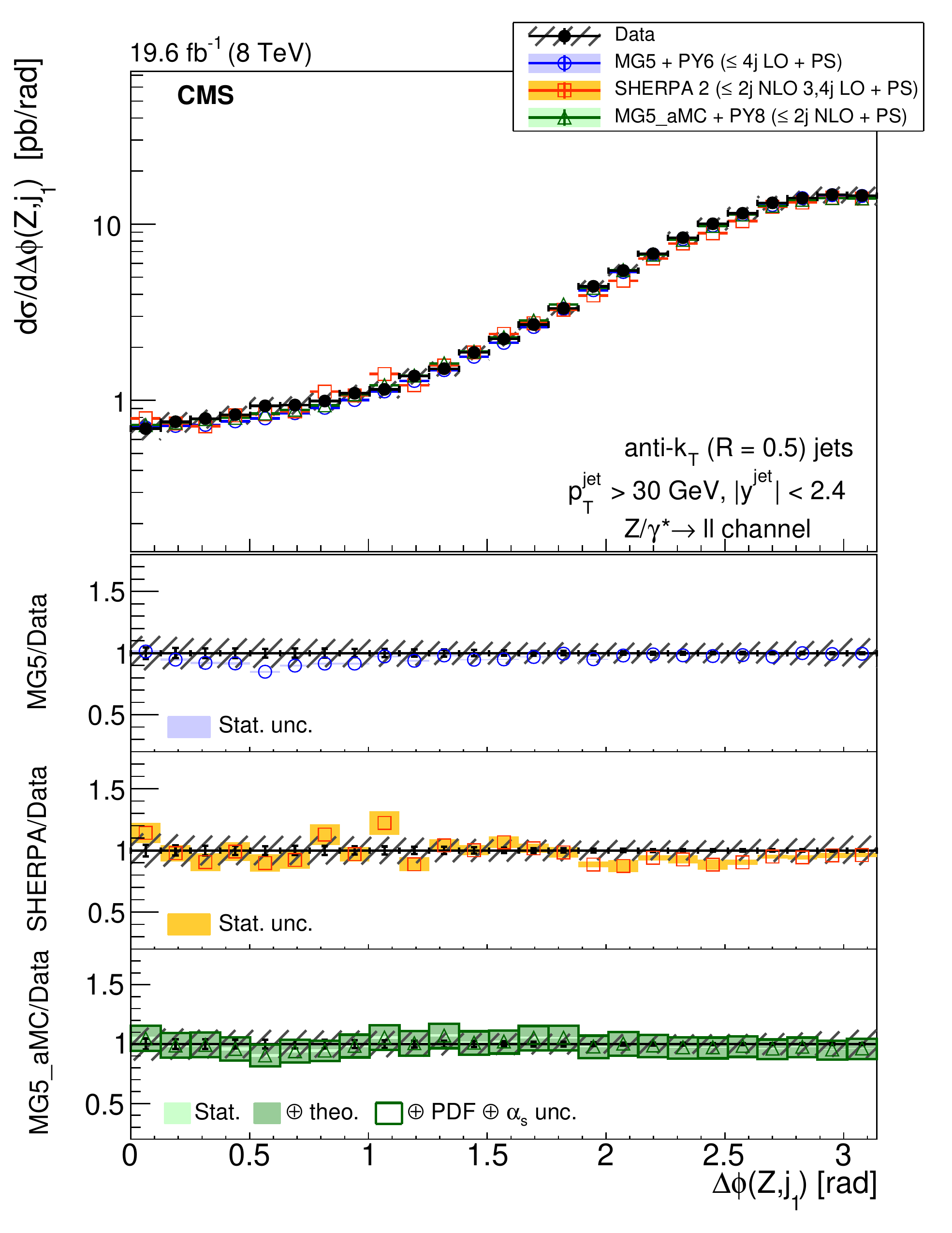}}
 {\includegraphics[width=0.32\textwidth]{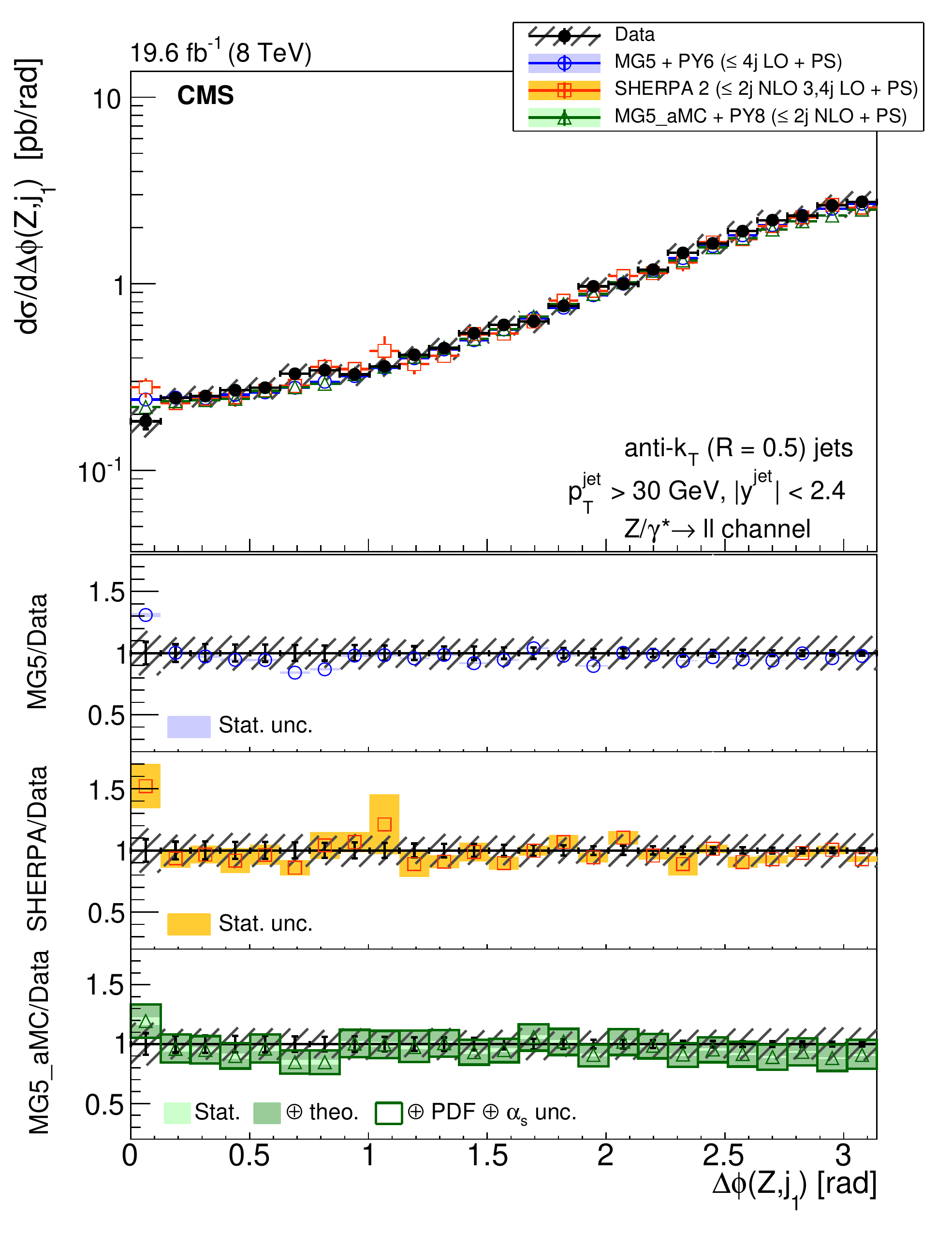}}
 \caption{The differential cross section as a function of the azimuthal angle between the $\cPZ$ boson and the leading jet for different jet multiplicities, (left) $N_{\text{jets}}\ge 1$, (middle) $N_{\text{jets}}\ge 2$, and (right) $N_{\text{jets}}\ge 3$. \plotstdcapt}
 \label{fig:dphizfj}
}
\end{figure}

Figure~\ref{fig:dphizfj} shows the differential cross section measurements as a function of the azimuthal angle between the $\cPZ$ boson and the leading jet for three different jet multiplicities.
The inclusion of several parton multiplicities in the ME calculations ensures that the Monte Carlo predictions model the data well even at tree level and small differences in azimuthal angles. Differences are observed between tree-level (\MADGRAPH~5 + \PYTHIA~6) and multileg NLO
(\SHERPA~2, and \MGaMC + \PYTHIA~8)
predictions, the latter being closer to the measurement, but the difference is smaller than one standard deviation in the experimental uncertainties. As the jet multiplicity increases, the $\Delta\phi(\cPZ,\text{ j}_1)$ distribution flattens out. In an event dominated by the leading jet, the jet is recoiling against the $\cPZ $ boson, resulting in a strong peak at $\Delta\phi(\cPZ,\text{ j}_1)\simeq\pi$. As the jet activity increases the $\cPZ$ boson recoils against a combination of several jets and this peak broadens, leading to an overall flattening of the distribution.

\begin{figure}[h!t]
{\centering
 {\includegraphics[width=0.49\textwidth]{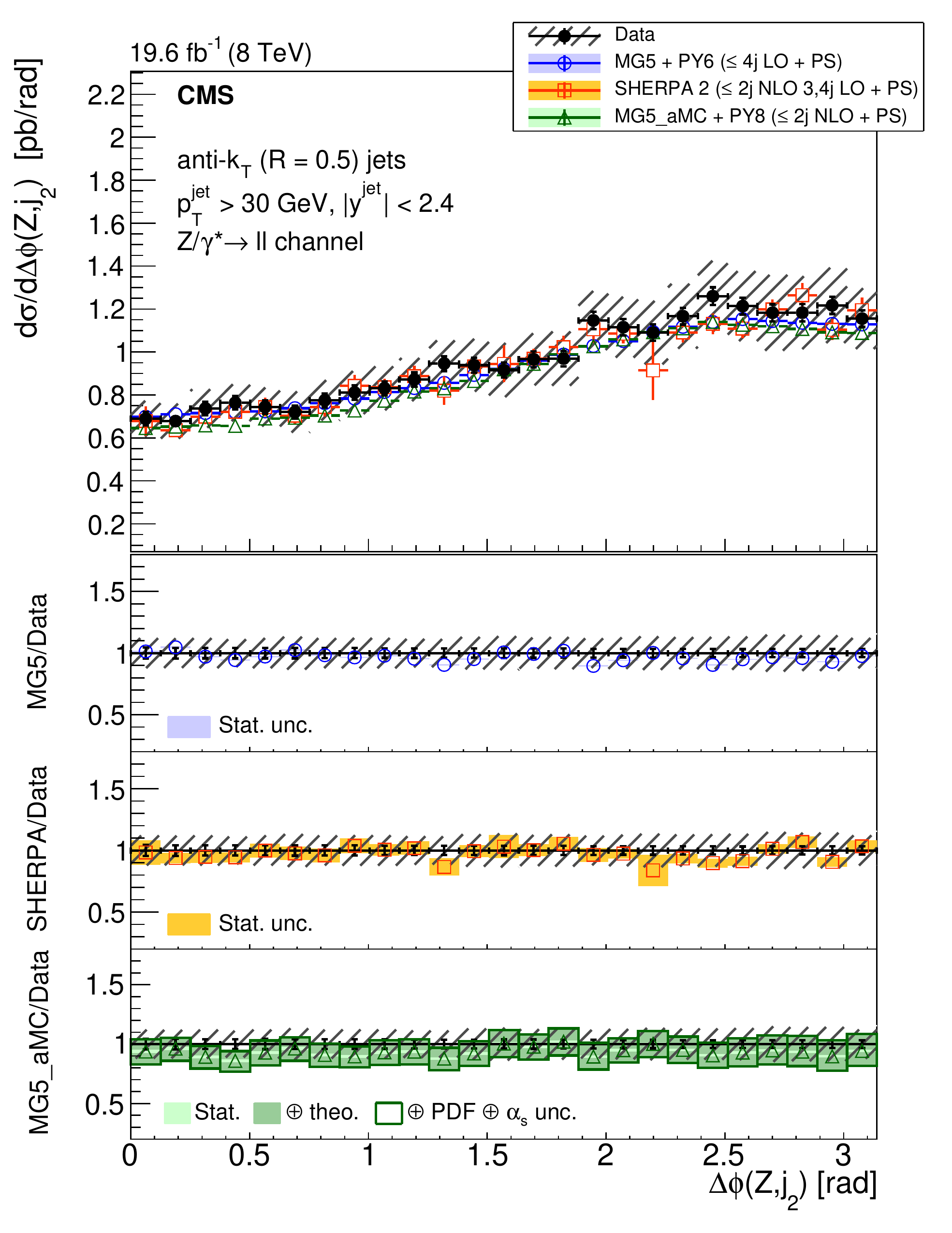}}
 {\includegraphics[width=0.49\textwidth]{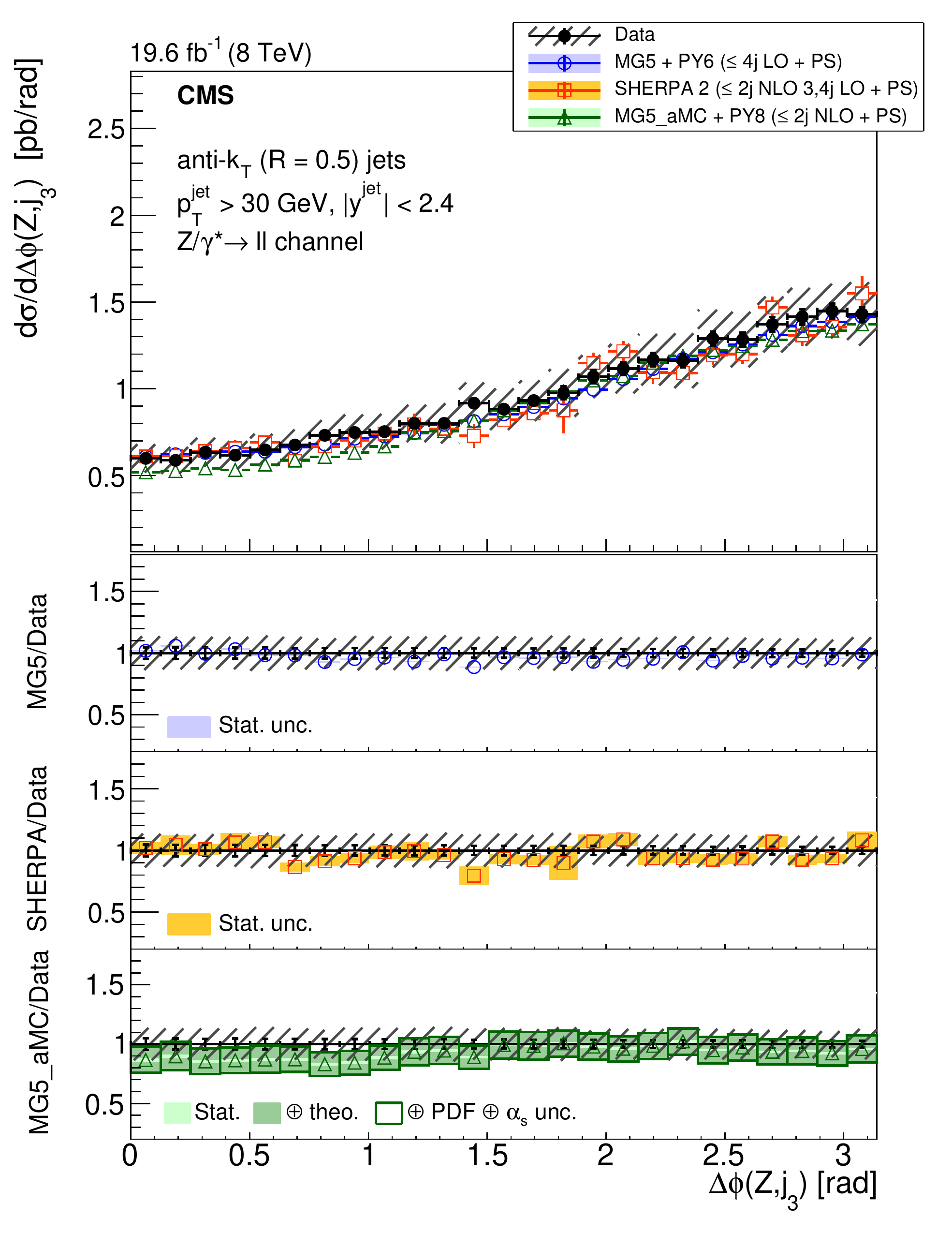}}
 \caption{The differential cross section for \zlljets production for $N_{\text{jets}}\ge 3$ as a function of the azimuthal angle between (left) the $\cPZ$ boson and the second leading jet, (right) the $\cPZ$ boson and the third leading jet. \plotstdcapt}
 \label{fig:dphizstj}
}
\end{figure}

\begin{figure}[h!t]
{\centering
 {\includegraphics[width=0.32\textwidth]{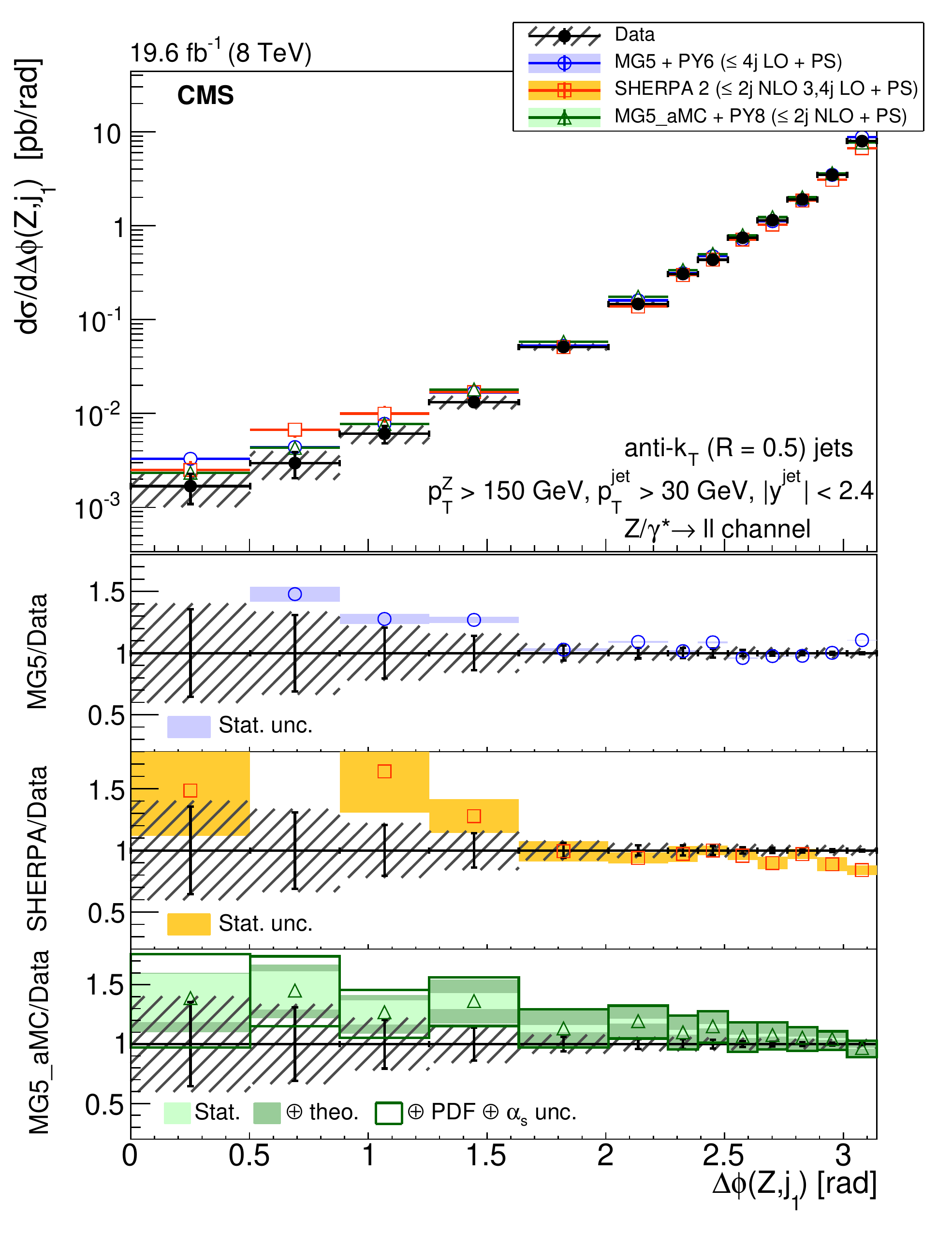}}
 {\includegraphics[width=0.32\textwidth]{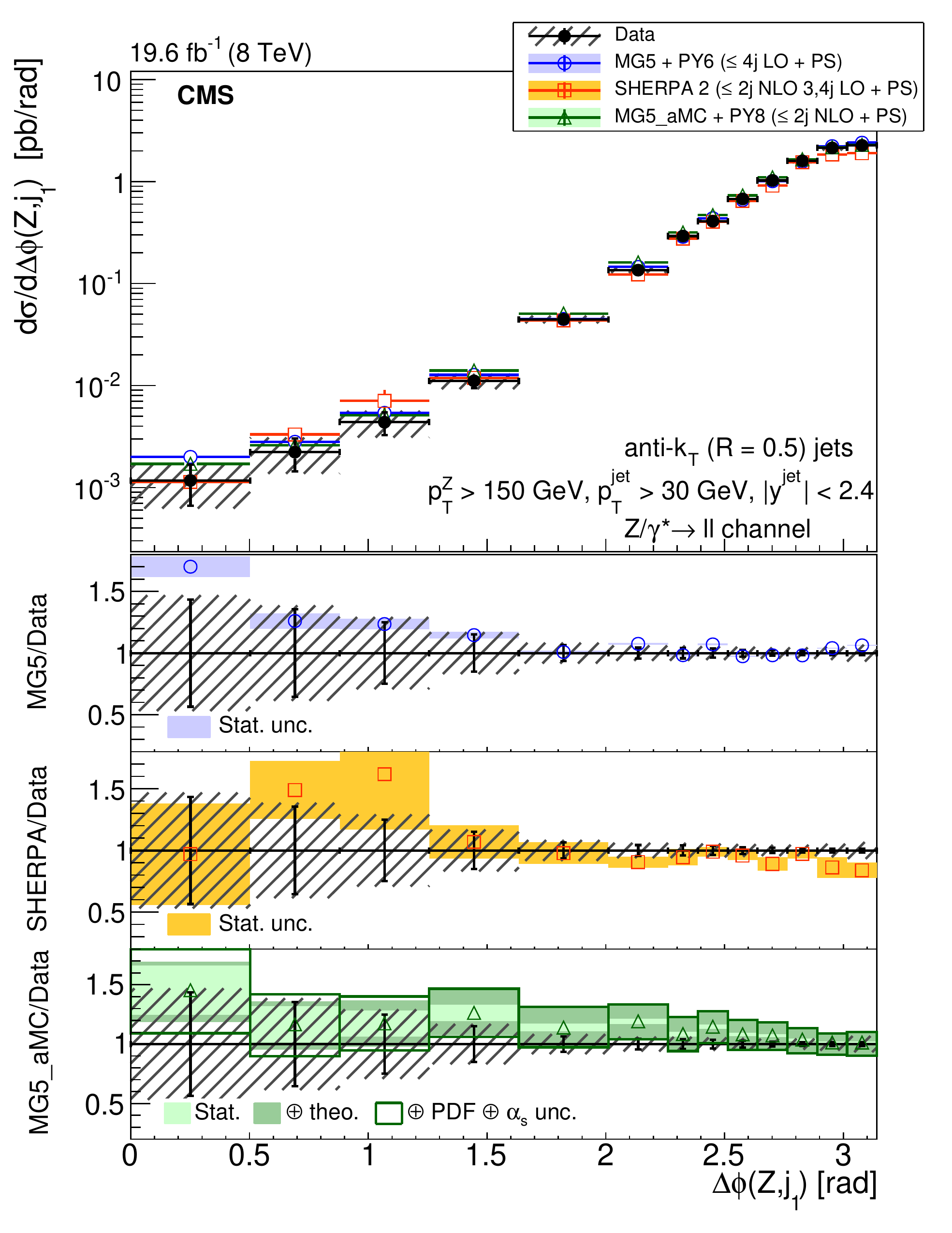}}
 {\includegraphics[width=0.32\textwidth]{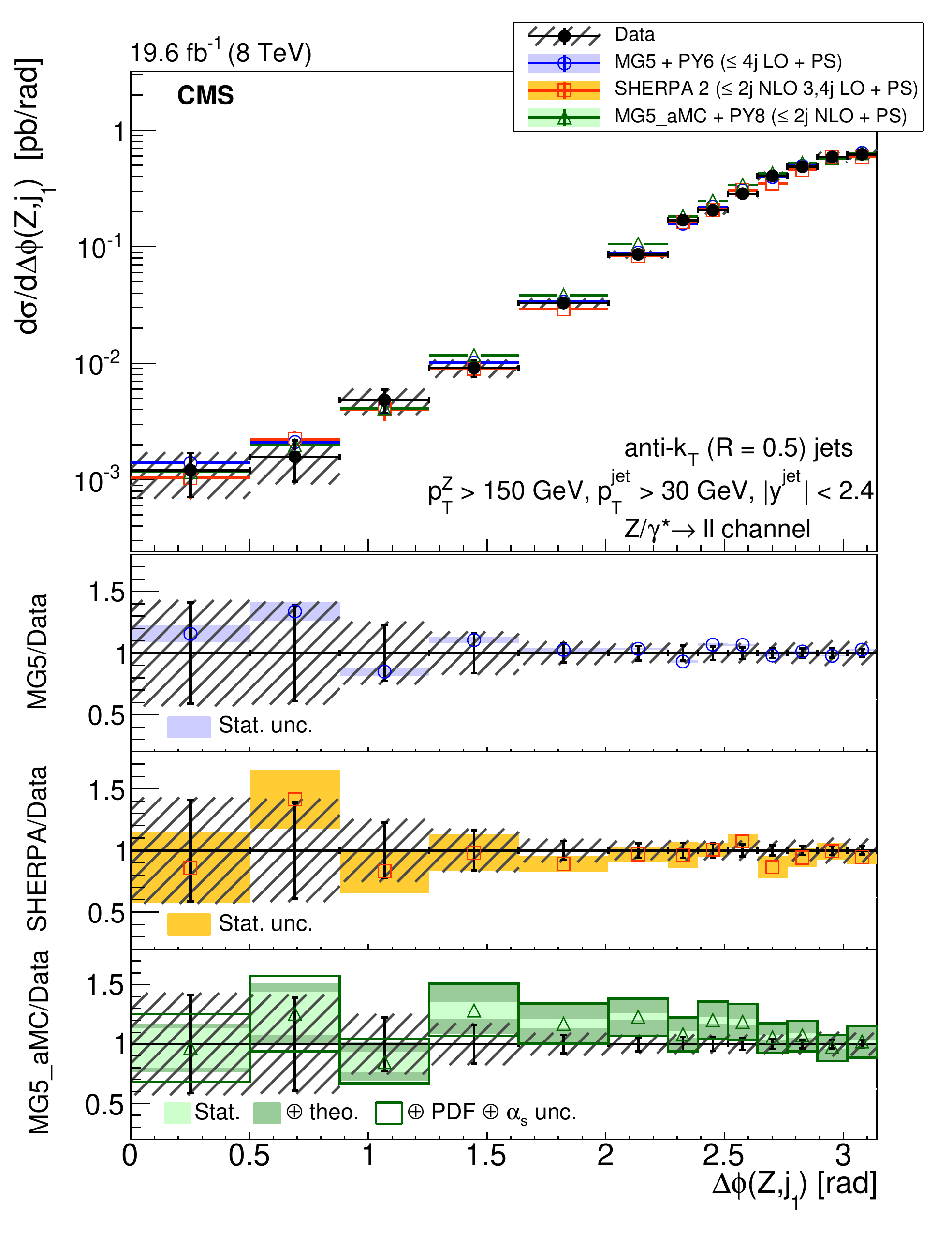}}
 \caption{The differential cross section as a function of the azimuthal angle between the $\cPZ$ boson and the leading jet, for $\pt(\cPZ)>150\GeV$ and (left) $N_{\text{jets}}\ge 1$, (middle) $N_{\text{jets}}\ge 2$, and (right) $N_{\text{jets}}\ge 3$. \plotstdcapt}
 \label{fig:dphizfjzpt150}
}
\end{figure}

\begin{figure}[h!t]
{\centering
 {\includegraphics[width=0.49\textwidth]{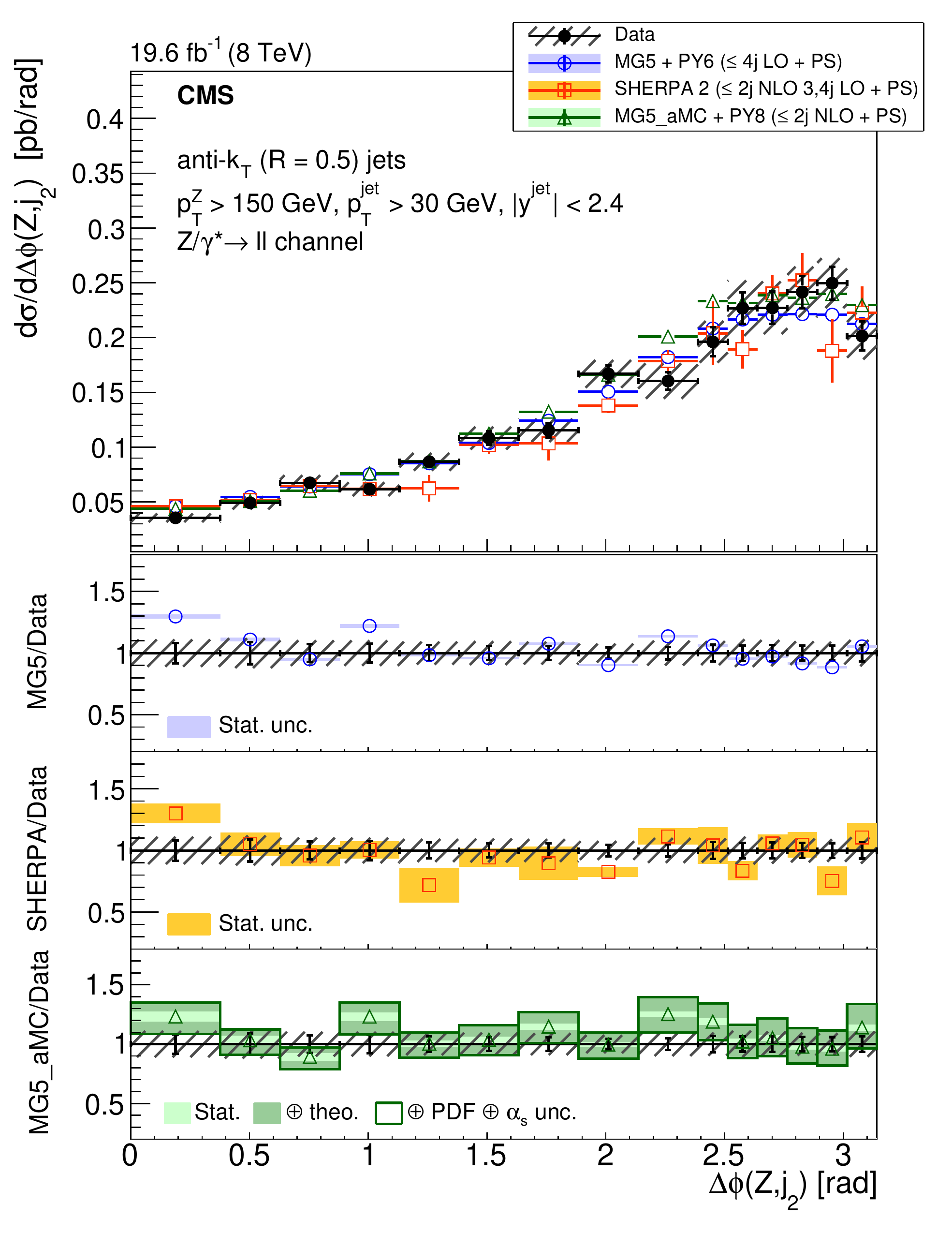}}
 {\includegraphics[width=0.49\textwidth]{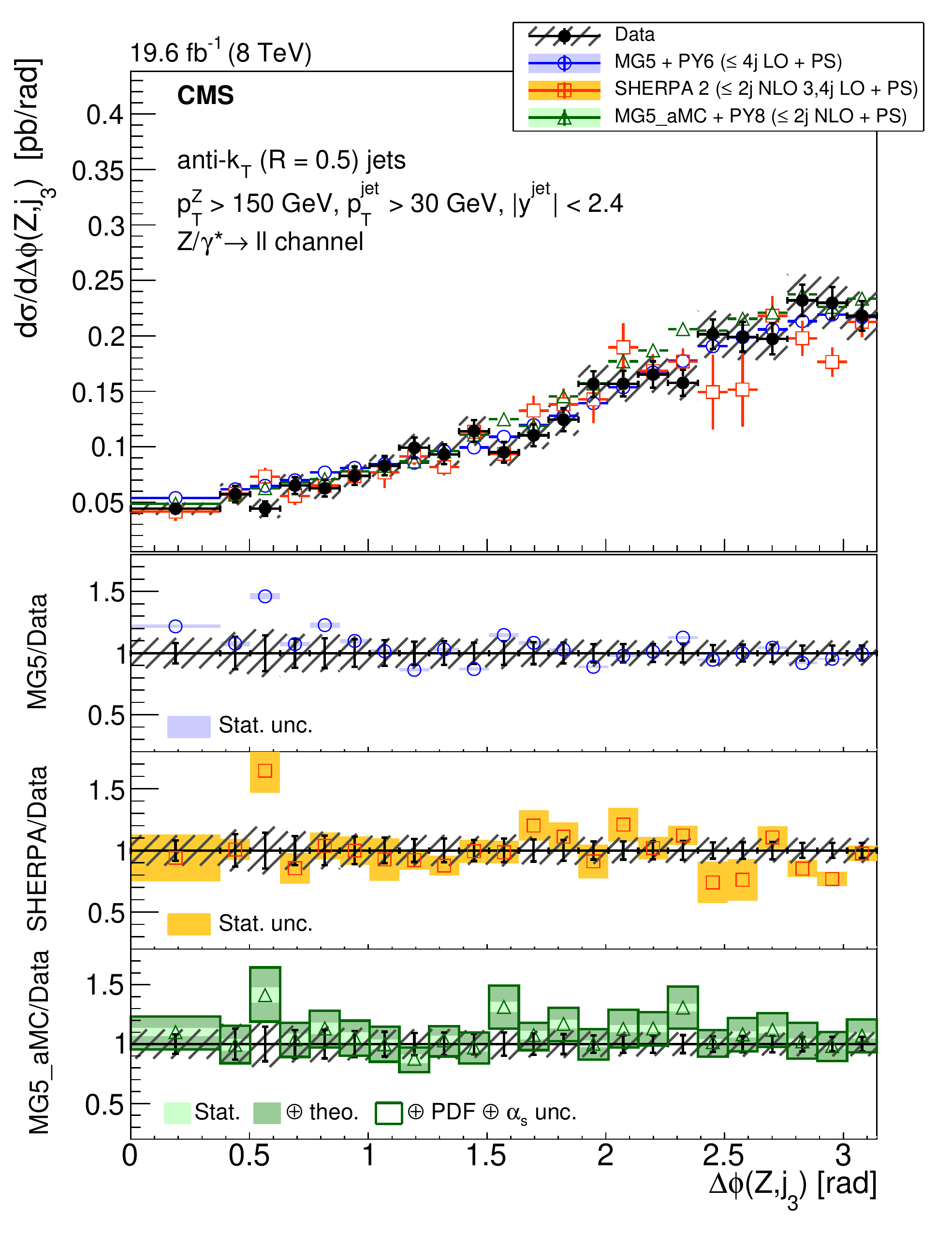}}
 \caption{The differential cross section for \zlljets production for $N_{\text{jets}}\ge 3$ and $\pt(\cPZ)>150\GeV$ as a function of the azimuthal angle (left) between the $\cPZ$ boson and the second leading jet and (right) between $\cPZ$ boson and third-leading jet. \plotstdcapt}
 \label{fig:dphizstjzpt150}
}
\end{figure}

\begin{figure}[h!t]
{\centering
 {\includegraphics[width=0.32\textwidth]{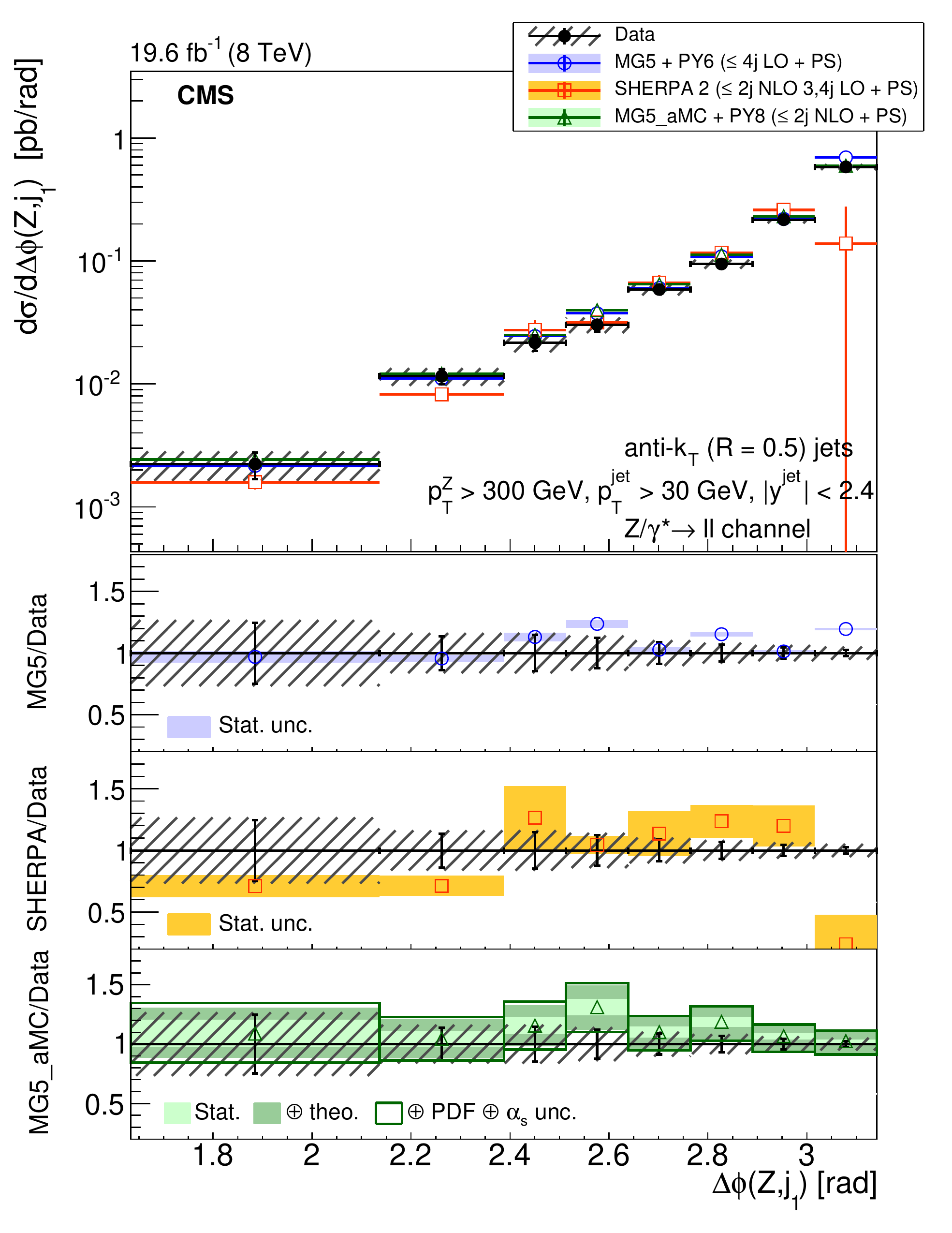}}
 {\includegraphics[width=0.32\textwidth]{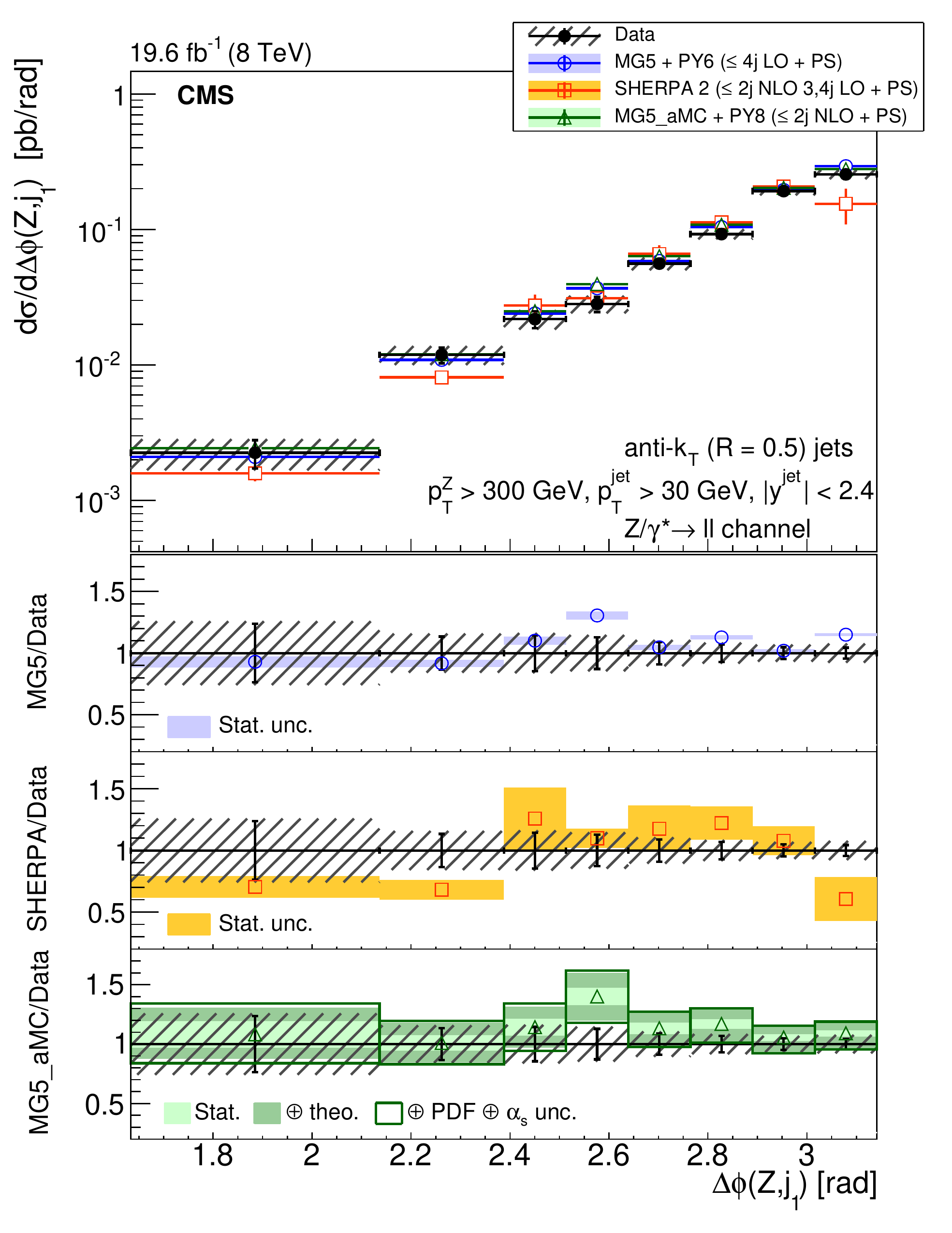}}
 {\includegraphics[width=0.32\textwidth]{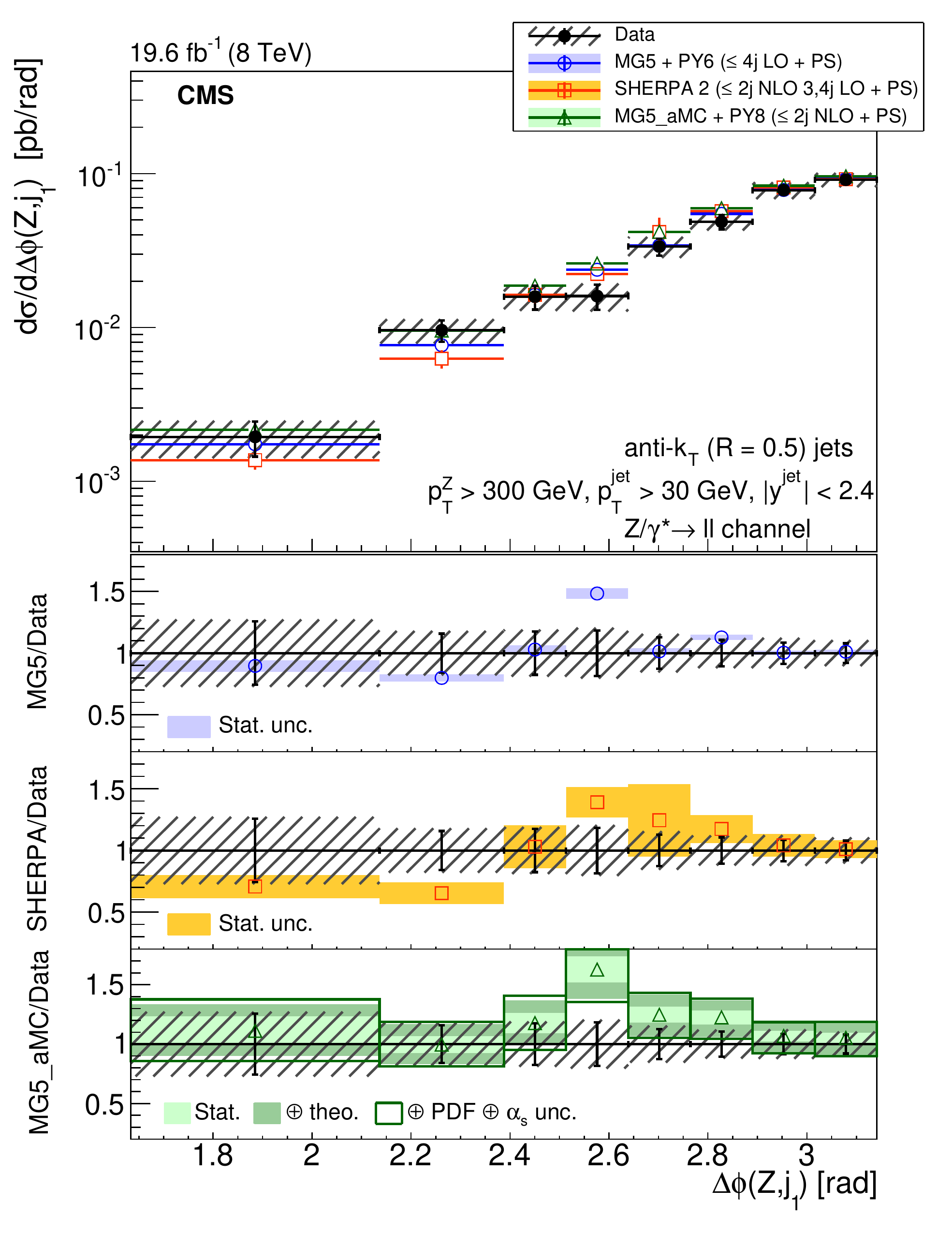}}
 \caption{The differential cross section as a function of the azimuthal angle between the $\cPZ$ boson and the leading jet, for $\pt(\cPZ)>300\GeV$ and (left) $N_{\text{jets}}\ge 1$, (middle) $N_{\text{jets}}\ge 2$, and (right) $N_{\text{jets}}\ge 3$. \plotstdcapt}
 \label{fig:dphizfjzpt300}
}
\end{figure}

\begin{figure}[h!t]
{\centering
 {\includegraphics[width=0.49\textwidth]{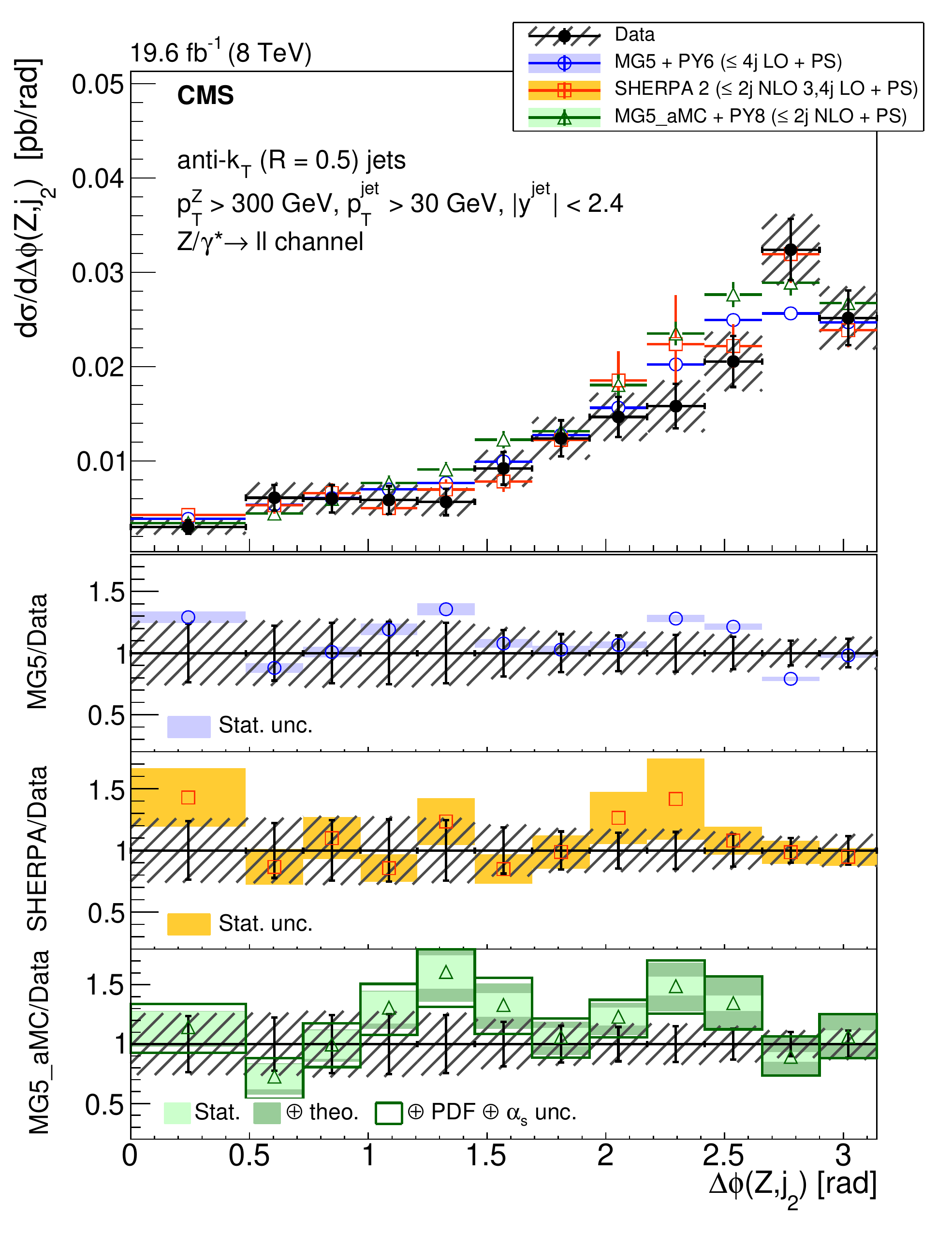}}
 {\includegraphics[width=0.49\textwidth]{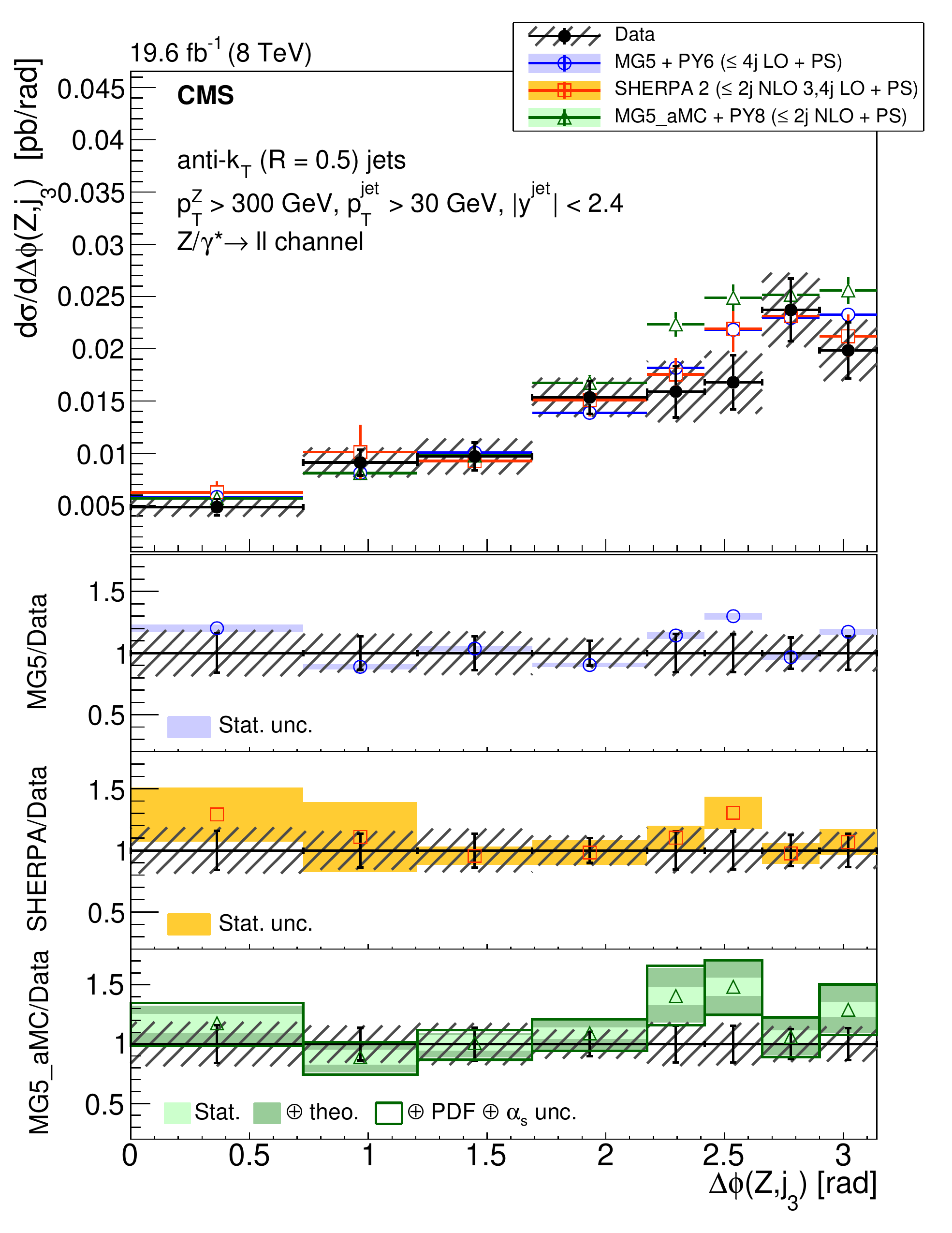}}
 \caption{The differential cross section for \zlljets production for $N_{\text{jets}}\ge 3$ and $\pt(\cPZ) > 300\GeV$ as a function of the azimuthal angle (left) between the $\cPZ$ boson and the second-leading jet and (right) between the $\cPZ$ boson and the third-leading jet. \plotstdcapt}
 \label{fig:dphizstjzpt300}
}
\end{figure}

For the azimuthal angle between the $\cPZ$ boson and the second- and third-leading jets, as shown in Fig.~\ref{fig:dphizstj}, predictions and measurement agree very well. The differential cross sections are measured for the phase space regions with $\pt(\Z)>150\GeV$ and $\pt(\Z)>300\GeV$. The results are shown in Figs.~\ref{fig:dphizfjzpt150}--\ref{fig:dphizstjzpt300}. The agreement of the predictions with the data is preserved, but the tree-level prediction computed with \MADGRAPH~5 + \PYTHIA~6
is an overestimate compared to the data at low azimuthal angle for the leading jet. The distributions are more uniform than in the $\Delta\phi(\cPZ,\text{j}_1)$ case, but retain a peak close to $\pi$. In the $\Delta\phi(\Z, \text{j}_2)$ case, we also see that the distributions show a larger correlation and a peak emerges at approximately $\Delta\phi(\cPZ,\text{j}_2)\approx2.6$. This peak becomes more pronounced as the $\pt(\cPZ)$ threshold
increases. A similar trend is seen in the $\Delta\phi(\Z, \text{j}_3)$ distribution: selecting a high $\cPZ$ boson \pt increases the fraction of events where the jets recoil against the boson.

\begin{figure}[h!t]
{\centering
{\includegraphics[width=0.32\textwidth]{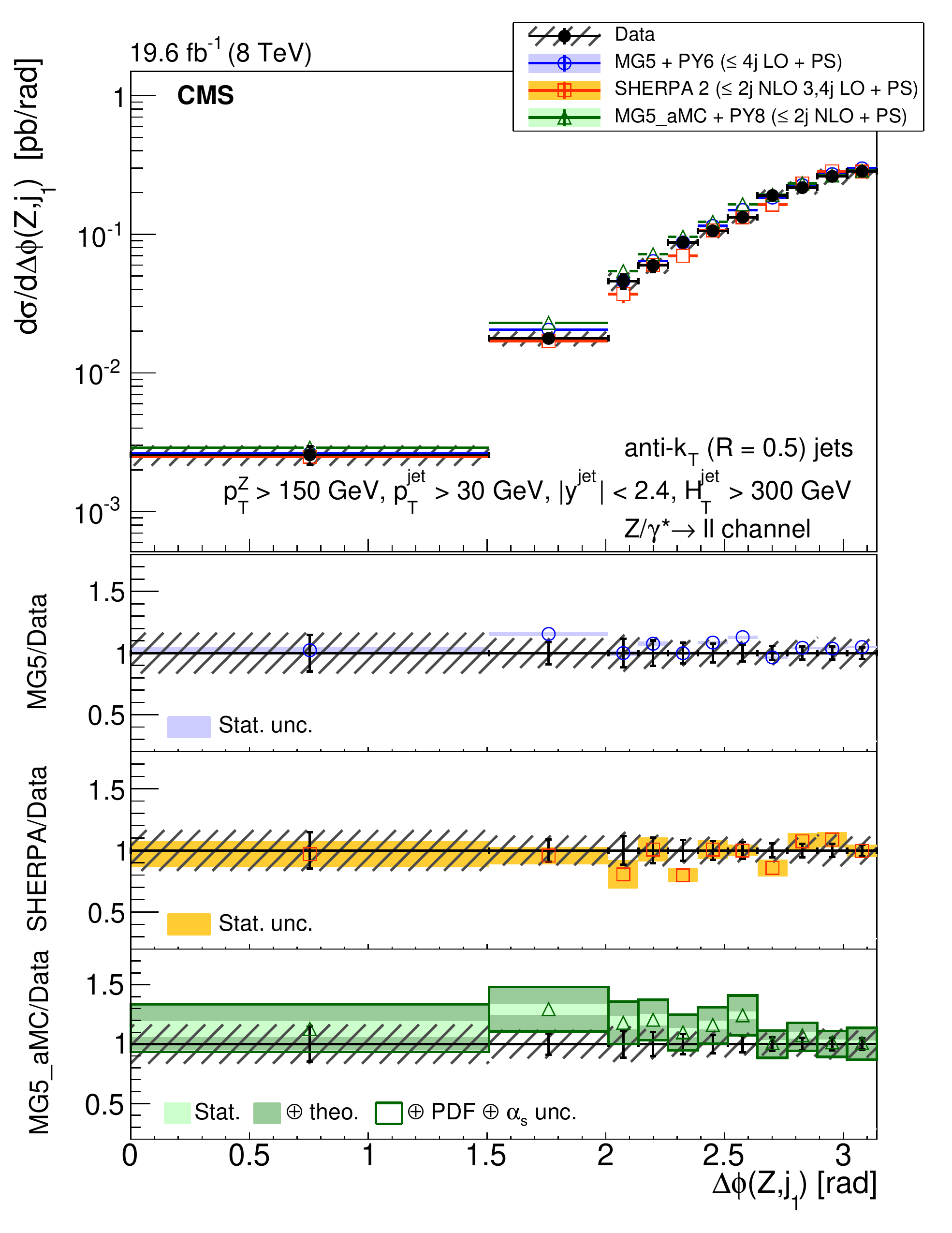}}
{\includegraphics[width=0.32\textwidth]{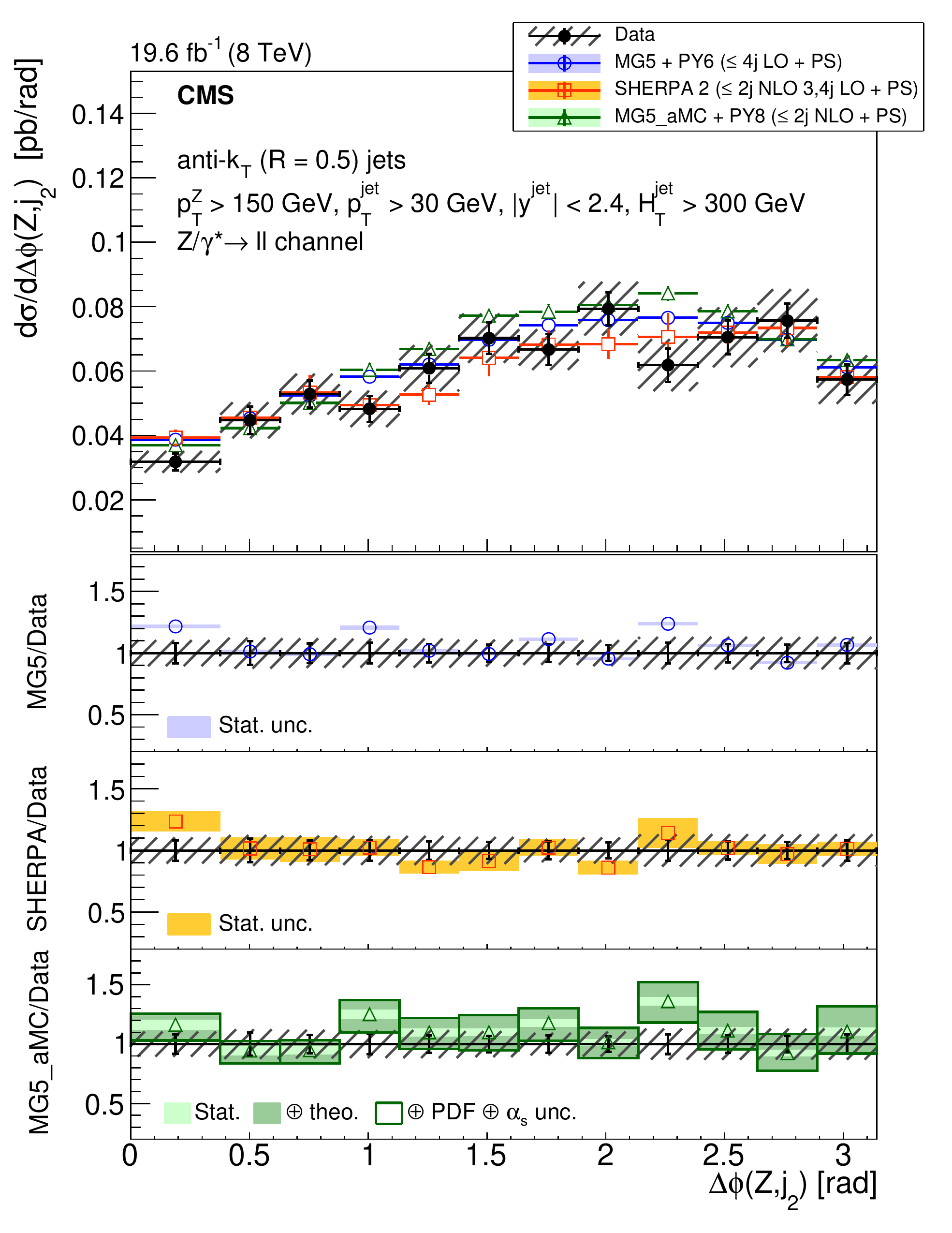}}
{\includegraphics[width=0.32\textwidth]{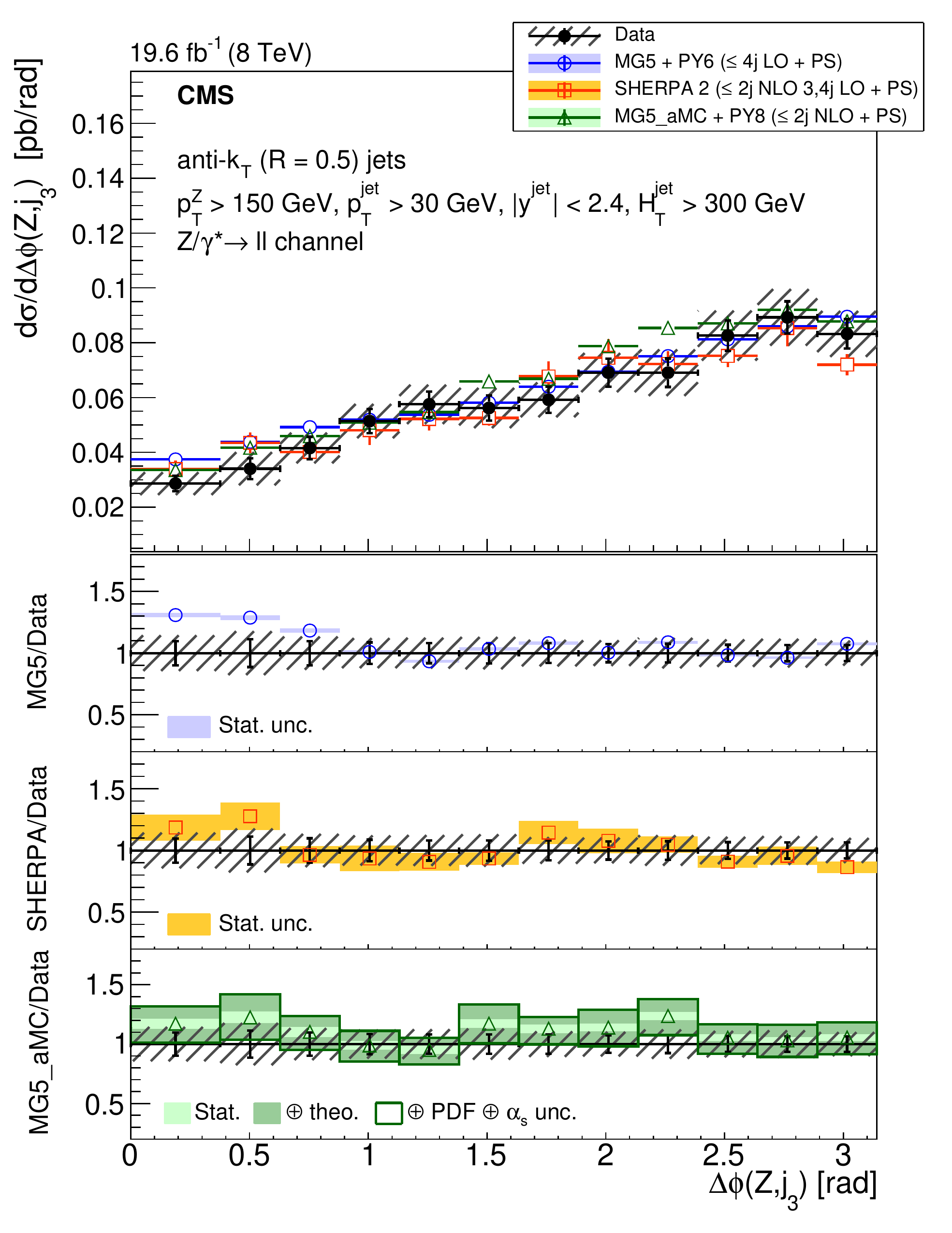}}
\caption{The differential cross section for \zlljets production for $N_{\text{jets}}\ge 3$, $\pt^{\text{Z}} > 150\GeV$, and $\HT^{\text{jet}}>300\GeV$ as a function of the azimuthal angle between the $\cPZ$ boson and the (left) first-, (middle) second-, and (right) third-leading jet.  \plotstdcapt}
\label{fig:dphizfstjzpt150ht300}
}
\end{figure}

Inclusive three-jet production is investigated in regions where both \HT and the $\pt(\cPZ)$ are large. Good agreement between data and predictions is also present here, as shown in Fig.~\ref{fig:dphizfstjzpt150ht300}. In this high-$\pt(\Z)$, high-$\HT$ regime, we see a similar behaviour to the other high-$\pt(\Z)$ selections. The $\Delta\phi\left(\cPZ,\text{j}_2\right)$ and  $\Delta\phi\left(\cPZ,\text{j}_3\right)$ distributions are also flatter than the corresponding distributions with no $\HT$ cut.

\begin{figure}[h!t]
{\centering
 {\includegraphics[width=0.32\textwidth]{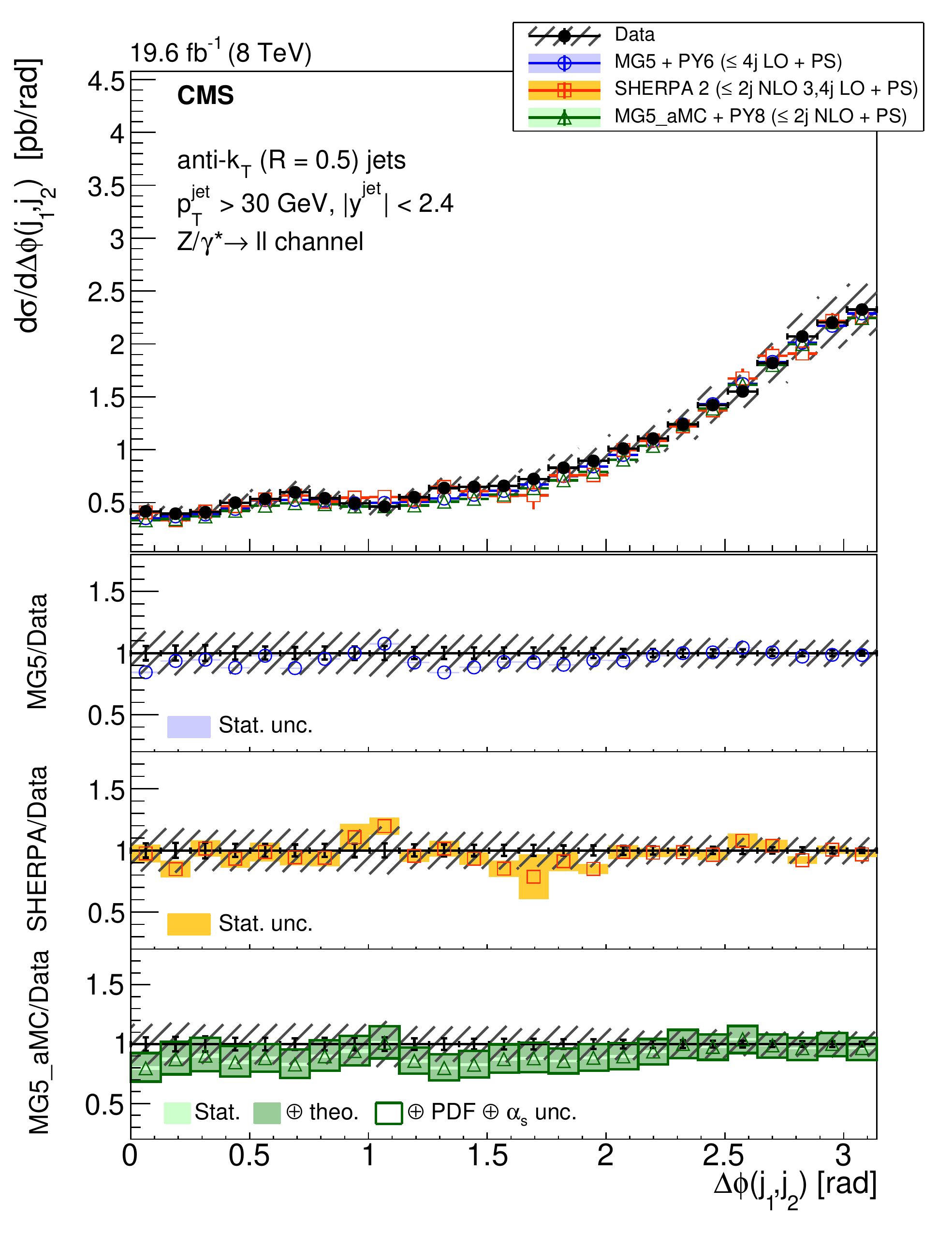}}
 {\includegraphics[width=0.32\textwidth]{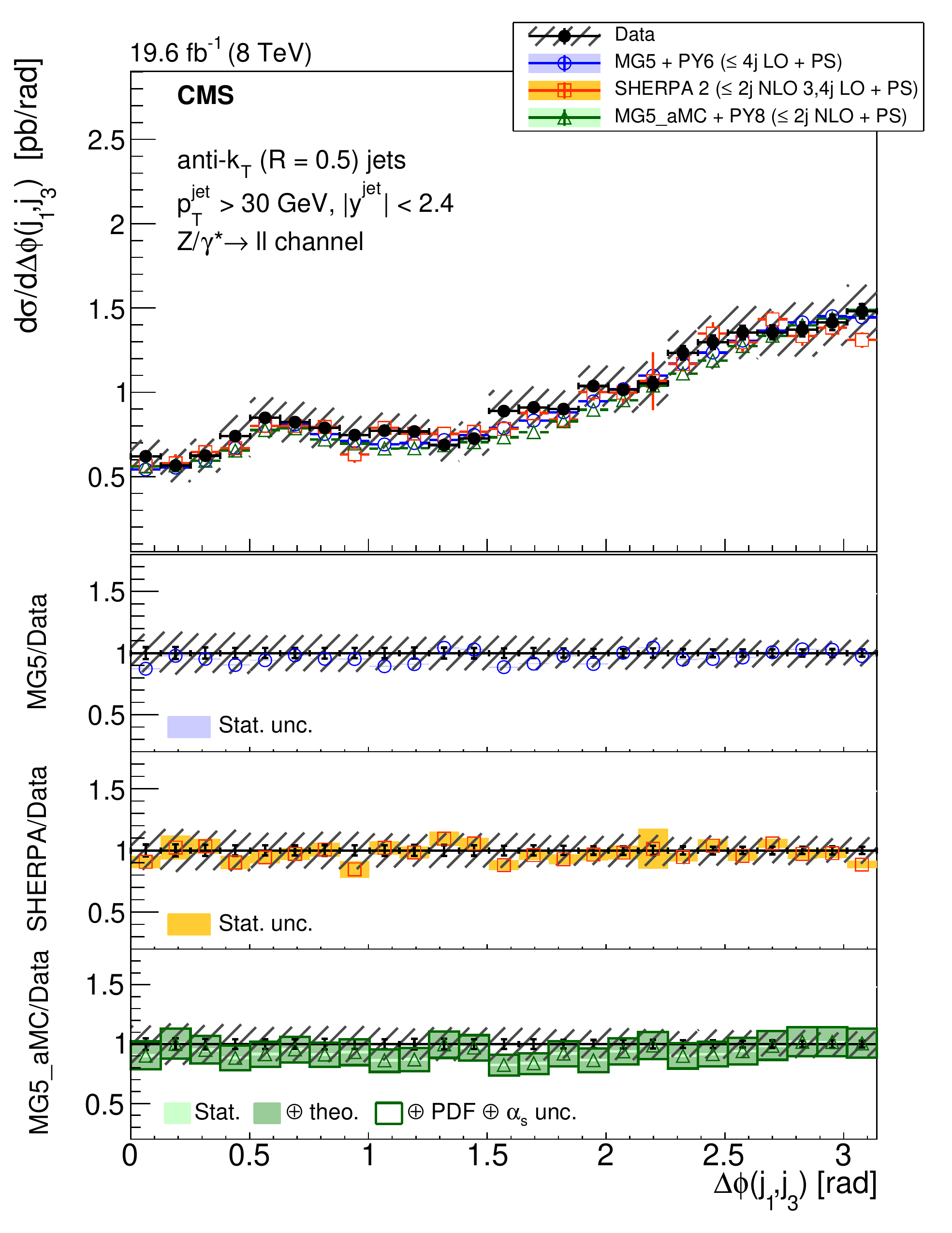}}
 {\includegraphics[width=0.32\textwidth]{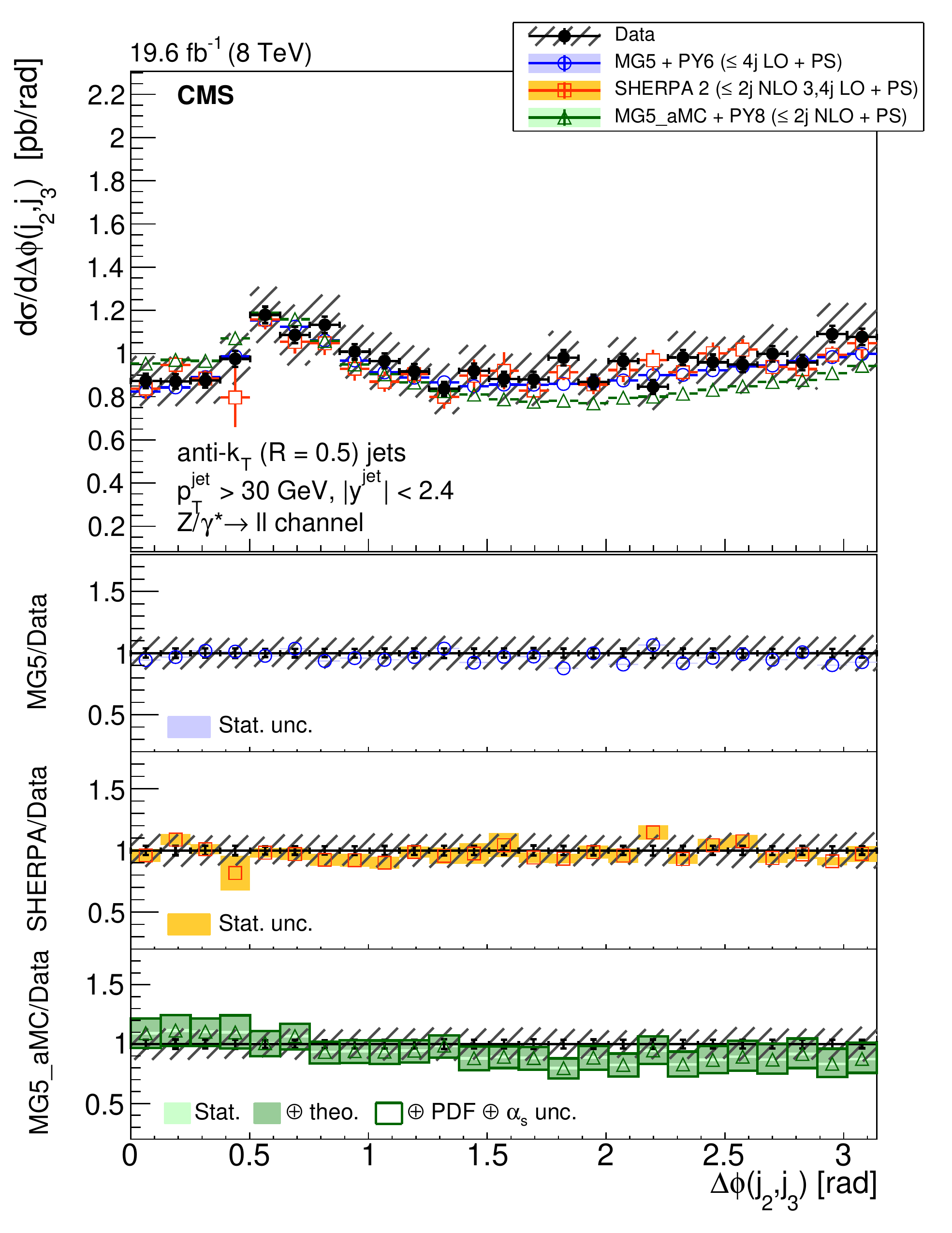}}
 \caption{The differential cross section for \zlljets production for $N_{\text{jets}}\ge 3$ as a function of the azimuthal angle between (left) the first- and second-, (middle) the first- and third-, and (right) the second- and third-leading jets. \plotstdcapt}
 \label{fig:dphifstj}
}
\end{figure}

\begin{figure}[h!t]
{\centering
 {\includegraphics[width=0.32\textwidth]{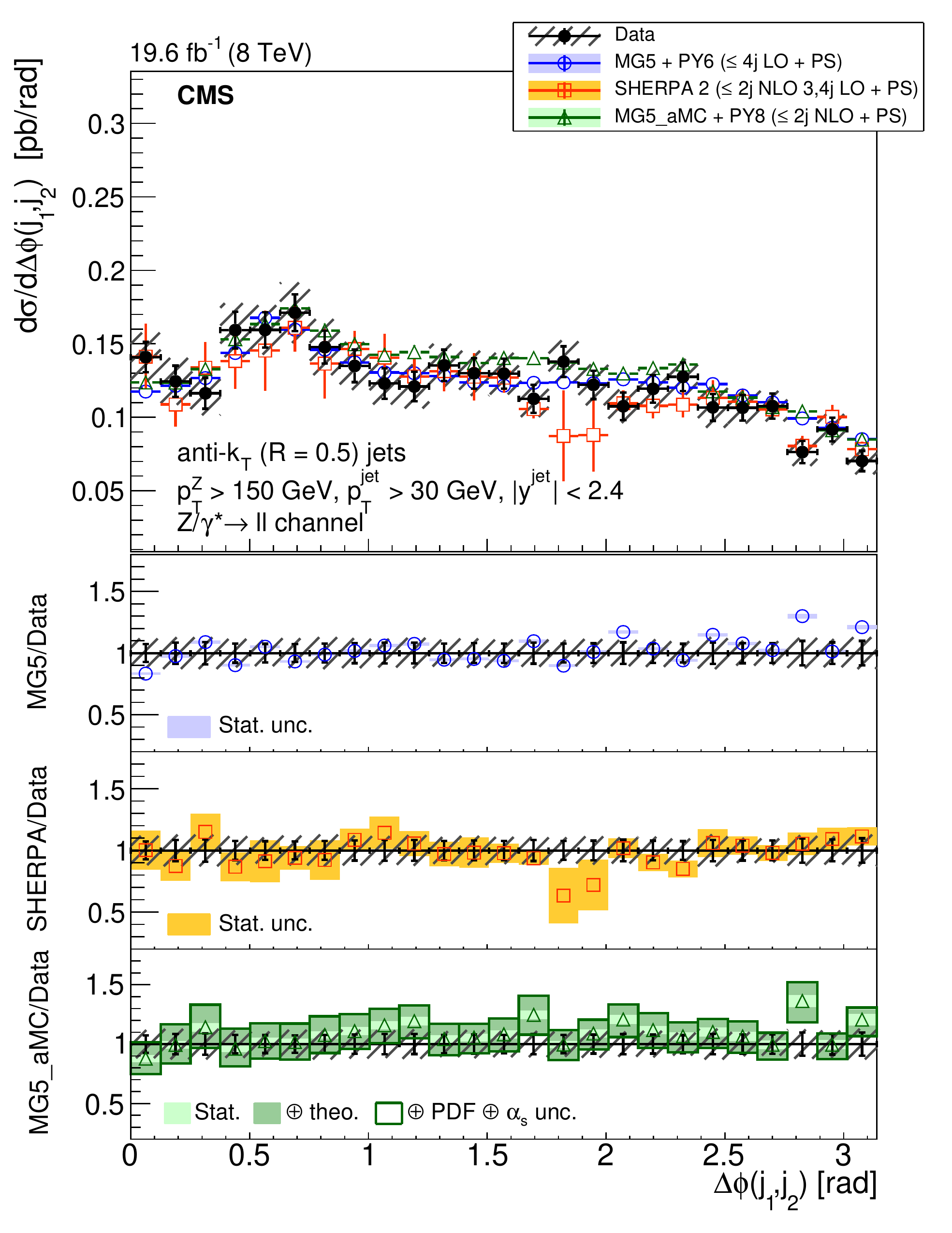}}
 {\includegraphics[width=0.32\textwidth]{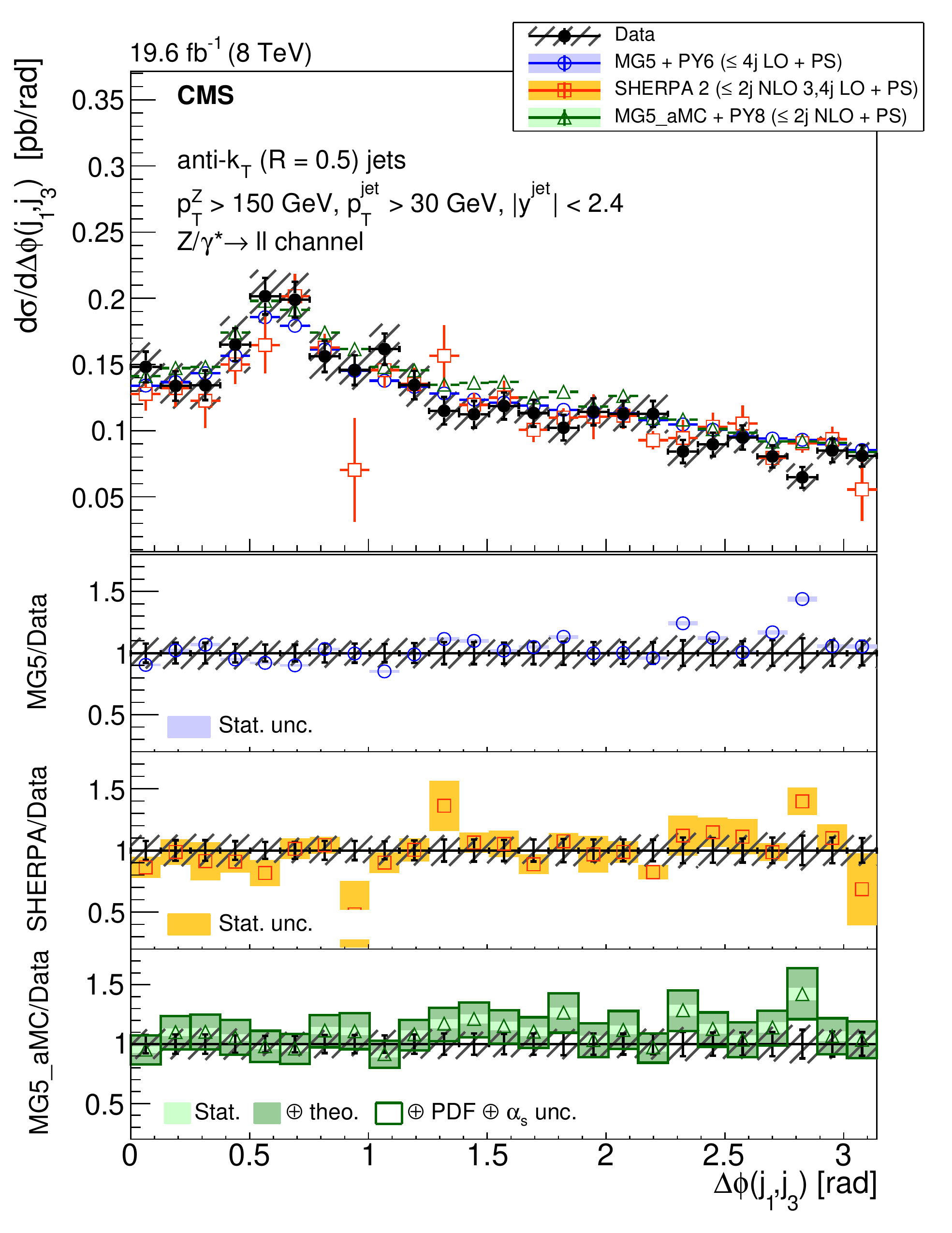}}
 {\includegraphics[width=0.32\textwidth]{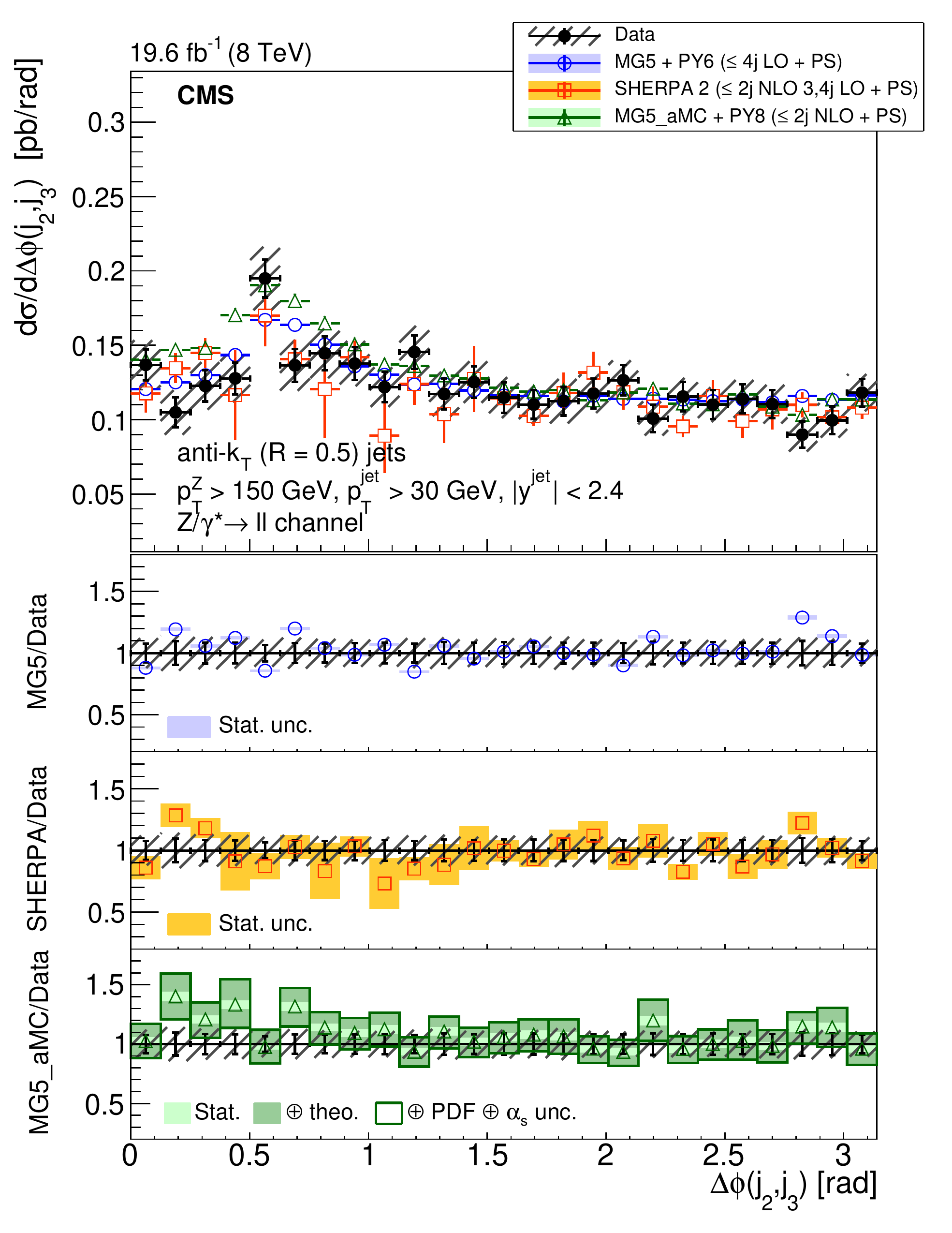}}
 \caption{The differential cross section for \zlljets production for $N_{\text{jets}}\ge 3$ and $\pt(\cPZ) > 150\GeV$ as a function of the azimuthal angle between (left) the first- and second-, (middle) the first- and third-, and (right) the second- and third-leading jets. \plotstdcapt}
 \label{fig:dphifstjzpt150}
}
\end{figure}

\begin{figure}[h!t]
{\centering
 {\includegraphics[width=0.32\textwidth]{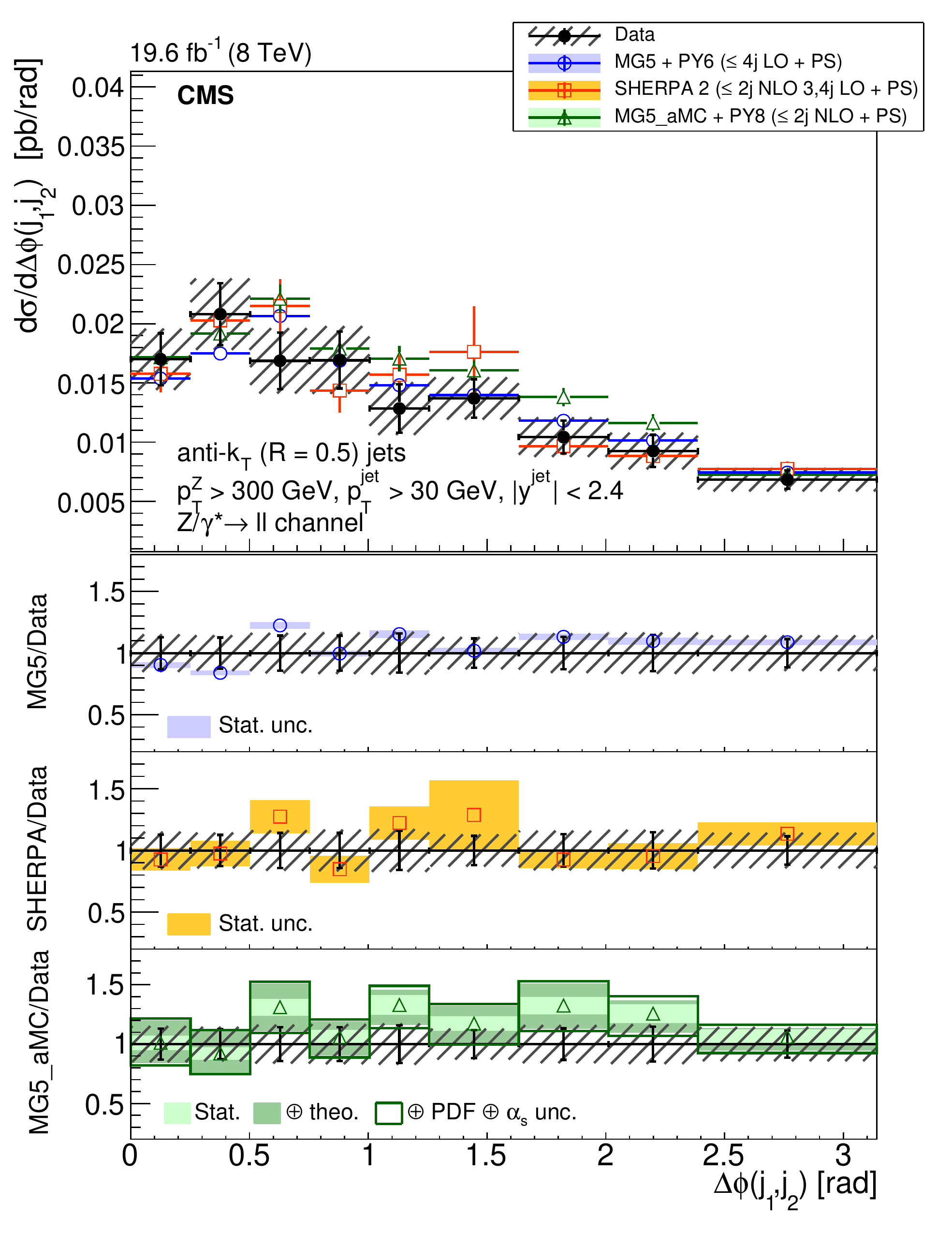}}
 {\includegraphics[width=0.32\textwidth]{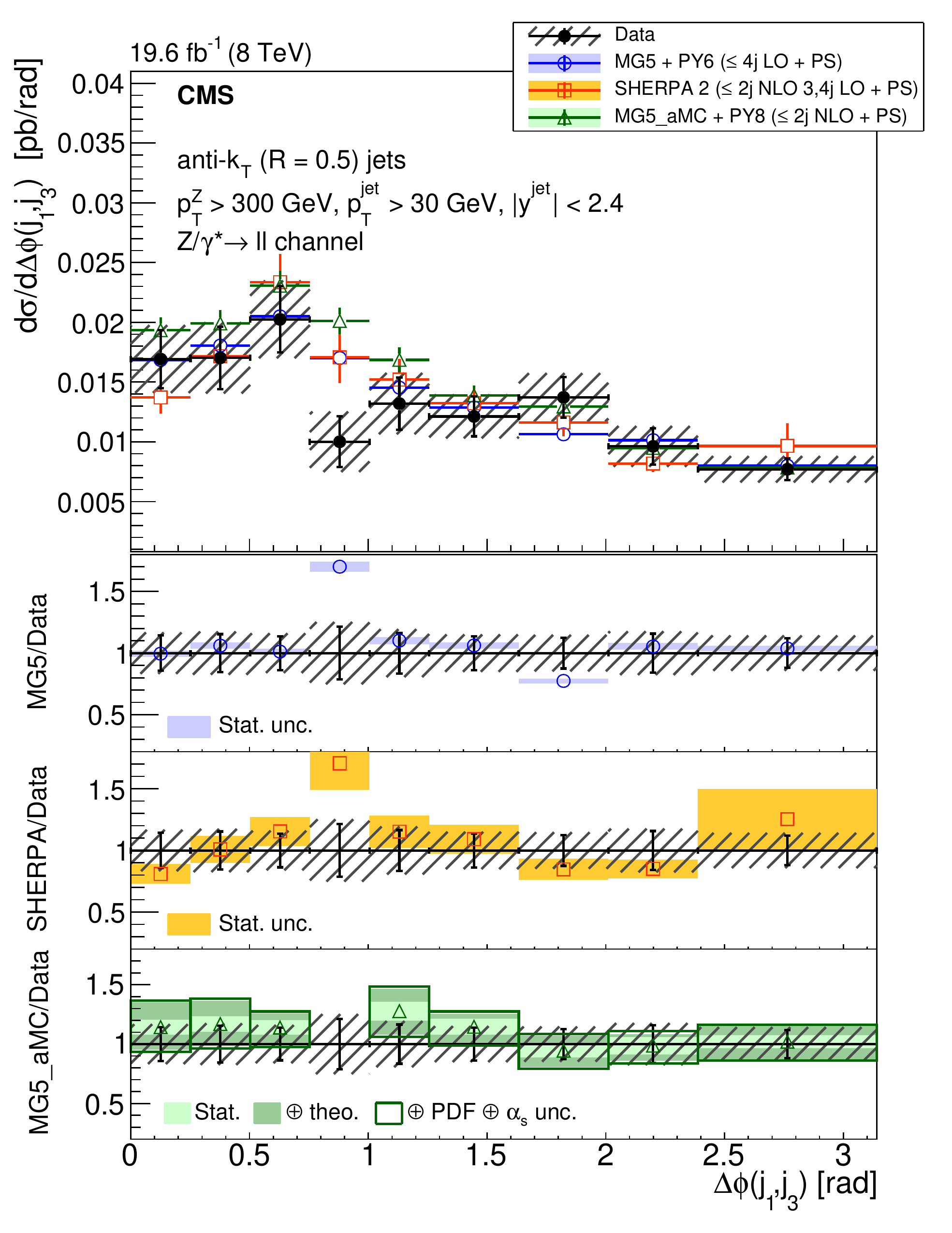}}
 {\includegraphics[width=0.32\textwidth]{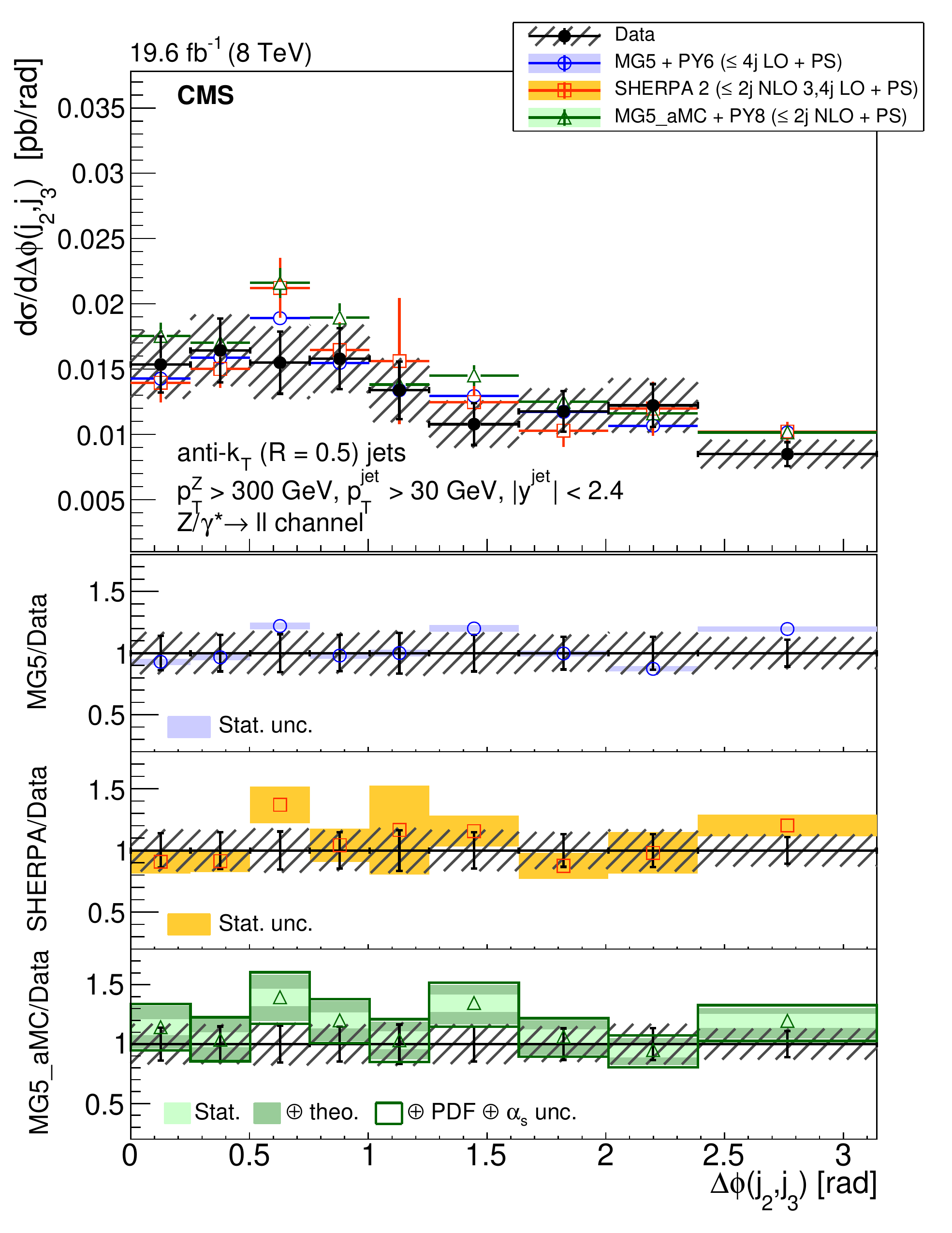}}
 \caption{The differential cross section for \zlljets production for $N_{\text{jets}}\ge 3$ and $\pt(\Z) > 300\GeV$ as a function of the azimuthal angle between (left) the first- and second-, (middle) the first- and third-, and (right) the second- and third- leading-jets. \plotstdcapt}
 \label{fig:dphifstjzpt300}
}
\end{figure}

Figure~\ref{fig:dphifstj} shows the azimuthal angle between the jets in the three-jet inclusive selections. The bumps seen at $\Delta\phi\sim0.5$ come from events with the two leading jets close in rapidity, $\abs{\Delta y} \lesssim 2 R$, where $R$ is the radius parameter of the jet anti-$k_{\text{t}}$ clustering algorithm, $R=0.5$. This region is sensitive to the transition from an area of hadronic activity being resolved as one jet to being resolved as two jets. Increasing the $\pt(\cPZ)$ threshold value
to 150\GeV (Fig.~\ref{fig:dphifstjzpt150}) shows that the splitting of jets in this case is the dominant feature in the three distributions.  Events where a dijet system radiates a $\cPZ$ boson are thus largely suppressed and this is most evident in the $\Delta\phi\left(\text{j}_1, \text{j}_2\right)$ distribution, where the peak at $\pi$ is gone. A further increase in the $\pt(\Z)$ threshold to $300\GeV$ (Fig.~\ref{fig:dphifstjzpt300}) continues this trend. In all cases, the agreement between the measurement and the prediction is still very good.
\clearpage
Overall, the measurements show that Monte Carlo predictions offer a very good description of the azimuthal angles between the jets and the $\cPZ$ boson, achieved when several parton multiplicities are included in the ME calculations and matched with parton showering.

\subsection{Differential cross section for the dijet invariant mass}

\begin{figure}[htbp!]
\centering
\includegraphics[width=0.48\textwidth]{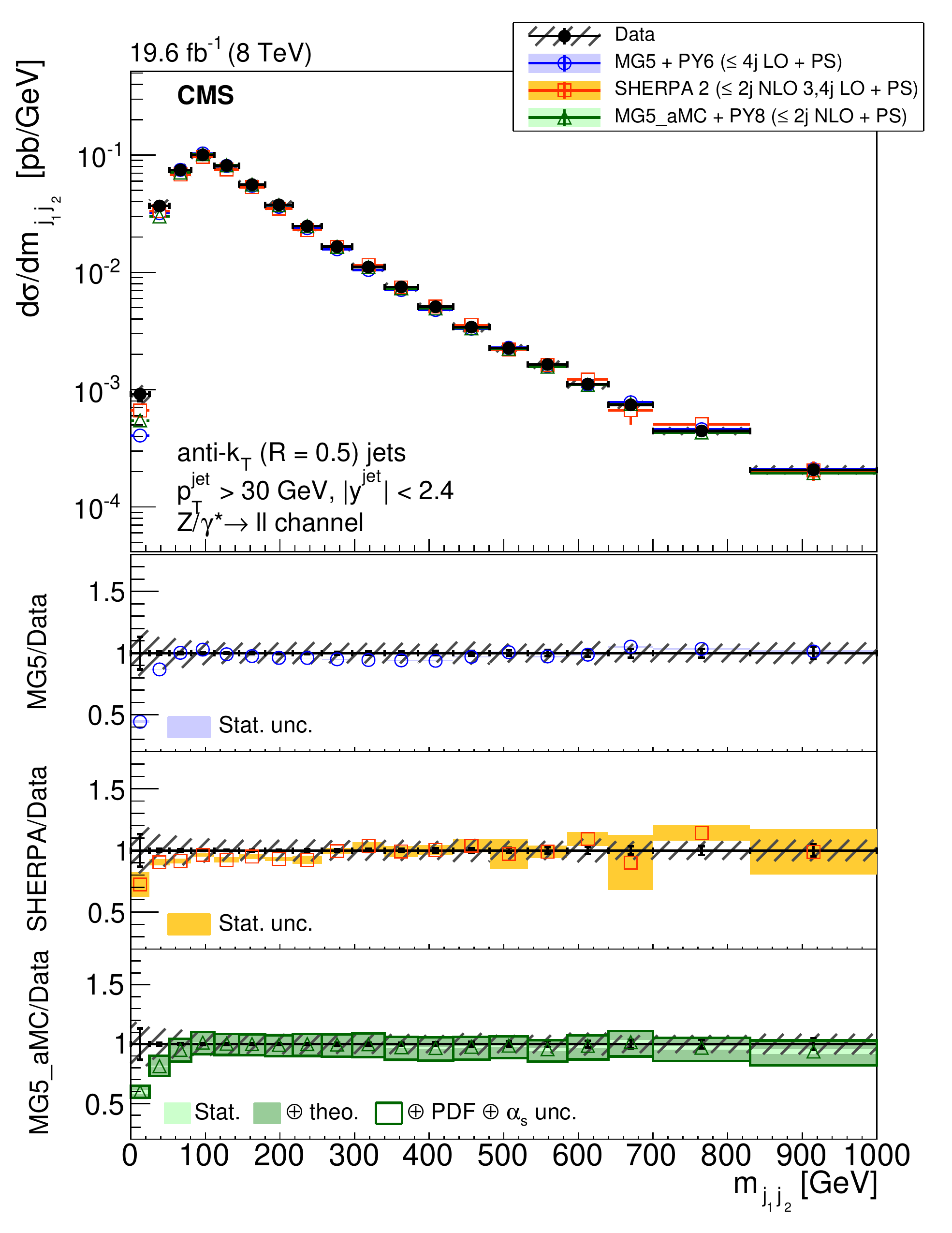}
\caption{The differential cross section for \zlljets production as a function of the dijet invariant mass for $N_{\text{jets}} \geq 2$ compared to the predictions calculated with \MADGRAPH~5 + \PYTHIA~6, \SHERPA~2, and \MGaMC + \PYTHIA~8. \plotstdcapt}
\label{fig:CombXSec_JetsMass}
\end{figure}

The dijet invariant mass is an important variable in the study of the production of a Higgs boson through vector boson fusion, since it can be used to select such events, which contain two jets well-separated in rapidity with a large dijet mass. For this measurement we consider all \zjets events with at least two jets. The measured cross section as a function of the dijet mass is shown in Fig.~\ref{fig:CombXSec_JetsMass}.

The three predictions considered here agree well with the measurement within the experimental uncertainties, except for a dijet mass below $\sim$50\GeV, where the predictions made with \MADGRAPH~5 + \PYTHIA~6 and \MGaMC + \PYTHIA~8 show a deficit with respect to the measurements, while \SHERPA~2 has a better agreement with the measurement in this region. In this region there is a relatively small angle between the two jets. The distribution of the
difference in the rapidities, which are directly linked to the polar angle $\theta$ for massless objects ($y=\eta=-\ln
[\tan(\theta/2)]$) is well reproduced by the three predictions. The distribution of the angle in the transverse plane between the two jets is also well reproduced by all three calculations.

  \subsection{Multidimensional differential cross sections}\label{subsec:multi}

The large number of $\Z +\ge1\ \text{jet}$ events allows the measurement of multidimensional cross sections. We focus on three observables, $\pt(\text{j}_1)$, $y(\cPZ)$, and $y(\text{j}_1)$, that describe the kinematics of the events. Three differential cross sections are measured: $\rd^2\sigma/\rd\pt(\text{j}_1)\rd y(\text{j}_1)$, $\rd^2\sigma/\rd y(\Z)\rd y(\text{j}_1)$, and $\rd^3\sigma/\rd\pt(\text{j}_1)\rd y(\text{j}_1)\rd y(\Z)$. The symmetry with respect to the transverse plane $y=0$ is used to
minimise the statistical uncertainties: $\rd^2\sigma/\rd\pt(\text{j}_1)\rd y(\text{j}_1)$ is obtained from a two-dimensional histogram of $(\pt$, $\abs{y(\text{j}_1)})$ and $\rd^2\sigma/\rd y(\Z)\rd y(\text{j}_1)$ from a histogram of $(|y({\text{j}_1})|$, $\abs{y(\Z)} \sign(y(\Z)y(\text{j}_1))$, where $\sign(x) = 1$ for $x\ge0$ and $\sign(x) = -1$, for $x<0$. The three-dimensional differential cross section is calculated similarly.

 \begin{figure}[htb]
  \centering \includegraphics[width=0.9\textwidth]{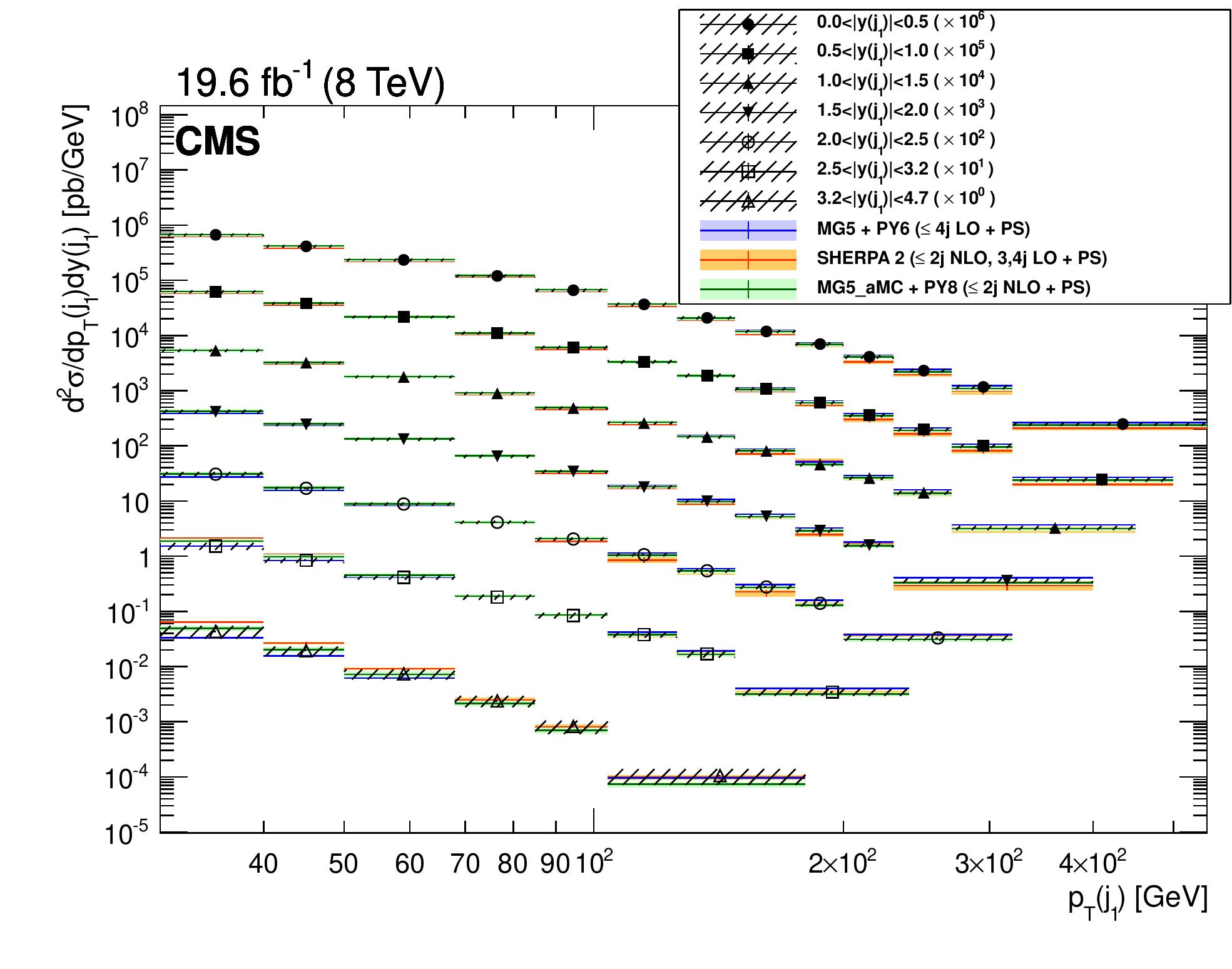}
  \caption{The differential cross section for \zlljets production as a function of the leading jet transverse momentum and rapidity. The bands around the measurement points represent the total measurement uncertainties. The bands around the prediction points represent the total uncertainty, and its statistical, theoretical\alphasunc{$\alpha_S$}, and PDF components for \MGaMC + \PYTHIA~8, and the statistical uncertainty alone for \MADGRAPH~5 + \PYTHIA~6 and \SHERPA~2.}
  \label{fig:ptj_yj}
\end{figure}

 \begin{figure}[!hptb]
   \centering \includegraphics[width=0.57\textwidth]{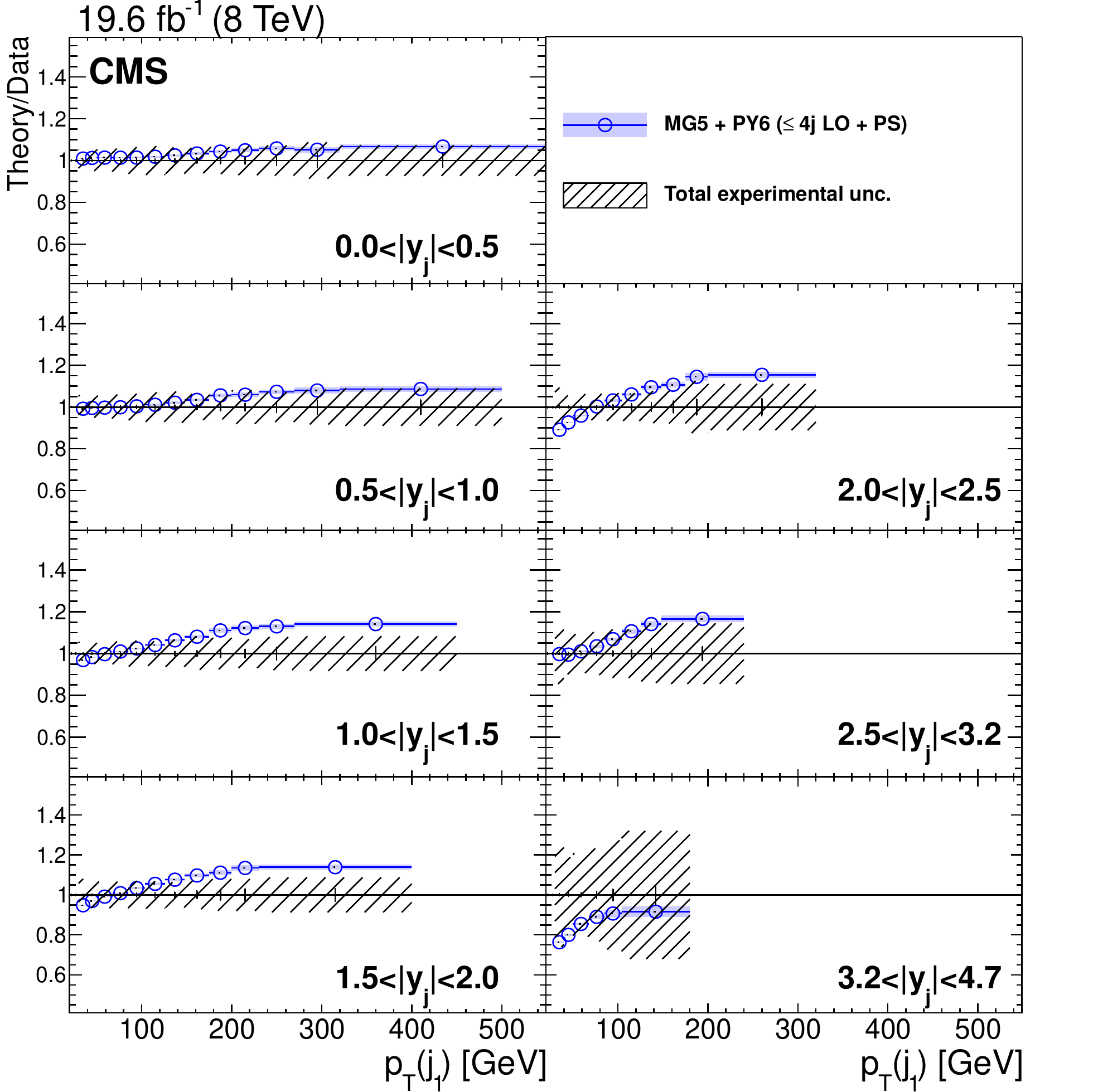}
   \caption{Ratio to the measurement of the differential cross section $\rd^2\sigma/\rd\pt(\text{j}_1)\rd y(\text{j}_1)$ obtained with \MADGRAPH~5 + \PYTHIA~6, with up to four jets at LO. The total experimental uncertainty is shown as a band around 1. Uncertainties in the predictions are shown on the ratio points and include the statistical uncertainty only.}
   \label{fig:ptj_yj_ratio1}
 \end{figure}

 \begin{figure}[!hptb]
   \centering \includegraphics[width=0.58\textwidth]{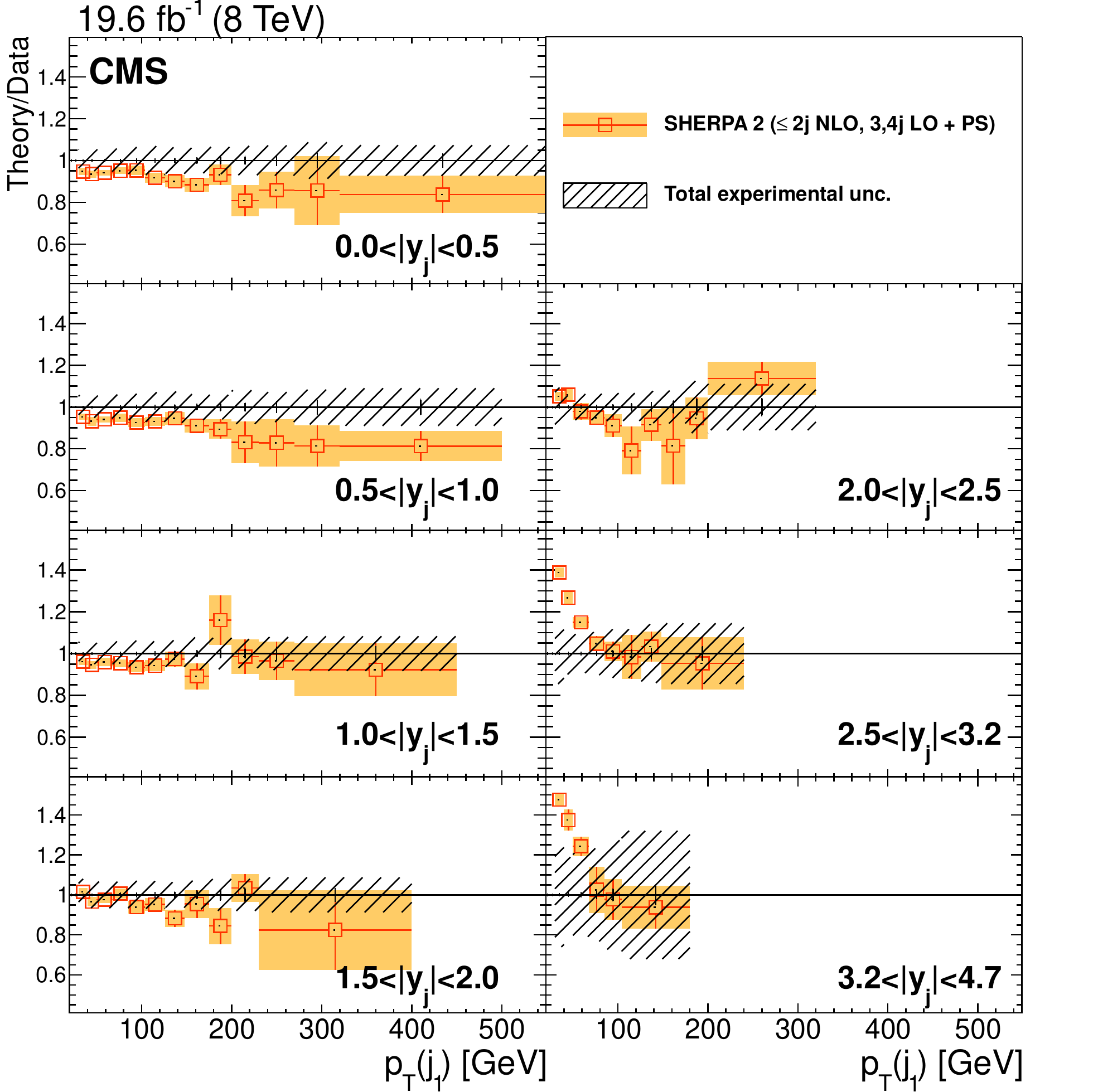}
   \caption{Ratio to the measurement of the differential cross section $\rd^2\sigma/\rd\pt(\text{j}_1)\rd y(\text{j}_1)$ obtained with \SHERPA~2, with up to two jets at NLO and up to four jets at LO. The total experimental uncertainty is shown as a band around 1. Uncertainties in the predictions are shown on the ratio points and include the statistical uncertainty only.}
   \label{fig:ptj_yj_ratio3}
 \end{figure}

 \begin{figure}[!hptb]
   \centering \includegraphics[width=0.57\textwidth]{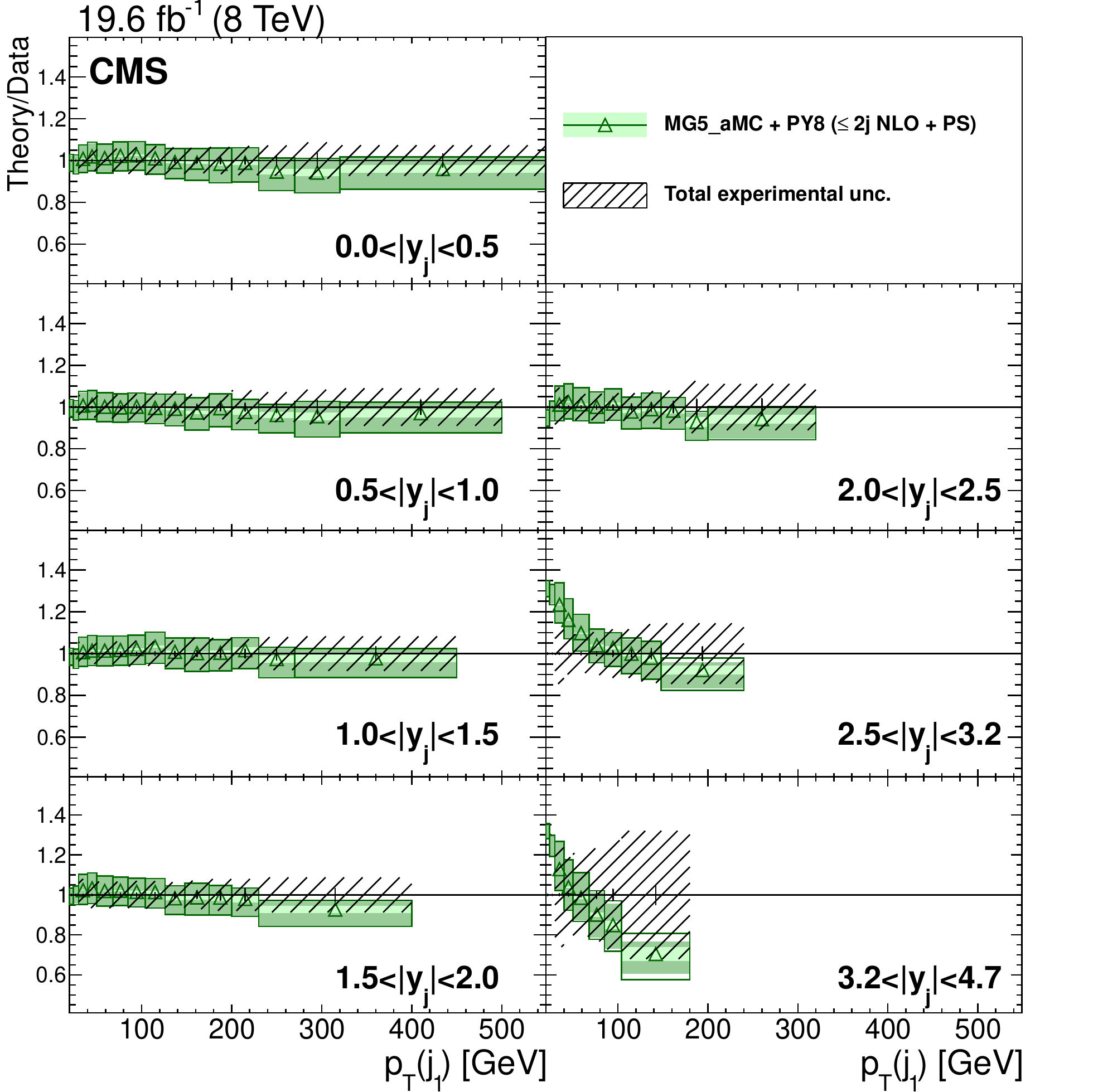}
   \caption{Ratio to the measurement of the differential cross section $\rd^2\sigma/\rd\pt(\text{j}_1)\rd y(\text{j}_1)$ obtained with \MGaMC + \PYTHIA~8, with up to two jets at NLO. The total experimental uncertainty is shown as a band around 1. Uncertainties in the predictions are shown on the ratio points and include the statistical, theoretical\alphasunc{, $\alpha_S$}, and PDF uncertainties. The dark green area represents the statistical and theoretical uncertainties only, while the light green area represents the statistical uncertainty alone.}

   \label{fig:ptj_yj_ratio2}
 \end{figure}

The $\rd^2\sigma/\rd\pt(\text{j}_1)d\abs{y(\text{j}_1)}$ measurement, shown in Fig.~\ref{fig:ptj_yj}, corresponds to the range $\pt<550\GeV$ of the $\rd\sigma/\rd\pt$ measurement shown in Fig.~\ref{fig:CombXSec_FirstJetPt} and extends the jet absolute rapidity range up to $4.7$. The ratios of the theoretical predictions obtained from \MADGRAPH~5 + \PYTHIA~6, \SHERPA~2, and \MGaMC + \PYTHIA~8 to the measurement are presented in Figs.~\ref{fig:ptj_yj_ratio1}--\ref{fig:ptj_yj_ratio2}. The difference in the shapes of the
$\rd\sigma/\rd\pt$ spectrum between the measurement and the predictions computed with \MADGRAPH~5 + \PYTHIA~6 increases when moving from the central region, $\abs{y(\text{j}_1)}=0$ to the more forward region, $\abs{y(\text{j}_1)}=2.5$. The comparison of the \SHERPA~2, and \MGaMC + \PYTHIA~8 predictions with the measurement does not show any dependence on the rapidity of the jet for the region $\abs{y(\text{j}_1)}<2.5$, within the statistical uncertainty of the prediction, that is larger than for the \MADGRAPH~5 + \PYTHIA~6
sample. In the region beyond $\abs{y(\text{j}_1)}=2.5$ the \MADGRAPH~5 + \PYTHIA~6 prediction-to-measurement ratio shows the same feature as for $\abs{y(\text{j}_1)}<2.5$, despite the large experimental uncertainties due to a larger jet energy scale uncertainty. The \SHERPA~2 prediction shows a significant difference with the spectrum of the jet transverse momentum being narrower than in data. The \MGaMC + \PYTHIA~8 shows a similar feature, but less pronounced and covered by the experimental uncertainties.

 \begin{figure}[p!]
  \centering \includegraphics[width=0.9\textwidth]{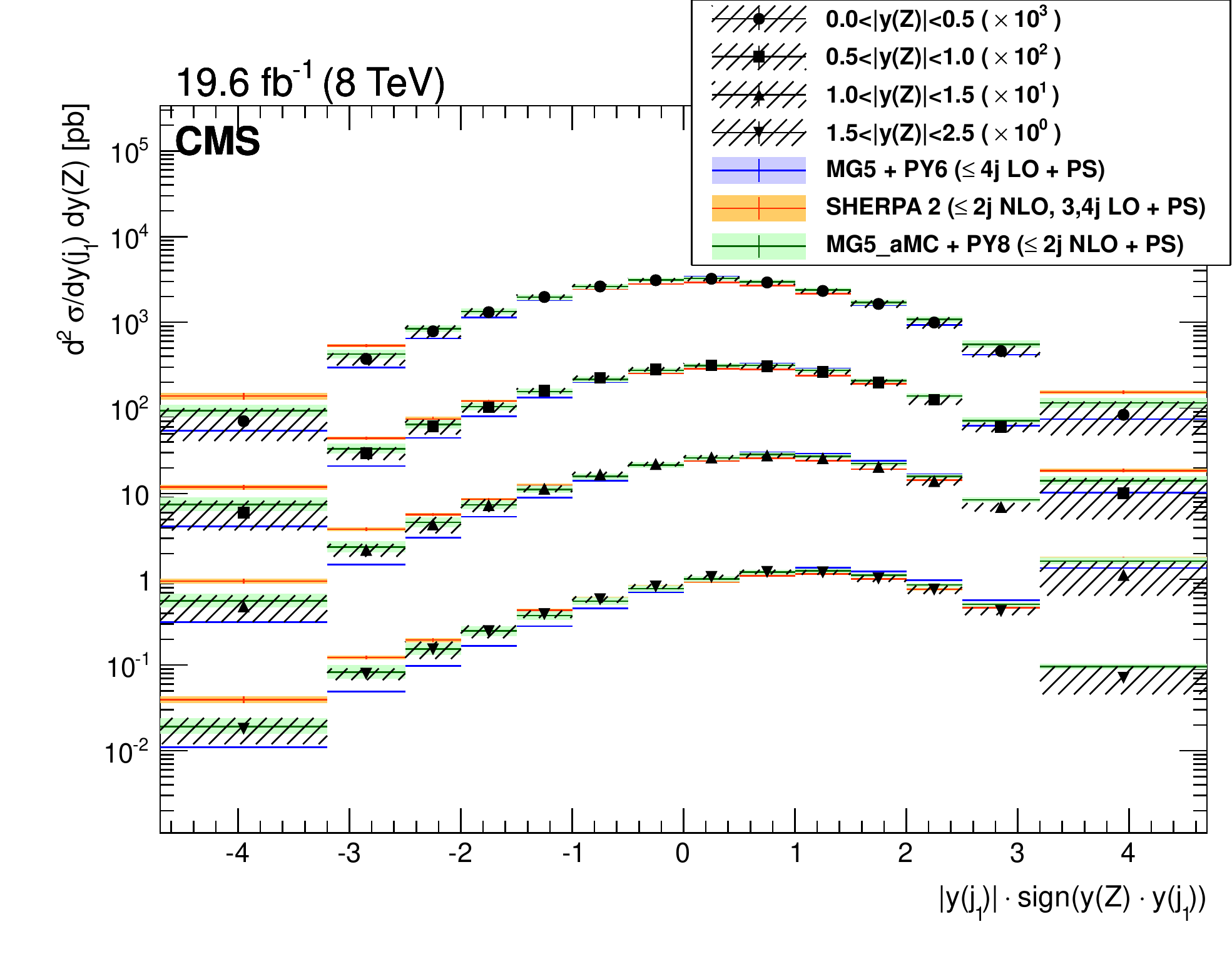}
  \caption{The differential cross section for \zlljets production as a function of the $\cPZ$ boson and leading jet rapidity. The bands around the measurement points represent the total measurement uncertainties. The bands around the prediction points represent the total uncertainty, statistical, theoretical\alphasunc{$\alpha_S$}, and PDF components for \MGaMC + \PYTHIA~8, and the statistical uncertainty alone for \MADGRAPH~5 + \PYTHIA~6 and \SHERPA~2.}
  \label{fig:yj_yZ}
\end{figure}

\begin{figure}[!hptb]
   \centering \includegraphics[width=0.5\textwidth]{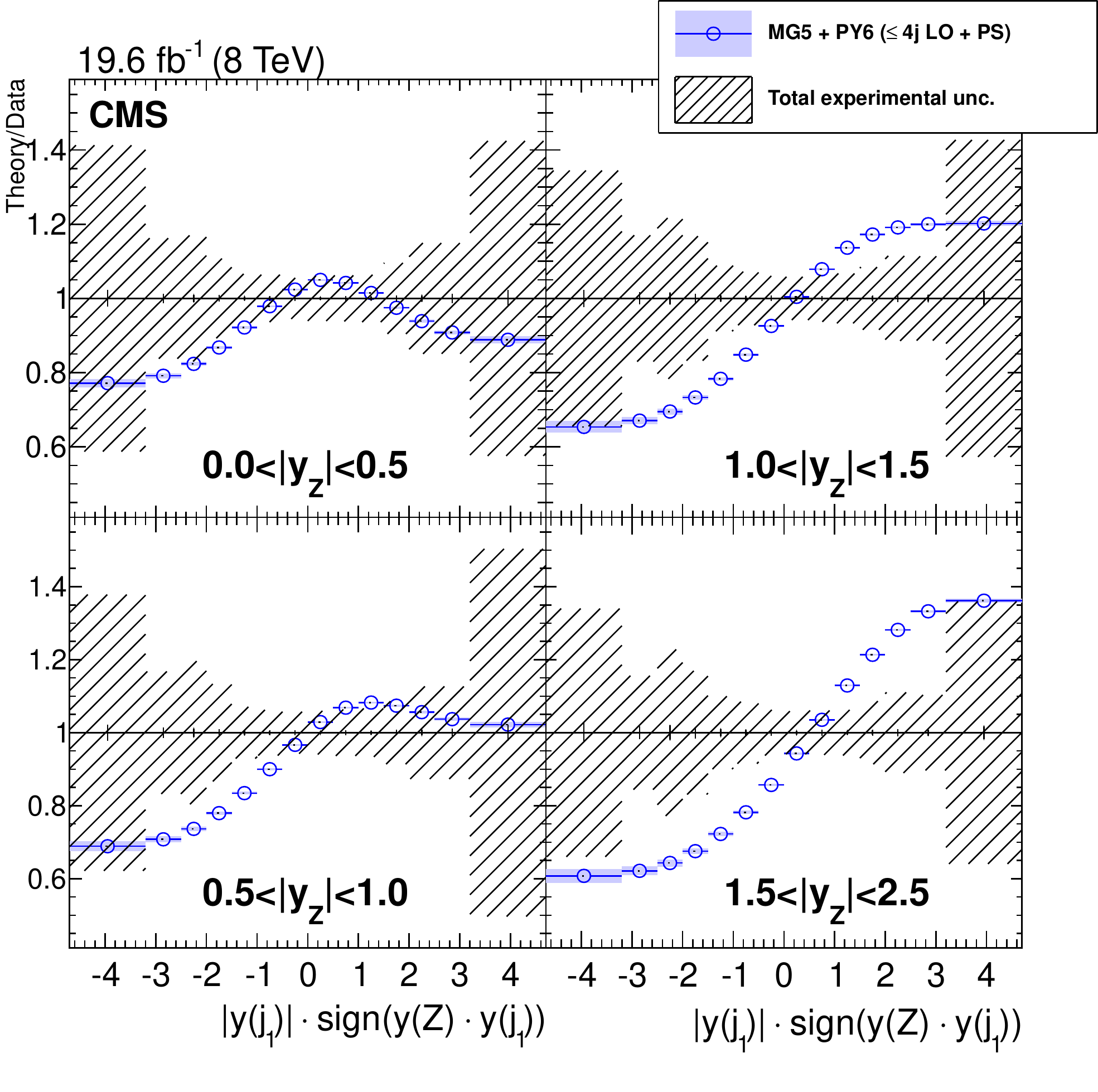}
   \caption{Ratio to the measurement of the differential cross section $\rd^2\sigma/\rd y(\cPZ)\rd y(\text{j}_1)$ obtained with \MADGRAPH~5 + \PYTHIA~6, with up to four jets at LO. The total experimental uncertainty is shown as a band around 1. Uncertainties in the predictions are shown on the ratio points and represent the statistical uncertainty alone.}
   \label{fig:yj_yZ_ratio1}
\end{figure}

\begin{figure}[ptbh]
   \centering \includegraphics[width=0.5\textwidth]{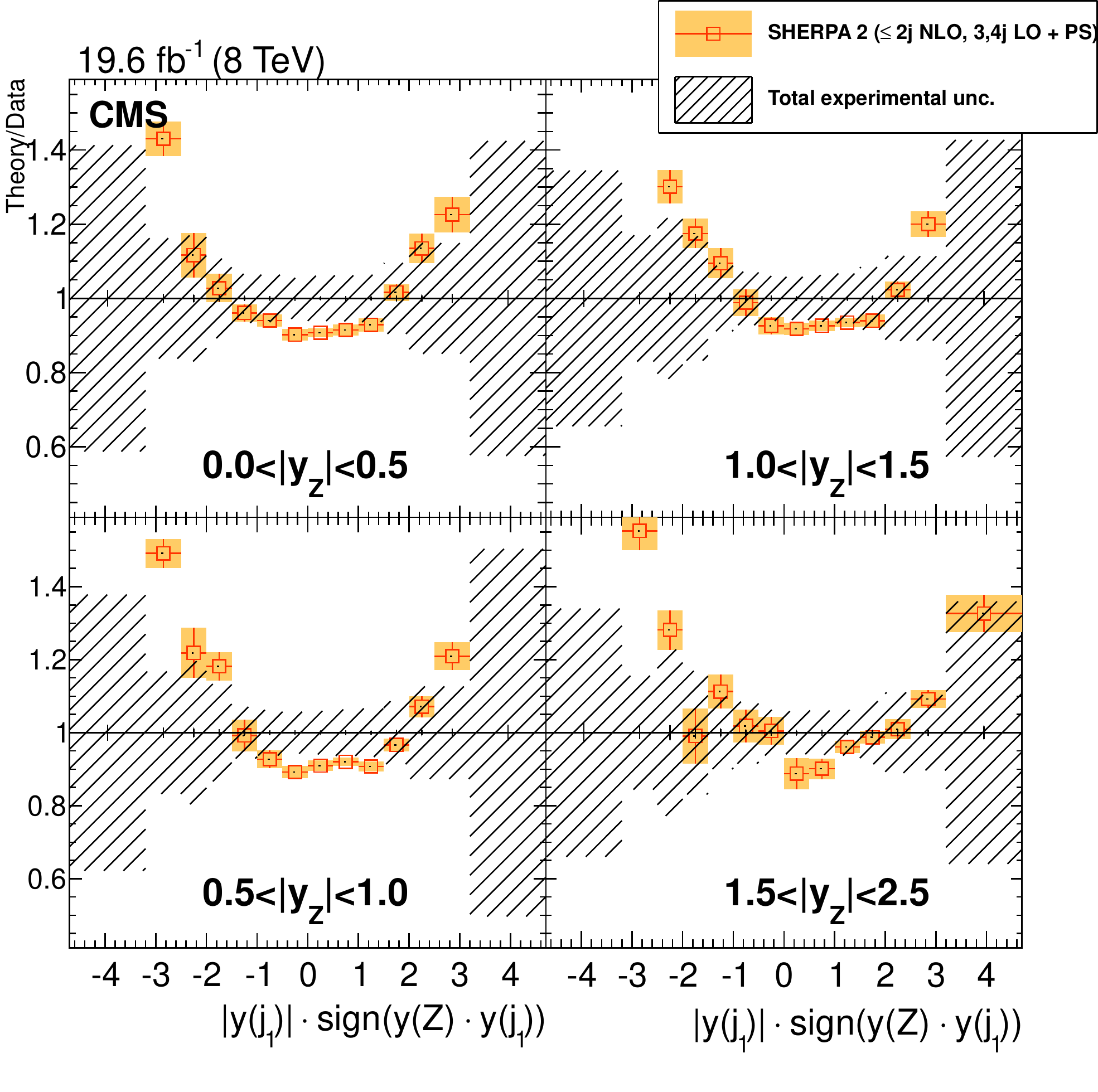}
   \caption{Ratio to the measurement of the differential cross section $\rd^2\sigma/\rd y(\cPZ)\rd y(\text{j}_1)$ obtained with \SHERPA~2, with up to two jets at NLO and up to four jets at LO. The total experimental uncertainty is shown as a band around 1. Uncertainties in the predictions are shown on the ratio points and include the statistical uncertainty only.}
   \label{fig:yj_yZ_ratio3}
\end{figure}

While the $\cPZ$ boson and jet rapidity distributions are independently well modelled by the simulation, we see in Section~\ref{sec:rap} that it is not the case with the tree-level calculations for the correlations between these two observables. Figures~\ref{fig:yj_yZ}--\ref{fig:yj_yZ_ratio3} show the two-dimensional cross section with respect to both rapidities. When the $\cPZ$ boson is central, the \MADGRAPH~5 + \PYTHIA~6 calculation predicts a more central leading jet, while when it is forward, it
predicts a more forward leading jet in the same hemisphere ($y(\cPZ) \, y(\text{j}_{1})>0$). These results are consistent with the measurement presented in Section~\ref{sec:rap} which showed that \MADGRAPH~5 + \PYTHIA~6 predicts a smaller $\ydiff$ (Fig.~\ref{fig:CombXSec_FirstJetEta}). The predictions from \SHERPA~2, and \MGaMC + \PYTHIA~8 agree well with the measurement when the jet is in the central region $\abs{y(\text{j}_1)} < 2.5$, while discrepancies start to appear when it is more forward. The tail of the jet
rapidity is larger in the prediction, especially when the $\cPZ$ boson and the jets are well separated in rapidity: the discrepancy for $y(\cPZ) \, y(\text{j}_1)<0$ is larger for higher $\abs{y(\cPZ)}$. The discrepancies are more pronounced for the prediction obtained with \SHERPA~2.

 \begin{figure}[ptbh]
   \centering \includegraphics[width=0.5\textwidth]{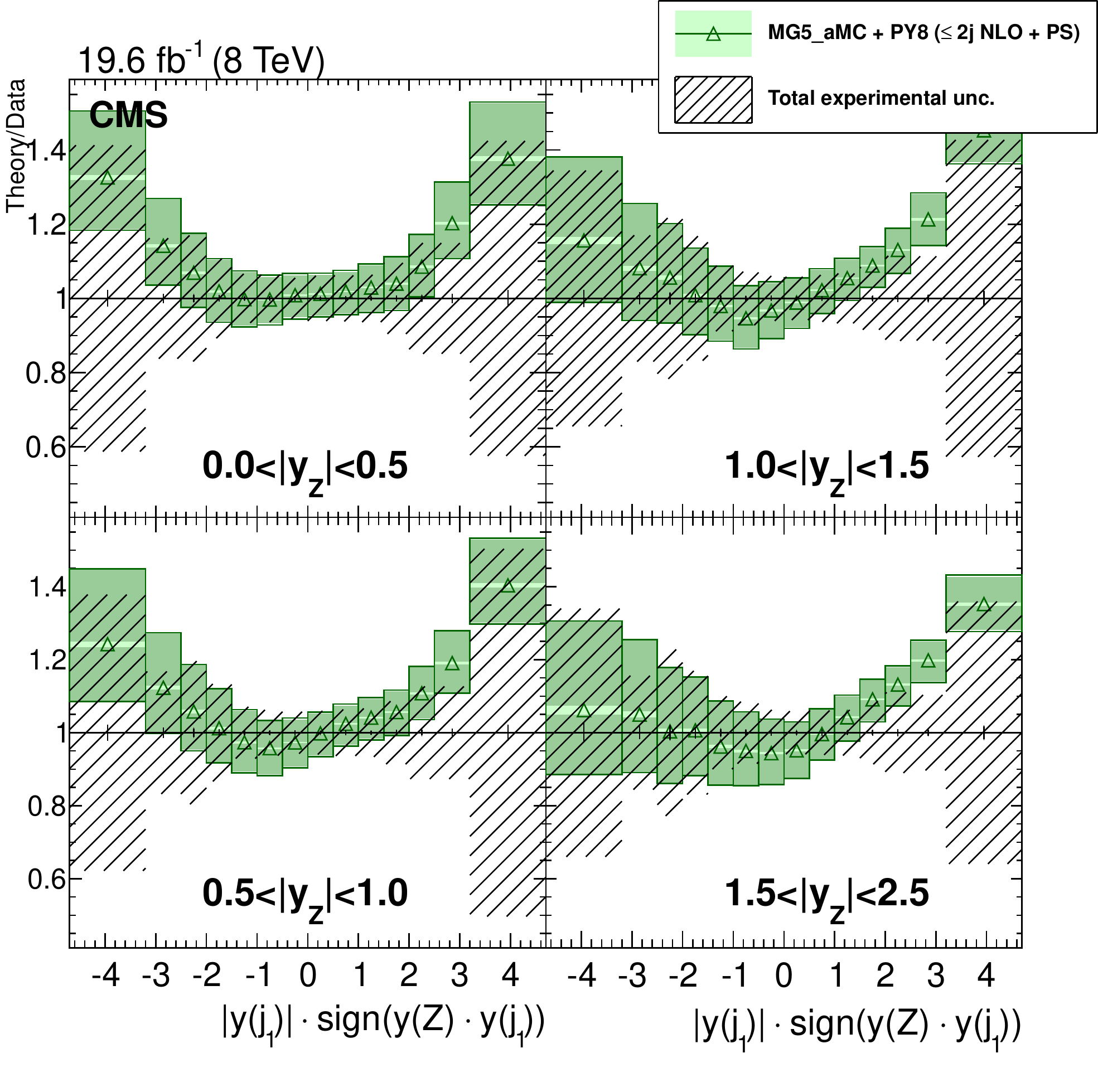}
   \caption{Ratio to the measurement of the differential cross section $\rd^2\sigma/\rd y(\cPZ)\rd y(\text{j}_1)$ obtained with \MGaMC + \PYTHIA~8, with up to two jets at NLO. The total experimental uncertainty is shown as a band around 1. Uncertainties in the predictions are shown on the ratio points and include the statistical, theoretical\alphasunc{, $\alpha_S$}, and PDF uncertainties. The dark green area represents the statistical and theoretical uncertainties only, while the light green area represents the statistical uncertainty alone.}
   \label{fig:yj_yZ_ratio2}
\end{figure}

 \begin{figure}[ptbh]
\centering\includegraphics[width=0.9\textwidth]{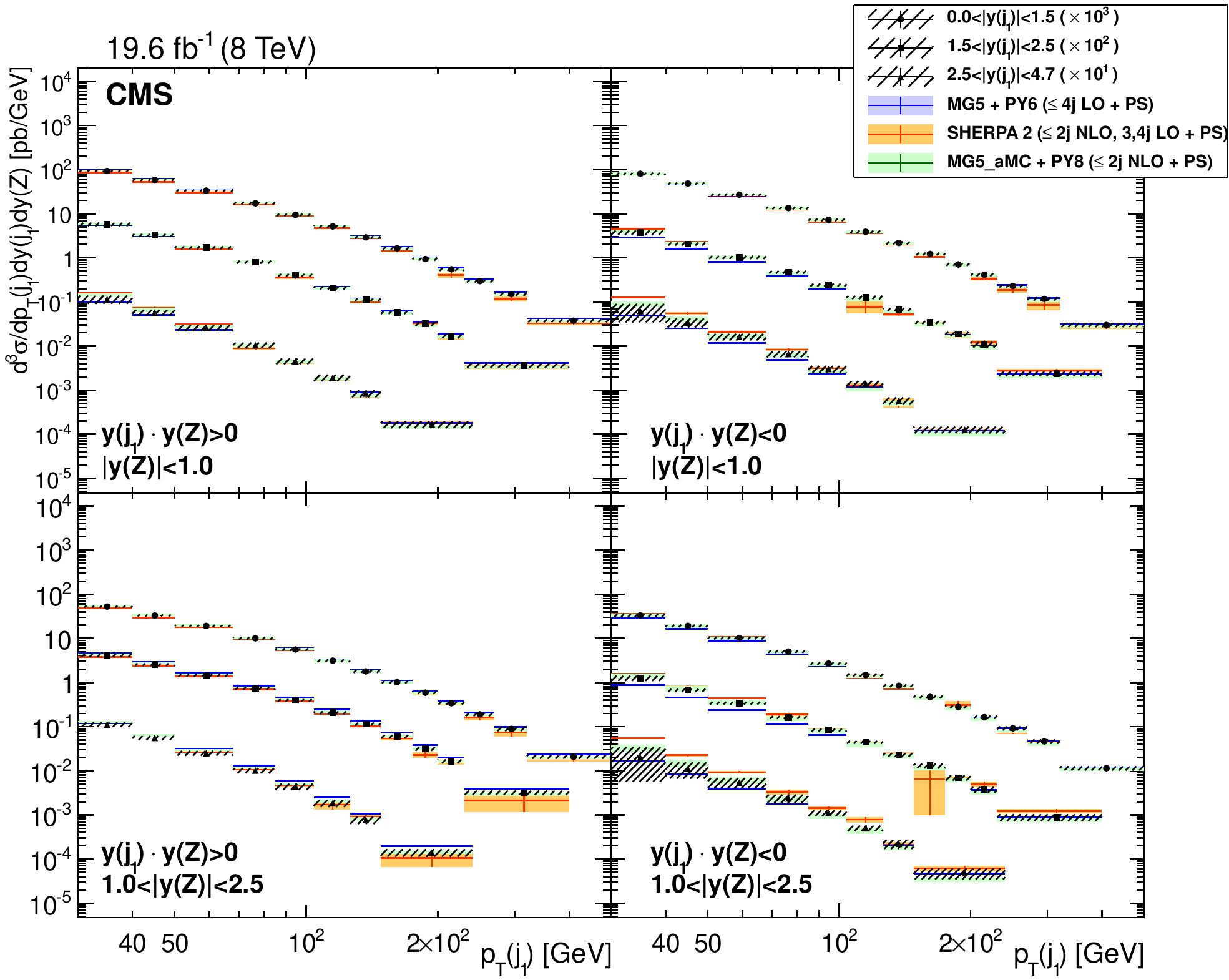}

  \caption{The differential cross section for \zlljets production as a function of the rapidities of the $\cPZ$ boson and leading jet, and of the transverse momentum of the jet for the configuration. The bottom plots correspond to the configuration where the boson and the jet are in different hemispheres ($y(\cPZ)y(\text{j}_1)<0$), while the top plots correspond to both objects in the same hemisphere. The left and right plots show the respective $\cPZ$ boson rapidity ranges, $\abs{y(\cPZ)}<1$ and
  $\abs{y(\cPZ)}\in(1,2.5)$. The bands around the measurement points represent the total measurement uncertainties. The bands around the prediction points represent the total uncertainty, statistical, theoretical\alphasunc{$\alpha_S$}, and PDF components for \MGaMC + \PYTHIA~8, and the statistical uncertainty alone for \MADGRAPH~5 + \PYTHIA~6 and \SHERPA~2.}
  \label{fig:pt_yj_yZ}
\end{figure}

\begin{figure}[ptbh]
  \centering \includegraphics[width=0.56\textwidth]{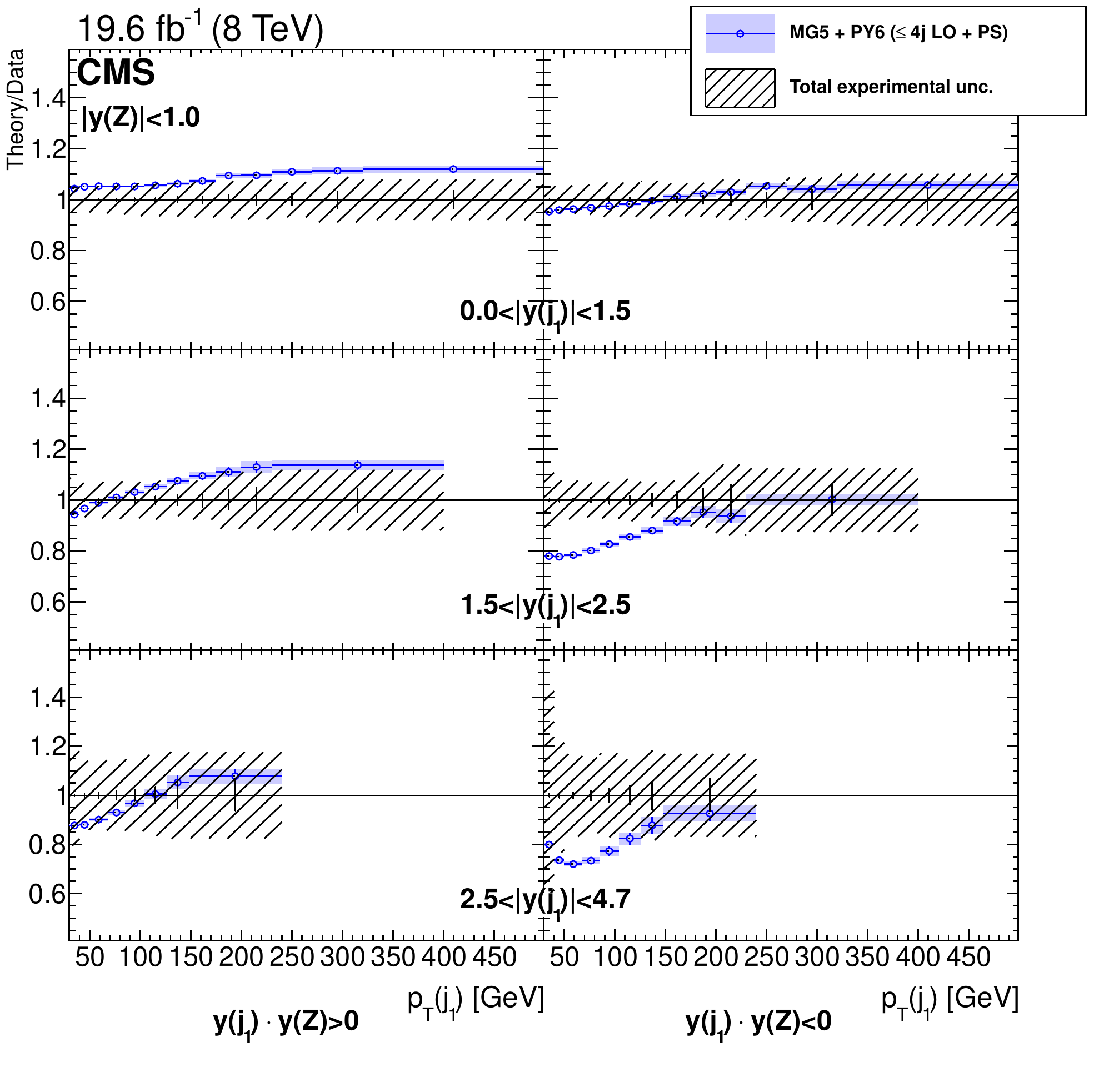}
 \caption{Ratio to the measurement of the differential cross section $\rd^3\sigma/\rd\pt(\text{j}_1)\rd y(\text{j}_1)\rd y(\cPZ)$ obtained with \MADGRAPH~5 + \PYTHIA~6, with up to four jets at LO, for $\abs{y(\cPZ)}<1$. Left column corresponds to $y(\text{j}_1)y(\cPZ)>0$ and right column to $y(\text{j}_1)y(\cPZ)<0$. The total experimental uncertainty is shown as a band around 1. Uncertainties in the predictions are shown on the ratio points and include the statistical uncertainty only.}
   \label{fig:ptj_yj_yZ_ratio_pos1}
 \end{figure}

\begin{figure}[ptbh]
  \centering \includegraphics[width=0.56\textwidth]{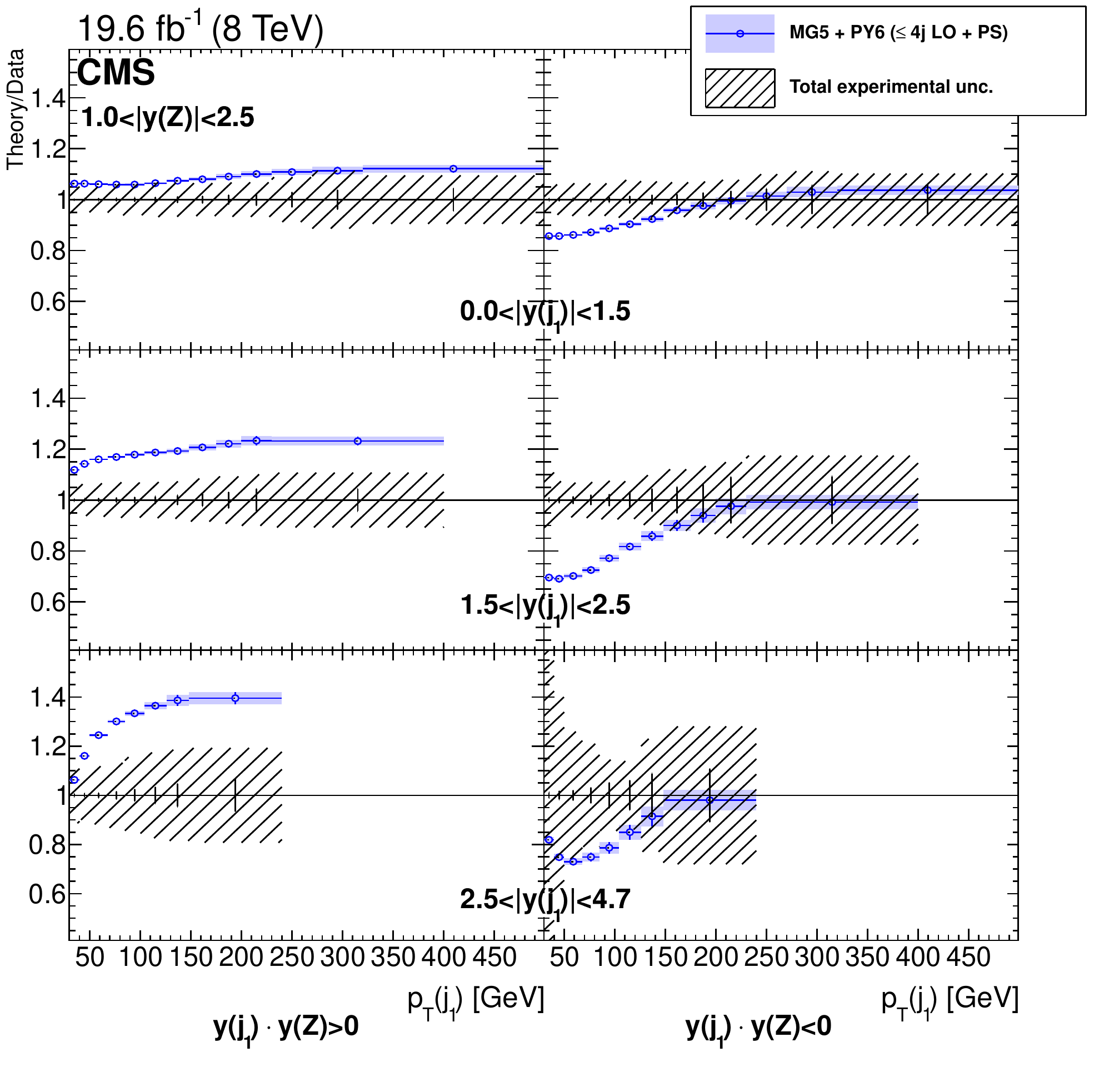}
  \caption{Ratio to the measurement of the differential cross section $\rd^3\sigma/\rd\pt(\text{j}_1)\rd y(\text{j}_1)\rd y(\text{Z})$ obtained with \MADGRAPH~5 + \PYTHIA~6, with up to four jets at LO, for $\abs{y(\cPZ)}\in(1,2.5)$. Left column corresponds to $y(\text{j}_1)y(\cPZ)>0$ and right column to $y(\text{j}_1)y(\cPZ)<0$. The total experimental uncertainty is shown as a band around 1. Uncertainties in the predictions are shown on the ratio points and include the statistical uncertainty only.}
   \label{fig:ptj_yj_yZ_ratio_neg1}
 \end{figure}

\begin{figure}[ptbh]
  \centering \includegraphics[width=0.56\textwidth]{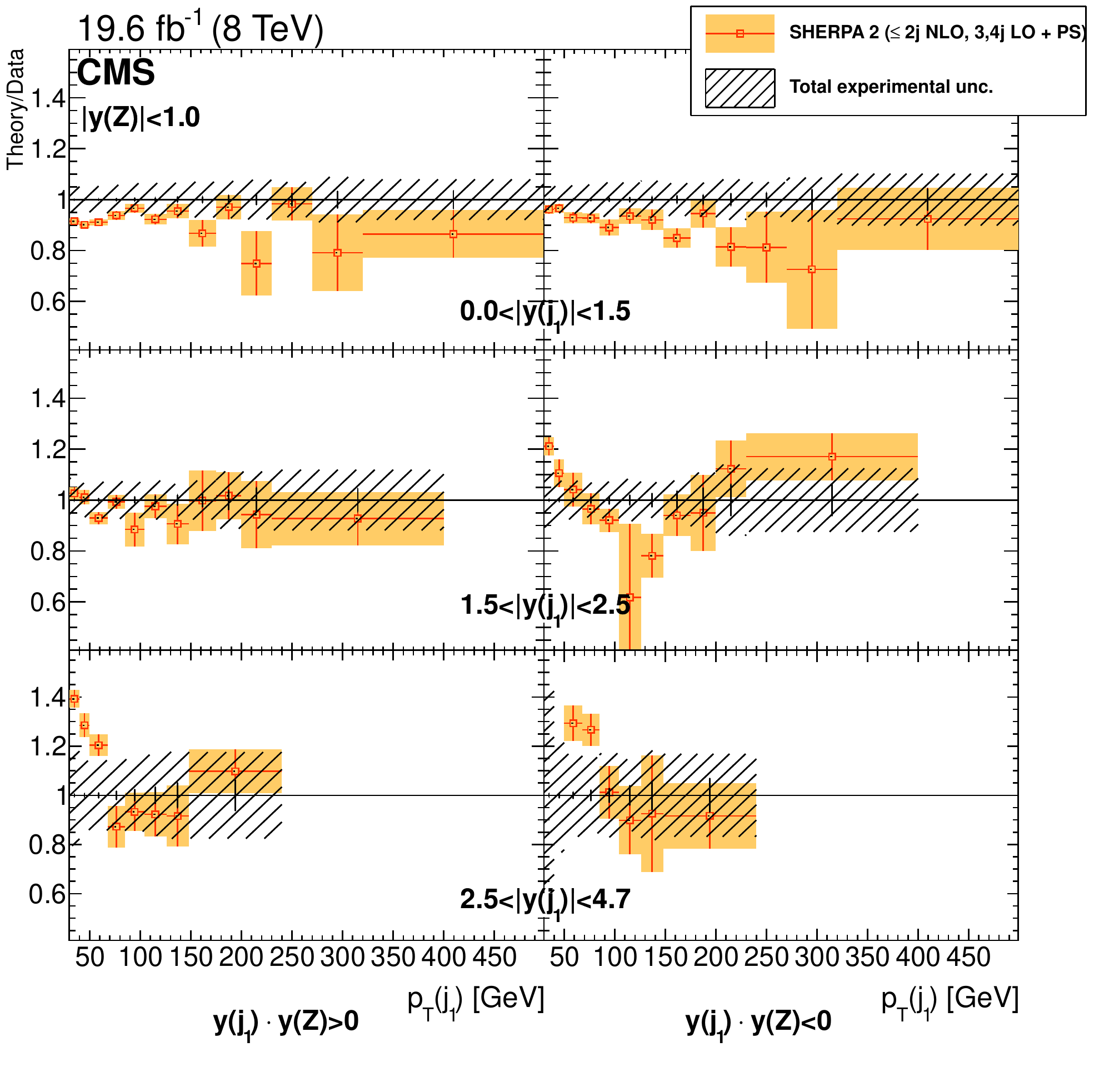}
 \caption{Ratio to the measurement of the differential cross section $\rd^3\sigma/\rd\pt(\text{j}_1)\rd y(\text{j}_1)\rd y(\cPZ)$ obtained with \SHERPA~2, with up to two jets at NLO and up to four jets at LO, for $\abs{y(\cPZ)}<1$. Left column corresponds to $y(\text{j}_1)y(\cPZ)>0$ and right column to $y(\text{j}_1)y(\cPZ)<0$. The total experimental uncertainty is shown as a band around 1. Uncertainties in the predictions are shown on the ratio points and include the statistical uncertainty only.}
   \label{fig:ptj_yj_yZ_ratio_pos3}
 \end{figure}

\begin{figure}[ptbh]
  \centering \includegraphics[width=0.56\textwidth]{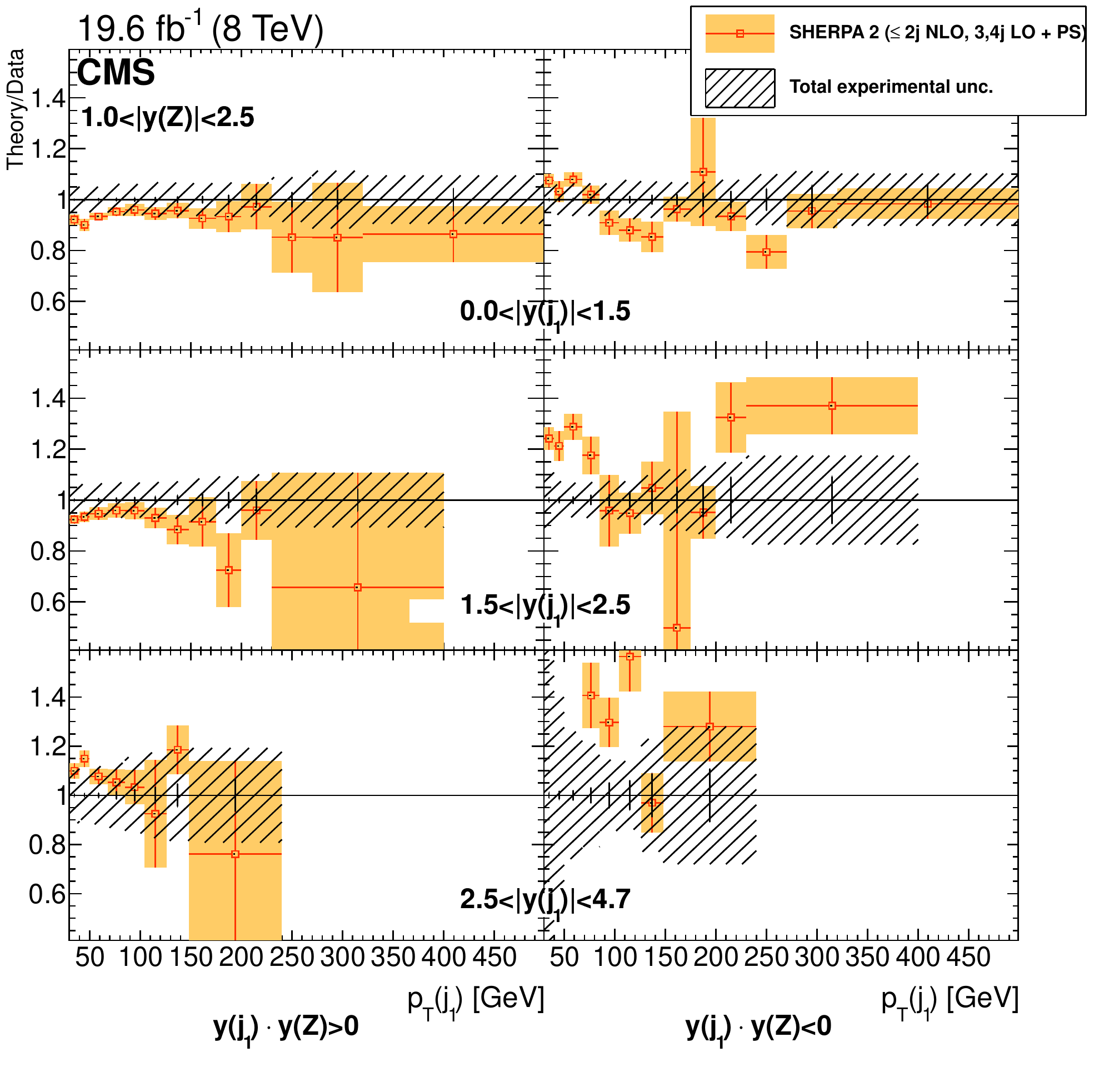}
 \caption{Ratio to the measurement of the differential cross section $\rd^3\sigma/\rd\pt(\text{j}_1)\rd y(\text{j}_1)\rd y(\cPZ)$ obtained with \SHERPA~2, with up to two jets at NLO and up to four jets at LO, for $\abs{y(\cPZ)}\in(1,2.5)$. Left column corresponds to $y(\text{j}_1)y(\cPZ)>0$ and right column to $y(\text{j}_1)y(\cPZ)<0$. The total experimental uncertainty is shown as a band around 1. Uncertainties in the predictions are shown on the ratio points and include the statistical uncertainty only.}
   \label{fig:ptj_yj_yZ_ratio_neg3}
 \end{figure}

\begin{figure}[ptbh]
  \centering \includegraphics[width=0.5\textwidth]{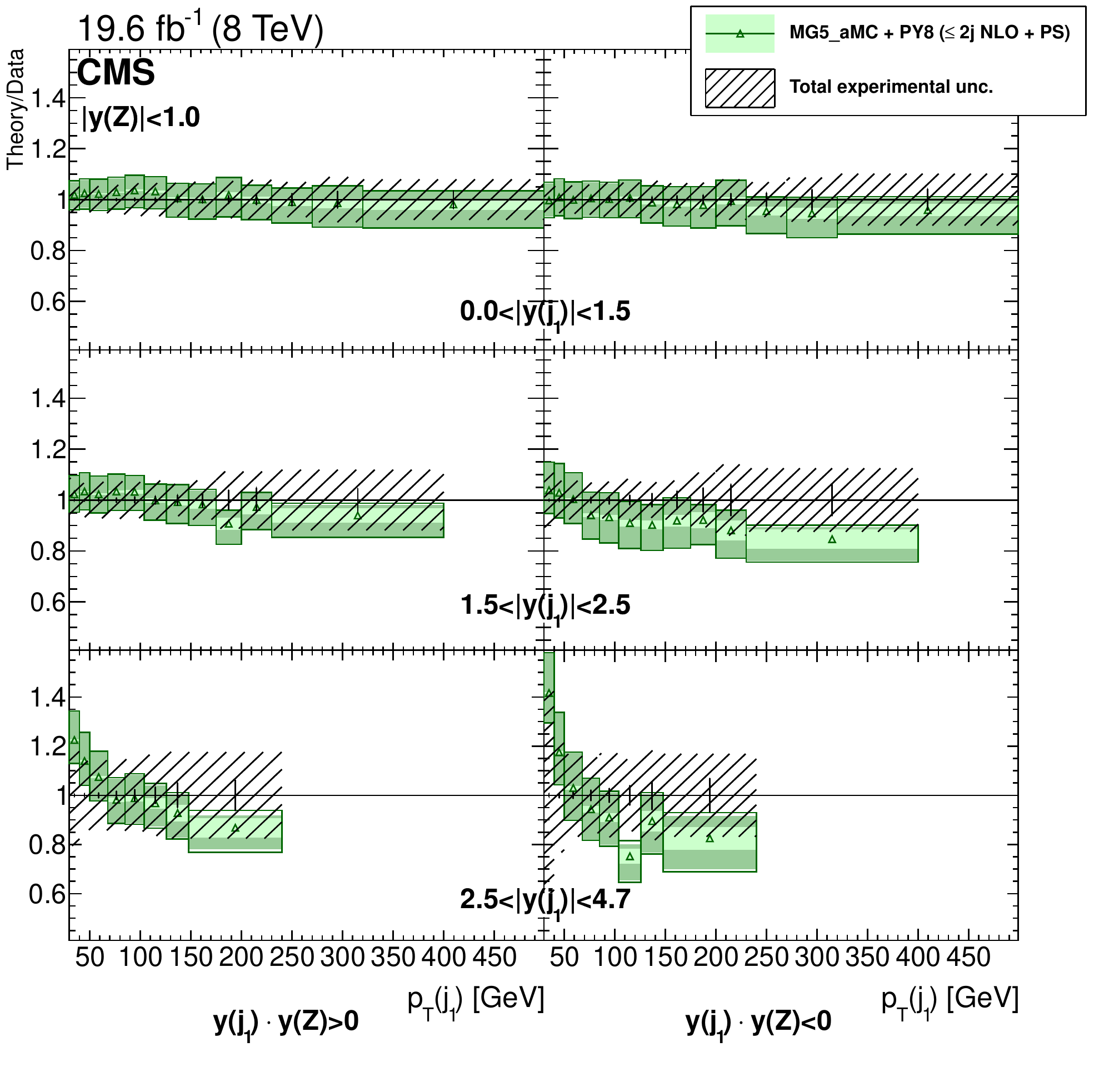}
 \caption{Ratio to the measurement of the differential cross section $\rd^3\sigma/\rd\pt(\text{j}_1)\rd y(\text{j}_1)\rd y(\cPZ)$ obtained with \MGaMC + \PYTHIA~8, with up to two jets at NLO, for $\abs{y(\cPZ)}<1$. Left column corresponds to $y(\text{j}_1)y(\cPZ)>0$ and right column to $y(\text{j}_1)y(\cPZ)<0$. The total experimental uncertainty is shown as a band around 1. Uncertainties in the predictions are shown on the ratio points and include the statistical, theoretical\alphasunc{,
 $\alpha_S$}, and PDF uncertainties. The dark green area represents the statistical and theoretical uncertainties only, while the light green area represents the statistical uncertainty alone.}
   \label{fig:ptj_yj_yZ_ratio_pos2}
 \end{figure}

\begin{figure}[ptbh]
  \centering \includegraphics[width=0.51\textwidth]{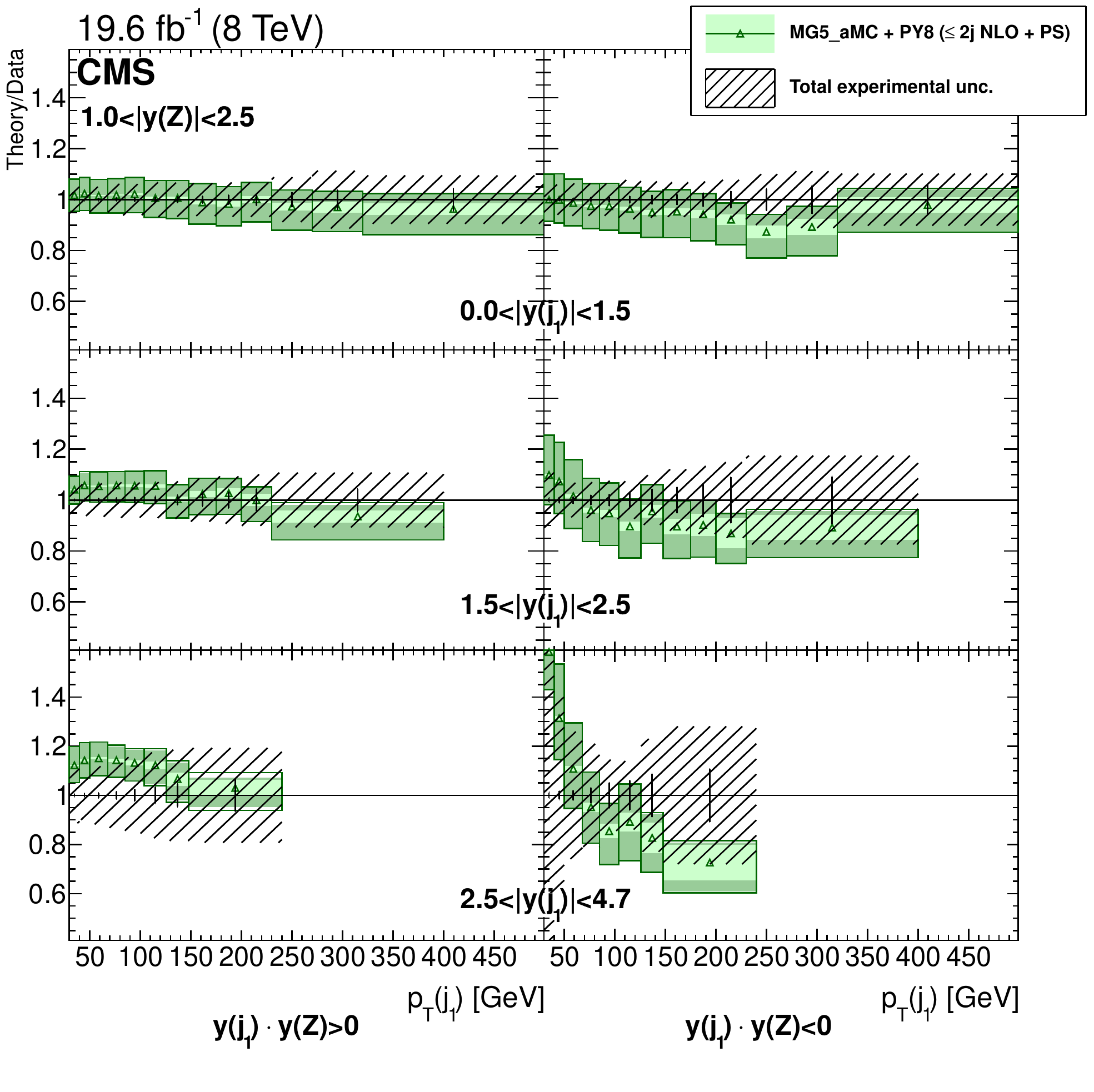}
 \caption{Ratio to the measurement of the differential cross section $\rd^3\sigma/\rd\pt(\text{j}_1)\rd y(\text{j}_1)\rd y(\text{Z})$ obtained with \MGaMC + \PYTHIA~8, with up to two jets at NLO, for $\abs{y(\cPZ)}\in(1,2.5)$. Left column corresponds to $y(\text{j}_1)y(\cPZ)>0$ and right column to $y(\text{j}_1)y(\cPZ)<0$. The total experimental uncertainty is shown as a band around 1. Uncertainties in the predictions are shown on the ratio points and include the statistical, theoretical\alphasunc{,
 $\alpha_S$}, and PDF uncertainties. The dark green area represents the statistical and theoretical uncertainties only, while the light green area represents the statistical uncertainty alone.}
   \label{fig:ptj_yj_yZ_ratio_neg2}
 \end{figure}

Finally, the measurement of the differential cross section with respect to both jet transverse momentum and rapidity is repeated for two different intervals of the $\cPZ$ boson rapidity as shown in Figs.~\ref{fig:pt_yj_yZ}--\ref{fig:ptj_yj_yZ_ratio_neg3}. The shape of the ratio of the \MADGRAPH~5 + \PYTHIA~6 prediction to the measurement of the leading jet transverse momentum spectrum is similar in both intervals, although it shows a more pronounced discrepancy when the boson is in the most forward
region. In the jet rapidity region $\abs{y(\text{j}_1)}\in(1,2.5)$ with $y(\text{j}_1)\, y(\cPZ) > 0$, the ratios actually differ between the two $\cPZ$ boson rapidity intervals. However, in view of the measurement uncertainties this discrepancy cannot be considered significant. The behaviour seen previously for $\rd^{2}\sigma/(\rd y(\cPZ)\rd y(\text{j}_1))$ translates into global shifts of the ratio distributions depending on the $y(\cPZ) \, y(\text{j}_1)$ interval. The bottom plots of
Figs.~\ref{fig:ptj_yj_yZ_ratio_pos2} and \ref{fig:ptj_yj_yZ_ratio_neg2} give more insight for the discrepancy with respect to the measurement observed previously for the \SHERPA~2, and \MGaMC + \PYTHIA~8 predictions when the jet is in the forward region, $\abs{y(\text{j}_1)}\in(2.5,4.7)$. The observed deficit in the cross section can be attributed to soft jets, since more events with the leading jet below $90\GeV$ are expected from the prediction. The discrepancy is larger when the $\cPZ$
boson and the leading jet are well separated in rapidity. Indeed, the discrepancy is the smallest for the region $\abs{y(\cPZ)}\in(1,2.5)$ and $y(\cPZ)\, y(\text{j}_1) > 0$, corresponding to the region where the rapidities of the boson and the jet are the closest in the jet rapidity range considered.

\section{Summary}~\label{sec:summary}
The kinematics of \zjets events in $\Pp\Pp$ collisions at the centre-of-mass energy  of $\sqrt{s}=8\TeV$ have been studied and the differential cross sections have been measured as a function of numerous observables. Multidimensional cross section measurements have been performed with respect to up to three variables. The results have been compared with predictions from several multileg generators at different fixed-order accuracies, tree-level and NLO up to 2 partons, and employing
different showering algorithms, as implemented in \PYTHIA~6, \PYTHIA~8, and \SHERPA~2.

The comparisons show that it is essential to include a large number of final-state partons in the matrix element calculations in order to correctly describe the kinematics of the leading jets. Besides the individual jet \pt, the observable $\HT$, used in searches for physics beyond the SM and defined in this measurement for jets with $\pt>30\GeV$, is modelled correctly at low values of $\HT$ only when
a sufficiently large number of partons is included in the matrix element calculations. The discrepancies found for large values of the jet momentum, first observed in the $\sqrt{s}=7\TeV$ measurements~\cite{Khachatryan:2014zya,Aad:2013ysa}, are confirmed at $\sqrt{s}=8\TeV$ with a larger data set. Such discrepancies are not seen when including the NLO corrections. The differences observed between tree-level predictions and the measurements of the leading jet are larger when the jet is more forward
($\abs{y}>2.5$). Discrepancies with LO and NLO predictions have been observed for the dijet mass spectrum at low mass in the region where the angle between the two jet directions is smaller than $\pi/2$. Nevertheless, the azimuthal angles between the $\cPZ$ boson and the jet and between the jets are very well reproduced by the predictions including the tree-level one. The excellent agreement remains when restricting the phase space by applying a threshold on the $\cPZ$ boson \pt,
on $\HT$, or on both. The rapidity distributions of the $\cPZ$ boson and jets are fairly well modelled by the generators, but the correlations between the rapidities, which have been studied by measuring multidimensional differential cross sections and distributions of rapidity differences and sums, are not well reproduced by the multileg tree-level calculation. We have shown that the multileg event generators including NLO terms reproduce the rapidity difference distributions very well. The rapidity
sum is also successfully described. For this variable the discrepancy with the tree-level calculation could also be due to a different choice of the parton distribution functions.

In summary, kinematics of \zjets events have been studied in detail and apart from a few discrepancies, the measurements show a very good agreement with the considered NLO multileg predictions.

\clearpage
\begin{acknowledgments}
\hyphenation{Bundes-ministerium Forschungs-gemeinschaft Forschungs-zentren Rachada-pisek} We congratulate our colleagues in the CERN accelerator departments for the excellent performance of the LHC and thank the technical and administrative staffs at CERN and at other CMS institutes for their contributions to the success of the CMS effort. In addition, we gratefully acknowledge the computing centres and personnel of the Worldwide LHC Computing Grid for delivering so effectively the computing infrastructure essential to our analyses. Finally, we acknowledge the enduring support for the construction and operation of the LHC and the CMS detector provided by the following funding agencies: the Austrian Federal Ministry of Science, Research and Economy and the Austrian Science Fund; the Belgian Fonds de la Recherche Scientifique, and Fonds voor Wetenschappelijk Onderzoek; the Brazilian Funding Agencies (CNPq, CAPES, FAPERJ, and FAPESP); the Bulgarian Ministry of Education and Science; CERN; the Chinese Academy of Sciences, Ministry of Science and Technology, and National Natural Science Foundation of China; the Colombian Funding Agency (COLCIENCIAS); the Croatian Ministry of Science, Education and Sport, and the Croatian Science Foundation; the Research Promotion Foundation, Cyprus; the Secretariat for Higher Education, Science, Technology and Innovation, Ecuador; the Ministry of Education and Research, Estonian Research Council via IUT23-4 and IUT23-6 and European Regional Development Fund, Estonia; the Academy of Finland, Finnish Ministry of Education and Culture, and Helsinki Institute of Physics; the Institut National de Physique Nucl\'eaire et de Physique des Particules~/~CNRS, and Commissariat \`a l'\'Energie Atomique et aux \'Energies Alternatives~/~CEA, France; the Bundesministerium f\"ur Bildung und Forschung, Deutsche Forschungsgemeinschaft, and Helmholtz-Gemeinschaft Deutscher Forschungszentren, Germany; the General Secretariat for Research and Technology, Greece; the National Scientific Research Foundation, and National Innovation Office, Hungary; the Department of Atomic Energy and the Department of Science and Technology, India; the Institute for Studies in Theoretical Physics and Mathematics, Iran; the Science Foundation, Ireland; the Istituto Nazionale di Fisica Nucleare, Italy; the Ministry of Science, ICT and Future Planning, and National Research Foundation (NRF), Republic of Korea; the Lithuanian Academy of Sciences; the Ministry of Education, and University of Malaya (Malaysia); the Mexican Funding Agencies (BUAP, CINVESTAV, CONACYT, LNS, SEP, and UASLP-FAI); the Ministry of Business, Innovation and Employment, New Zealand; the Pakistan Atomic Energy Commission; the Ministry of Science and Higher Education and the National Science Centre, Poland; the Funda\c{c}\~ao para a Ci\^encia e a Tecnologia, Portugal; JINR, Dubna; the Ministry of Education and Science of the Russian Federation, the Federal Agency of Atomic Energy of the Russian Federation, Russian Academy of Sciences, and the Russian Foundation for Basic Research; the Ministry of Education, Science and Technological Development of Serbia; the Secretar\'{\i}a de Estado de Investigaci\'on, Desarrollo e Innovaci\'on and Programa Consolider-Ingenio 2010, Spain; the Swiss Funding Agencies (ETH Board, ETH Zurich, PSI, SNF, UniZH, Canton Zurich, and SER); the Ministry of Science and Technology, Taipei; the Thailand Center of Excellence in Physics, the Institute for the Promotion of Teaching Science and Technology of Thailand, Special Task Force for Activating Research and the National Science and Technology Development Agency of Thailand; the Scientific and Technical Research Council of Turkey, and Turkish Atomic Energy Authority; the National Academy of Sciences of Ukraine, and State Fund for Fundamental Researches, Ukraine; the Science and Technology Facilities Council, UK; the US Department of Energy, and the US National Science Foundation.

Individuals have received support from the Marie-Curie programme and the European Research Council and EPLANET (European Union); the Leventis Foundation; the A. P. Sloan Foundation; the Alexander von Humboldt Foundation; the Belgian Federal Science Policy Office; the Fonds pour la Formation \`a la Recherche dans l'Industrie et dans l'Agriculture (FRIA-Belgium); the Agentschap voor Innovatie door Wetenschap en Technologie (IWT-Belgium); the Ministry of Education, Youth and Sports (MEYS) of the Czech Republic; the Council of Science and Industrial Research, India; the HOMING PLUS programme of the Foundation for Polish Science, cofinanced from European Union, Regional Development Fund, the Mobility Plus programme of the Ministry of Science and Higher Education, the National Science Center (Poland), contracts Harmonia 2014/14/M/ST2/00428, Opus 2013/11/B/ST2/04202, 2014/13/B/ST2/02543 and 2014/15/B/ST2/03998, Sonata-bis 2012/07/E/ST2/01406; the Thalis and Aristeia programmes cofinanced by EU-ESF and the Greek NSRF; the National Priorities Research Program by Qatar National Research Fund; the Programa Clar\'in-COFUND del Principado de Asturias; the Rachadapisek Sompot Fund for Postdoctoral Fellowship, Chulalongkorn University and the Chulalongkorn Academic into Its 2nd Century Project Advancement Project (Thailand); and the Welch Foundation, contract C-1845.
\end{acknowledgments}
 \bibliography{auto_generated}

 \appendix

\cleardoublepage \appendix\section{The CMS Collaboration \label{app:collab}}\begin{sloppypar}\hyphenpenalty=5000\widowpenalty=500\clubpenalty=5000\textbf{Yerevan Physics Institute,  Yerevan,  Armenia}\\*[0pt]
V.~Khachatryan, A.M.~Sirunyan, A.~Tumasyan
\vskip\cmsinstskip
\textbf{Institut f\"{u}r Hochenergiephysik,  Wien,  Austria}\\*[0pt]
W.~Adam, E.~Asilar, T.~Bergauer, J.~Brandstetter, E.~Brondolin, M.~Dragicevic, J.~Er\"{o}, M.~Flechl, M.~Friedl, R.~Fr\"{u}hwirth\cmsAuthorMark{1}, V.M.~Ghete, C.~Hartl, N.~H\"{o}rmann, J.~Hrubec, M.~Jeitler\cmsAuthorMark{1}, A.~K\"{o}nig, I.~Kr\"{a}tschmer, D.~Liko, T.~Matsushita, I.~Mikulec, D.~Rabady, N.~Rad, B.~Rahbaran, H.~Rohringer, J.~Schieck\cmsAuthorMark{1}, J.~Strauss, W.~Treberer-Treberspurg, W.~Waltenberger, C.-E.~Wulz\cmsAuthorMark{1}
% ...
\vskip\cmsinstskip
% ...
1:~~Also at Vienna University of Technology, Vienna, Austria\\
% ...

\end{sloppypar}
\end{document}